\pdfoutput=1
\documentclass[preprint2]{proto}
\usepackage{times}
\usepackage{xcolor}
\usepackage{graphicx}
\usepackage{wasysym}
\usepackage{float}
\usepackage{dblfloatfix}

\graphicspath{{./}{Figures/}}

\def\apgt{\ {\raise-.5ex\hbox{$\buildrel>\over\sim$}}\ }
\def\aplt{\ {\raise-.5ex\hbox{$\buildrel<\over\sim$}}\ }

\def\third{{\textstyle{1\over3}}}

\begin{document}

\title{\textbf{\LARGE AMORPHOUS ICE IN COMETS: \\ EVIDENCE AND CONSEQUENCES}}

\author {\textbf{\large Dina Prialnik}}
\affil{\small\em Department of Geosciences, Tel Aviv University, Israel}

\author {\textbf{\large David Jewitt}}
\affil{\small\em Department of Earth, Planetary and Space Sciences, University of California at Los Angeles, USA }

\begin{abstract}

\begin{list}{ } {\rightmargin 1in}
\parindent=1pc
{\small 
Ice  naturally forms in the disordered or ``amorphous'' state when accreted from vapor at  temperatures and pressures found in the interstellar medium and in the frigid, low density outer regions of the Sun's protoplanetary disk.  It is therefore the expected form of ice in comets and other primitive bodies that have escaped substantial heating since formation.  Despite expectations, however, the observational evidence for amorphous ice in comets remains largely indirect.  This is both because the spectral features of amorphous ice are subtle and because the solar system objects for which we possess high quality  data are mostly too close to the Sun and too hot for amorphous ice to survive near the surface, where it can be detected.  This chapter reviews the properties of amorphous ice, the evidence for its existence and its consequences for the behavior of comets.
\\~\\~
}
\end{list}
\end{abstract}  

\section{\textbf{INTRODUCTION}}
\label{sec:intro}
Almost half a century ago, the idea that cometary ice might be amorphous and thus explain the ubiquitous comet outbursts was proposed by \cite{Patashnick1974} in a short {\textit{Nature}} paper: ``Observational evidence indicates that comet outbursts require an internal energy source. If at least the surface of a comet nucleus contains a substantial percentage of amorphous ice, then the phase transition of the amorphous ice to a cubic structure provides a release of energy which may be responsible for the outbursts observed in many comets." The idea was pursued a few years later by \cite{Smoluchowski1981}, who computed temperature profiles through the nucleus and even alluded to the possibility of self-propagation of the crystallization front. 

\begin{figure*}[h]
\begin{center}
\includegraphics[width=13cm]{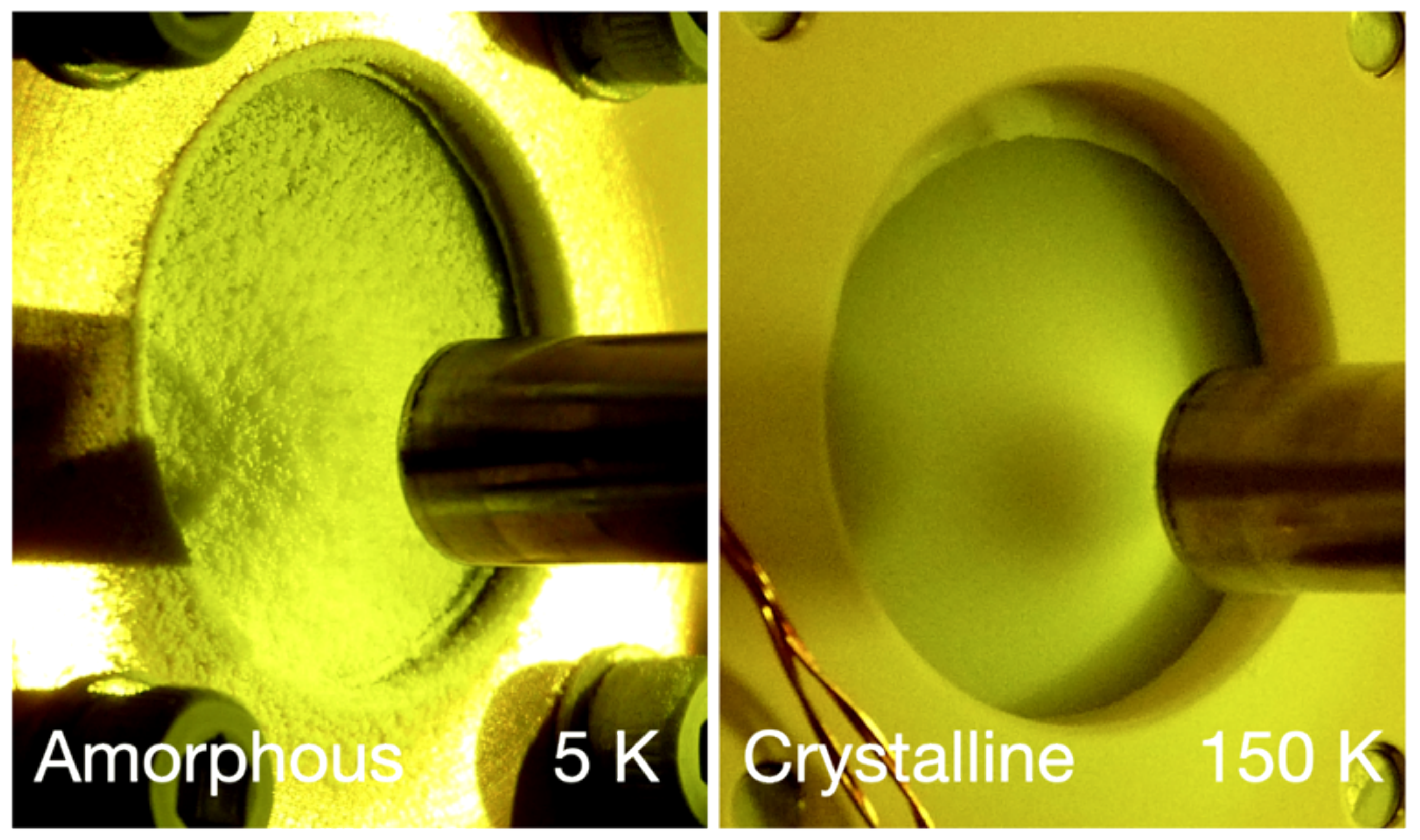}
\caption{Water ice samples prepared by vapor deposition onto a 2.5 cm diameter cold finger (left) at 5 K, where amorphous ice is the result and (right) at 150 K, where the ice is crystalline.   Note the granular surface of the amorphous ice indicating its fluffy nature relative to the much smoother surface of the crystalline ice.  Figure courtesy of Murthy Gudipati.}
\label{fig:murthy}
\end{center}
\end{figure*}

At the time, other volatiles observed in cometary comae were assumed to be included in the nucleus as ices or in the form of clathrates, although it was recognized that the formation of clathrates would require a much higher pressure upon formation than that prevailing in the solar nebula and that impurities could be occluded in limited amounts.
The first to consider gas trapping in amorphous water ice were \cite{BarNun1985}, by developing a dedicated experimental setup. The important result of their study was that gases could be trapped in large amounts and that they were released from the ice upon crystallization, typically at a temperature of $\sim140$~K, too cold for crystalline water ice sublimation, meaning that crystallization in comets may trigger activity at large heliocentric distances, {\textbf where solar radiation is too weak}.
This conclusion was soon confirmed by theoretical studies \citep{Prialnik1987}.  Since then,  numerous studies have addressed experimental, observational, and theoretical evidence concerning amorphous ice, as we shall briefly review in this chapter.  Related reviews  include those by \cite{Mastrapa13} concerning amorphous ice, by \cite{Gudipati15}, focused more generally on laboratory studies of ice and by Guilbert-Lepoutre et al. (in this volume) regarding the structure of cometary nuclei.

\subsection{What is amorphous ice?}
\label{ssec:whatis}

Amorphous water ice is a metastable form of ice, produced when ice is deposited at very low temperatures. The individual molecules have insufficient energy (mobility) to reorient themselves into more energetically favorable positions and a highly disordered solid is formed. Upon warming, the molecules rearrange themselves into lower energy orientations and somewhat more ordered structures are produced. This process is called annealing and it ends when a fully ordered crystalline lattice emerges. Macroscopic samples of amorphous and crystalline ice are shown in Fig.~\ref{fig:murthy}. 

Amorphous ice may include a large fraction of impurities by trapping them in the structureless and porous matrix. When it undergoes crystallization, these impurities are in part expelled from the crystal lattice, and in part, retained, more tightly bound.
Crystalline ice may occlude at most one foreign molecule per six H$_2$O molecules \citep{Whipple1976,Gudipati2015}, to form clathrate-hydrates. Trapping processes will be discussed in Section~\ref{sec:properties}.
\subsection{Is cometary ice amorphous?}
\label{ssec:isitamorphous}

There are several reasons why it is reasonable to assume that cometary ice is amorphous, or at least was amorphous when the nucleus formed. First, interstellar ice grains---the building blocks of comet nuclei---are made of amorphous ices \citep{Oberg2011, VanDishoeck2013}. Secondly, comets eject large amounts of volatiles of various species, and amorphous ice is more capable of trapping larger quantities of impurities than crystalline ice forming clathrates. Finally, comets are small porous bodies that have spent most of the last 4.5 Gyr in cold storage in the Kuiper belt and Oort cloud. Except perhaps for their surface layers, they have experienced little thermal or structural alteration and hence could have preserved their primordial composition.  \cite{AHearn2008}, for example, argue based on the {\textit{Deep Impact}} mission results, that pristine material may be found below a m-thick surface layer of short-period comets, indicating very low internal temperatures \citep[see also][]{Herman1987}. We return to this point in Section~\ref{ssec:early}.

Experiments carried out by \cite{Sanford1988} with H$_2$O and CO mixtures deposited at a low temperature and pressure have shown by infrared spectra that the ice formed is amorphous in structure, rather than crystalline or clathrate.
Furthermore, \cite{Ciesla14} has shown that even if ice condensed in the solar nebula at sufficiently high temperatures for a crystalline structure to form, dynamical evolution of the icy grains in the outer part of the nebula would cause this ice to be lost and reformed in the amorphous state.

Nevertheless, there is still some controversy in the literature regarding the question whether cometary ice is amorphous or crystalline and several studies have argued that the amorphous form of ice may not be a necessary prerequisite \citep{Marboeuf2012}. This point is thoroughly reviewed by \cite{Gudipati15}; 
we will return to it in the conclusions of Section~\ref{sec:outbursts}.
Observational evidence for amorphous ice in comets and related objects will be described in Section~\ref{sec:observational}. Consequences of the presence of amorphous ice in comet nuclei concerning cometary behavior will be considered in Section~\ref{sec:outbursts}, and prospects for the future, in Section~\ref{sec:conclusions}.  


\section{\textbf{PROPERTIES OF AMORPHOUS ICE -- \\ FROM LABORATORY EXPERIMENTS}}
\label{sec:properties}

\subsection{Volatile trapping and release}
\label{ssec:crystal}

\begin{figure*}[ht!]
\begin{center}
\includegraphics[width=17.0cm]{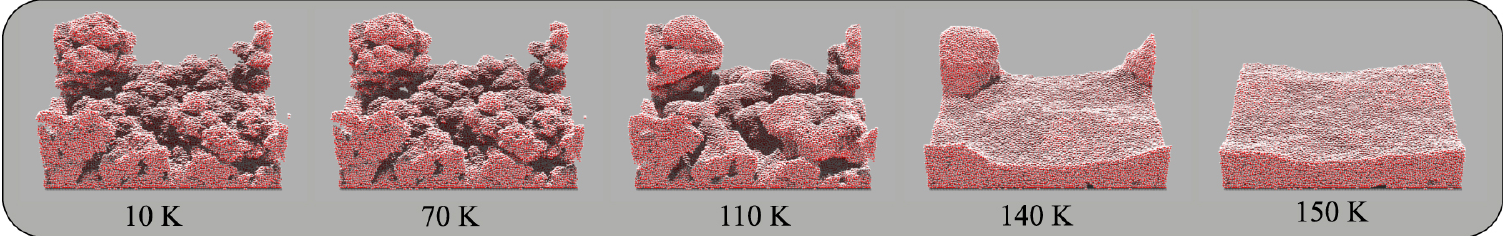}
\caption{The microstructure of amorphous water ice (25 monolayers thickness); the column-like structure at the lower temperatures becomes smoother with increasing temperature, until eventually an entirely smooth structure is obtained at 140–150 K. From \cite{He2019}.}
\label{fig:He19}
\end{center}
\end{figure*}

A distinguishing characteristic of amorphous ice is its large surface area per unit mass (c.f.~Fig.~\ref{fig:He19}), giving it the ability to trap substantial quantities of other gases present upon its formation.  This has been experimentally demonstrated by forming amorphous ice at low temperatures and pressures in the presence of other gases, then measuring the progressive {\textbf expulsion} of these gases as the sample temperature is raised \citep{BarNun1985, BarNun1987, BarNun1988, Notesco03}. These experiments reveal ice areas per unit mass from $\sim$90 m$^2$ g$^{-1}$ to $\sim$300 m$^2$ g$^{-1}$ \citep{Schmitt87, Yokochi12}, and even as large as $\sim$400 m$^2$ g$^{-1}$ \citep{Mayer1986}, and densities of 0.6-0.7 g cm$^{- 3}$ \citep{BarNun1985, Sanford1988, Kouchi2016}.

Very large trapping efficiencies have been reported, particularly at low temperatures of relevance to the formation of comets.  For example, \cite{BarNun1989} started at $T<30$~K with 1:1 CO:water ratios in gas, and so much CO was trapped in the amorphous ice formed, that in the emitted gas, CO/H$_2$O $>$1. Trapping in amorphous ice might be particularly important for the noble gases which, being chemically unreactive, are difficult to retain otherwise \citep[e.g.,][]{BarNun2007,Ciesla18}.  However,  laboratory measurements are taken under physical conditions quite different from those prevailing in the interstellar medium and the outer regions of the protoplanetary disk where cometary ice likely accreted.  Therefore, some consideration of the nature of the trapping mechanism is appropriate.  
\begin{figure}[tb]
\begin{center}
\includegraphics[width=8.5cm]{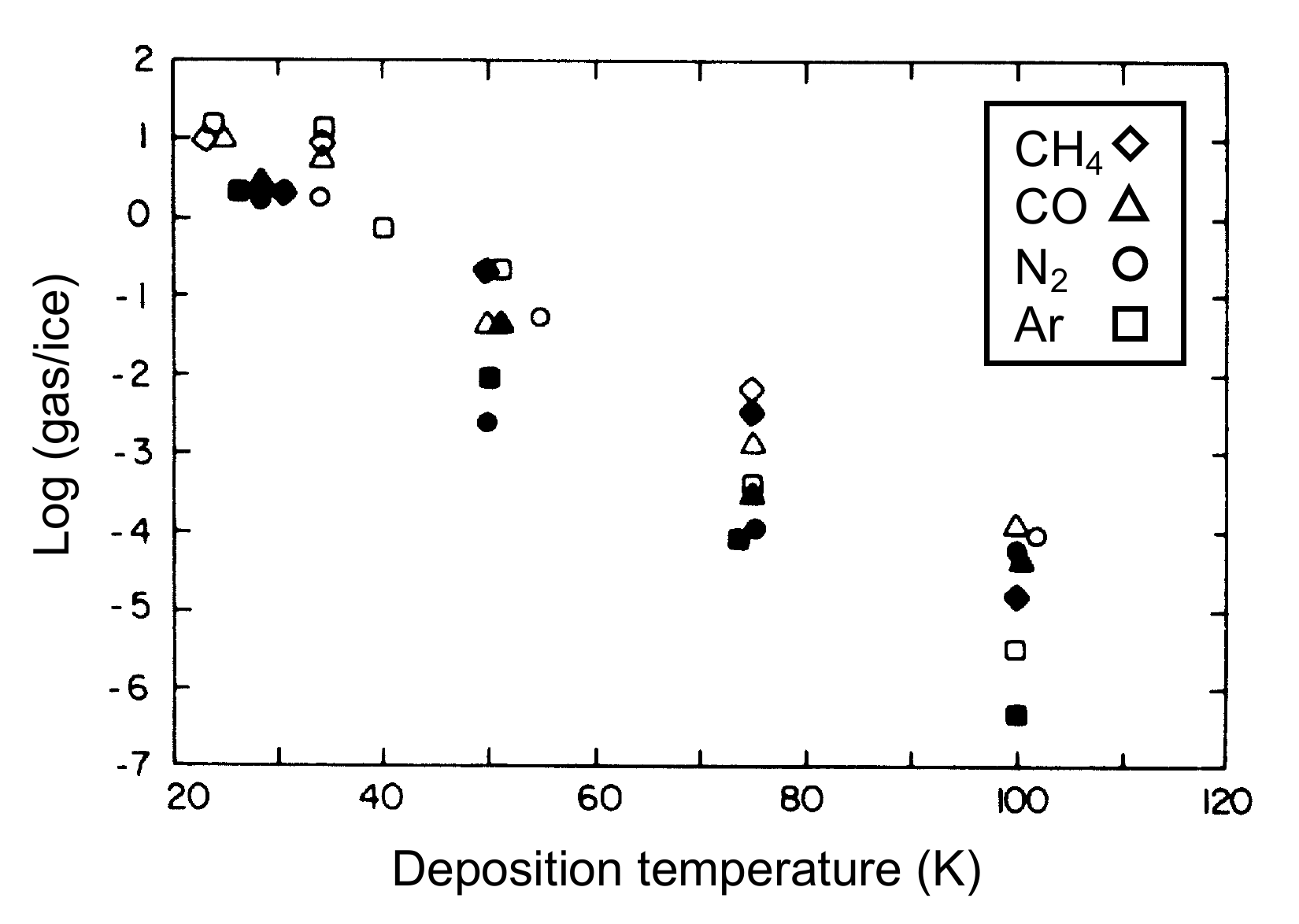}
\caption{Total amount of trapped gas vs. deposition temperature for water vapor/gas mixtures: open symbols -- deposition of a 1:1 ratio of a single gas species; filled symbols -- deposition of gas mixtures, H$_2$O:CH$_4$:Ar:CO(or N$_2$) = 1:0.33:0.33:0.33. Adapted from \cite{BarNun1988}}.
\label{fig:trapp}
\end{center}
\end{figure}

Two distinct modes of gas trapping in ice  are potentially important.  First, as described by \cite{Kouchi1992}, ice forms in the amorphous state when its hop length across the surface is small compared to the spacing of the ice lattice.  Equivalently, molecules are adsorbed on the ice surface at the point of impact and held for a residence time  $t_{r} = \nu^{-1} \exp(E/(kT))$, where $\nu \sim10^{12}$ s$^{-1}$ is the vibrational frequency of the molecule in the surface potential well that holds it, $E$ is the binding energy and $k$ is Boltzmann's constant. Adsorption thus strongly favors low temperatures, $T$. For example, for Argon on water ice, $E = 1.4\times10^{-20}$~J \citep{Ciesla18}, giving $t_{r} = 420$~s at $T=30$~K and falling by seven orders of magnitude to only $t_{r} = 2\times10^{-5}$~s when doubling the temperature to $T=60$~K. 
Secondly, molecules can also be physically trapped by burial under later-arriving monolayers of ice. These two modes of gas trapping have been investigated by \cite{Martin2002}. Whereas adsorption depends exponentially on the temperature of the sticking surface,  burial depends as well on the density-dependent rate of growth of the ice, which determines how long an enveloping monolayer takes to accrete.   
The efficiency of gas trapping in amorphous ice is therefore a function of both temperature and gas density. The strong dependence on temperature in one set of experiments is illustrated in Fig.~\ref{fig:trapp}. 

While a detailed balance treatment is needed to properly account for the impact and escape of water molecules at the growing ice surface, we can obtain a rough estimate from $t_r$, above. If we take, say, $\ell \sim$ 1\AA~as the order of magnitude scale of a molecule (the van der Waals radius of H$_2$O is 1.7\AA), then burial is possible if the ice accretes at a rate $\gg \ell/t_r$. For example, Argon at 30 K can be buried if the ice accretes at rates $\ell/t_r \gg 10^{-5}$ $\mu$m min$^{-1}$. The laboratory experiments by Bar-Nun and collaborators were conducted at astrophysically relevant temperatures as low as 10 K but at ice growth rates $10^{-5}$ to $10^{-1}$ $\mu$m min$^{-1}$, orders of magnitude larger than likely to be found in nature \citep{Cuppen2007}. \cite{Yokochi12} found that the trapping efficiency depends on the temperature and the partial pressure of the trapped species and suggested that the results of  \cite{BarNun1989} could not be easily interpreted because of experimental concerns.  In addition, we should expect strong abundance gradients with respect to the formation distance in the protoplanetary disk, since both temperature and pressure vary radially. High temperature in the inner disk and low pressure in the outer disk suggest the existence of a critical distance at which gas trapping efficiencies by amorphous ice should be maximized. 
\begin{figure}[tb]
\begin{center}
\includegraphics[width=8.5cm]{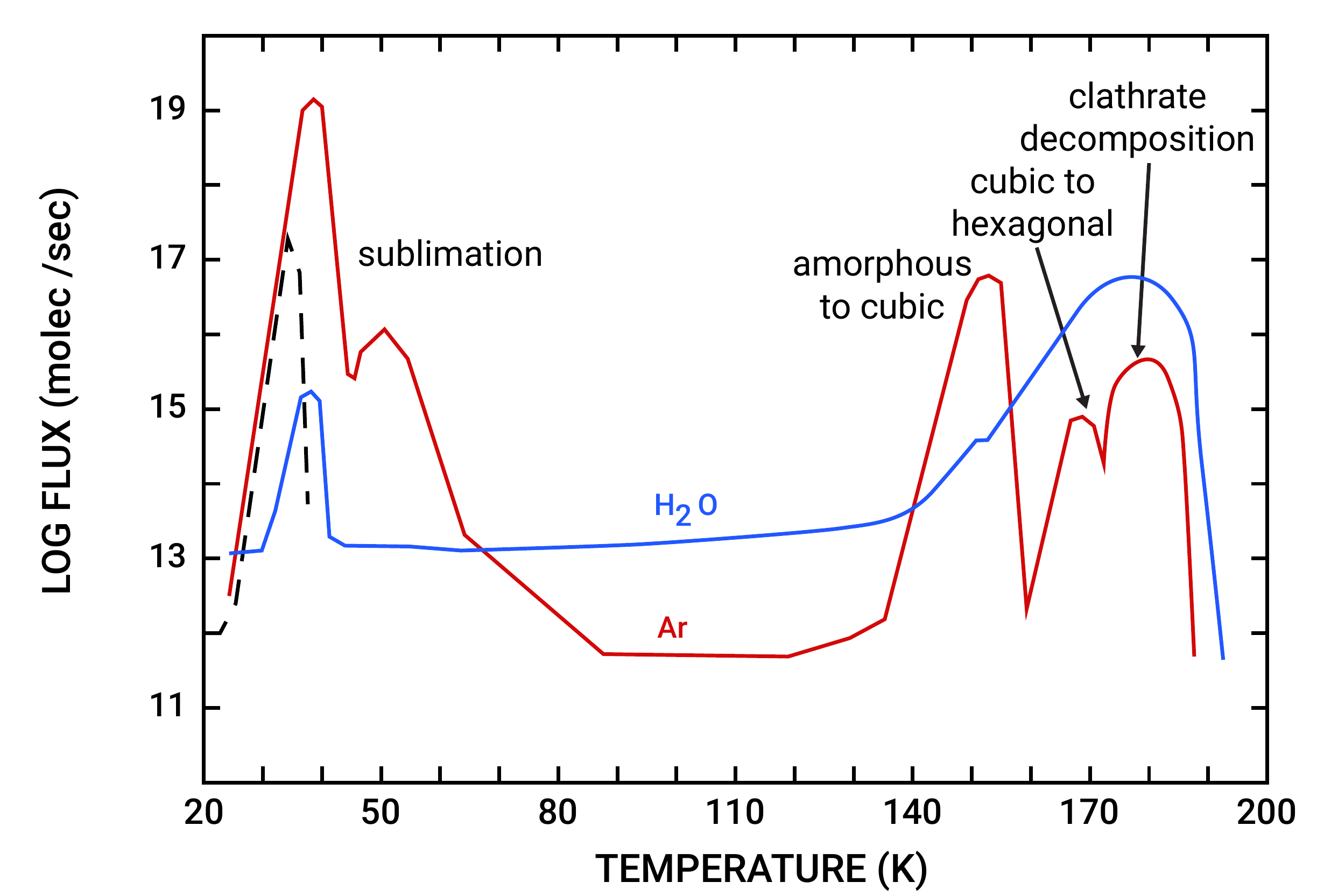}
\caption{Fluxes of evolved argon and water vs. temperature in various ranges. Dashed line: sublimation of frozen argon from an ice free plate, shown for comparison. Adapted from \cite{BarNun1987}.}
\label{fig:release}
\end{center}
\end{figure}

Experimentally, gas release from amorphous ice occurs at a strongly temperature-dependent but non-monotonic rate. The ice transforms into the cubic form at $\sim 140$~K and into the hexagonal form at 160~K \citep{Jenniskens94}, with each phase transition leading to a pulse of expelled gases. When molecules such as CO, CH$_4$, N$_2$, and Ar are trapped in the ice below 30~K,  they are released  upon warming in several temperature ranges (see Fig.~\ref{fig:release}), starting at 23~K, where the gas frozen on the surface evaporates, to 44~K, where the monolayer of adsorbed gas evaporates. At $\sim60$~K, gas release ceases altogether, as the remaining gas is locked in an impermeable amorphous ice matrix. It resumes only at 140~K, where the ice becomes temporarily more mobile during its transformation into the cubic form, but ceases when the transformation is completed.

The next chance for gas to escape occurs at 160~K, where the cubic ice transforms into a hexagonal crystal. The remaining trapped volatile forms a clathrate \citep{Sandford1990, Blake1991, Marboeuf2012}. 
When water ice undergoes crystallization, it is justified to assume
that the rate of clathrate formation is equal to that of crystallization \citep{Marboeuf2012}, because the mobility of the water molecules is high and the amount of volatile molecules released from the amorphous ice phase is sufficient to form cages in the crystalline phase.
Finally, at $\sim$180~K water vapor and gas evolve simultaneously, when the
clathrate evaporates. If the gas is trapped in the ice between 40~K and 80~K, the peaks between 140--180~K diminish by orders of magnitude, depending on the formation temperature and on the trapped species.
\subsection{Formation of hydrocarbon aggregates}
\label{ssub:hydrocarbons}

Besides volatiles, amorphous ice may also trap large hydrocarbon molecules. \cite{Lignell2015} used C$_{16}$H$_{10}$ (four, fused carbon rings, known as pyrene) to show that upon crystallization the hydrocarbons  form aggregates that remain trapped in the crystalline ice. Subsequent heating leads to emission of the aggregates. Experiments with even more massive C$_{60}$ (Buckminsterfullerene) show similar aggregation upon crystallization \citep{Hakukeerthi20}. In comets, the mobilization and aggregation of hydrocarbons at the phase transition may enhance the formation of an outer crust, when the organic molecules become mixed with silicate particles.  Chemical processing through micrometeorite impact heating \citep{Nelson16} adds to the complexity of this process.

\subsection{The latent heat of crystallization}
\label{ssec:latent}

The crystallization of amorphous ice that takes place upon heating, first into a cubic structure and then into a hexagonal lattice, is an irreversible process. The timescale for  crystallization, $\tau_{CR}$, is strongly temperature-dependent; it was determined experimentally by \cite{Schmitt1989b},

\begin{equation}
 \tau_{CR}(T)=9.7\times10^{-14}\exp^{5370/T} {\rm [s]}.  
 \label{lambda}
\end{equation}

When pure ice is involved, crystallization is clearly an exothermic process, as the entropy decreases. The first and widely used estimate of the latent heat released, $H_a=90$~kJ/kg, was obtained experimentally by \cite{Ghormley1968} based on temperature measurements (warming curves). A somewhat higher value, $118$kJ/kg, obtained by comparing the mutual binding energies of H$_2$O molecules in amorphous and crystalline ice, was derived experimentally by \cite{Sanford1988, Sandford1990}, based on measurements of release rates of occluded gases in amorphous ice samples (sticking coefficients). 

The question whether the process remains exothermic when impurities are occluded in the amorphous ice has been debated for many years, particularly since the experimental results of \cite{Kouchi2001}, who claimed that the process becomes endothermic if even only 3\% of occluded CO is released, an extreme conclusion that has been regarded with caution.

The basic question is whether the occluded gas is just trapped in cages of the highly porous amorphous ice structure, or else it is bound to water molecules. In the former case, there would be a small amount of energy release when the gas escapes, resulting from the difference in the specific internal energies of the gas and the amorphous ice. For CO, for example, it would amount to $\sim18f_{{\rm CO},t}$~kJ/kg, where $f_{{\rm CO},t}$ is the fraction of trapped CO. Many models of comet nuclei \citep[e.g.][]{Prialnik1990, Espinasse1991, Tancredi1994} have adopted this approach.

There is experimental evidence, however, that the impurities are bound to water molecules. Here we have to distinguish between the fraction of gas that remains trapped in the annealed ice and the fraction of escaping gas. Accordingly, in the case of CO, the energy absorbed in gas release should be proportional to $f_{{\rm CO},r}$, the fraction of released CO, and to the binding energy of CO on H$_2$O, $H_{a,r}=517$~kJ/kg \citep{Sandford1990} or 418~kJ/kg, as obtained by \cite{Manca2001} from quantum calculations for the absorption of small molecules on ice. At the same time, latent heat should be released due to the increased binding energy of the remaining CO. According to the experimental results of \cite{Sanford1988}, the difference between the volume binding energies of CO on H$_2$O in amorphous and crystalline ice yields a latent heat of $H_{a,c}\approx74f_{{\rm CO},c}$~kJ/kg, where $f_{{\rm CO},c}$, is the fraction of CO that remains trapped when the ice crystallizes (keeping in mind the large error bars).
As a simple, illustrative example, if 10\% of CO (by mass) is trapped in amorphous ice, half of which remains trapped when the ice crystallizes, while the other half escapes - in agreement with Fig.~\ref{fig:release} - the process will still be highly exothermic. The heat released 
\begin{equation}
    H=[1-(f_c+f_r)]H_a+f_cH_{a,c}-f_rH_{a,r}
\end{equation}
will be in this case about 60~kJ/kg. Some studies \citep[e.g.][]{Davidsson2021, Enzian1997} adopt the ad-hoc assumption that the amount of energy absorbed in gas release is equal to the latent heat of sublimation of the trapped species (for CO, $285f_{{\rm CO},t}$~kJ/kg), which results in a moderate reduction of the latent heat of pure ice crystallization.
\cite{Gonzalez2008} consider the effect of different values of latent heat.
The escape of CO (and other species) may be a combination of diffusion through the ice and removal from the ice surface \citep{Sanford1988}, which are energetically different, and may depend on the formation process of the mixture. 

In conclusion, reliable quantitative data for the heat released in crystallization of gas-laden amorphous ice are still needed from laboratory experiments. Nevertheless, for the typical amounts of trapped volatiles in amorphous ice that are released upon crystallization, it is safe to assume the process to be exothermic and to play a significant role in cometary outbursts, as we shall describe in Section~\ref{sec:outbursts}.

\subsection{Thermal properties of amorphous ice}
\label{ssec:thermal}

The thermal conductivity of amorphous ice has been determined both experimentally and from theoretical considerations.  The formulae provided by \cite{Klinger1980} and \cite{Klinger1981} have gained 
widespread use. \cite{Klinger1980} derived a theoretical expression from the classical phonon theory
\begin{equation}
K = \third v_s \ell_{ph} c \rho \ , 
\label{eq:conductivity}
\end{equation}
where $v_s = 2.5 \times 10^3$~m/s is the speed of sound in ice and
$\ell_{{\rm ph}} = 5 \times 10^{-10}$~m is the phonon mean free path, $\rho$ is the ice density, and $c$ is the heat capacity, given by an empirical relation $c=7.49T+90$~J~kg$^{-1}$~K$^{-1}$, derived from measurements \citep{Giauque1936}.  Substitution gives $K\approx 0.4$~W~m$^{-1}$~K$^{-1}$ in the temperature range $70\le T \le 135$~K. 
\cite{Kouchi1992} measured in their laboratory experiments a much lower conductivity of $K = (0.6 \textrm{~to~} 4.1) \times 10^{-5}$~W~m$^{-1}$~K$^{-1}$ in the range $125 \le T \le 135$~K. 
\cite{Andersson1994, Andersson2002} found an experimental value for the conductivity 
of nonporous, low density, amorphous water ice that is similar to the value derived by Klinger.
In a unique experiment with a relatively large sample of fluffy amorphous ice \cite{BarNun2003} measured an intermediate value between those of Klinger and Kouchi. This may suggest that differences between experimentally derived thermal conductivity coefficients may be caused by different porosity of the samples.

While additional measurements are clearly needed, the thermal conductivity of amorphous ice is one or more orders of magnitude smaller than that of crystalline ice (cubic or hexagonal). This has significant consequences for the thermal evolution of comet nuclei, since the thermal timescale is inversely proportional to $K$ and the skin depth is proportional to $\sqrt{K}$. 

\section{\textbf{OBSERVATIONAL EVIDENCE}}
\label{sec:observational}

\subsection{Stability}
\label{ssec:stability}

The radiation equilibrium temperature of a solar system body depends on the distance from the Sun, as well as optical properties (albedo, emissivity), thermodynamic properties (thermal conductivity, density and specific heat capacity) and also on the rotational properties (rotation period, spin axis orientation).  For most objects, these parameters are unknown or poorly constrained, and the surface temperature cannot be accurately defined.   As a guide, however, we consider $T_{\rm iso}$, the temperature of a spherical, isothermal body as a lower limit to the dayside temperature and $T_{SS}$, the temperature of a  flat plate oriented normal to the Sun, as an upper limit. 

Equating the power absorbed from the Sun to the power radiated in equilibrium, we have 

\begin{equation}
  T_{\rm iso} =   \left(\frac{L_{\odot}(1-A)}{16 \pi \epsilon \sigma r_H^2}\right)^{1/4}
  \label{tbb}
\end{equation}

\noindent and
\begin{equation}
  T_{SS} =   \sqrt{2} T_{\rm iso}
  \label{tss}
\end{equation}

\noindent in which $L_{\odot} = 3.83\times10^{26}$ W is the luminosity of the Sun, $\sigma$
is the Stefan-Boltzmann constant, $A$ and $\epsilon$ are the Bond albedo and emissivity, respectively, and $r_H$ is the heliocentric distance. Substituting $A$ = 0, $\epsilon$ = 1 into Eqs.~(\ref{tbb}) and (\ref{tss}) and expressing $r_H$ in AU (1 AU = 1.5$\times10^{11}$ m) gives blackbody temperatures (in Kelvin),

\begin{equation}
    T_{BB} = \frac{278}{\sqrt{r_H}}~~\textrm{ and }~T_{SS} = \frac{393}{\sqrt{r_H}}.
    \label{tlimits}
\end{equation}


\begin{figure}[tbp]
\begin{center}
\includegraphics[width=7cm]{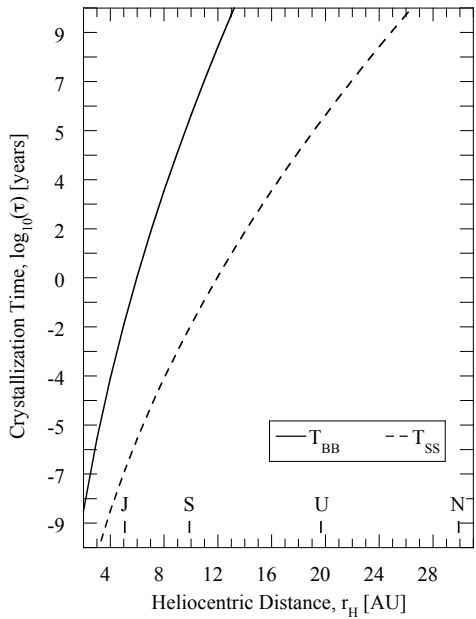}
\caption{Logarithm of the crystallization timescale (from Eq.~(\ref{lambda}) as a function of heliocentric distance for high ($T_{SS}$) and low ($T_{BB}$) temperature limits given by Eq.~(\ref{tlimits}).  The orbital radii of the giant planets are marked.}
\label{fig:tau}
\end{center}
\end{figure}

\noindent The crystallization time, $\tau_{CR}$,  is shown as a function of heliocentric distance for $T_{BB}$ and $T_{SS}$ in Fig.~\ref{fig:tau}.  The figure shows (a) a substantial range of crystallization timescales at any heliocentric distance caused by the exponential temperature dependence in Eq.~(\ref{lambda}),  and (b)  temperatures at Kuiper belt distances ($r_H \ge$ 30~AU)  are too low for crystallization to have occurred in the $4.5\times10^9$ year age of the solar system, no matter which temperature model is used.  On this basis, we should reasonably expect that the Kuiper belt objects, at least  those small enough to have escaped self-heating due to the energy of formation and to radioactive decay, could retain amorphous ice. The nuclei of comets, being derived from the Kuiper belt and the (even colder) Oort cloud reservoirs, are also expected to contain amorphous ice, subject to uncertainties about the temperature-time histories of these bodies. 

\subsection{Diagnostics}

For practical reasons, most spectroscopy of icy bodies has been done at optical and near infrared (1 to 2.5 $\mu$m) wavelengths.  The vibrational and overtone bands of water and other simple molecules are best accessed in the near infrared, but differences between amorphous and crystalline ice in this spectral region are, for the most part, subtle.  They occur because, while the vibrational and overtone bands due to the OH bond in water molecules are substantially the same, smaller differences exist in the degree of hydrogen bonding between water molecules. Unfortunately, many of the important differences occur in regions of the electromagnetic spectrum which are traditionally difficult to access.  This is because the  bands  in water ice largely overlap the corresponding features from water vapor in the atmosphere, rendering ground-based study difficult.  The optically thick 3 $\mu$m band is strongly affected, and can be studied only from the most stable and dry high altitude sites or, better, from space.  A weaker crystalline ice band at 1.65~$\mu$m falls in an atmospherically transparent region, and  is more frequently used as a convenient diagnostic.    

\begin{figure}[htb!]
\begin{center}
\includegraphics[width=7cm]{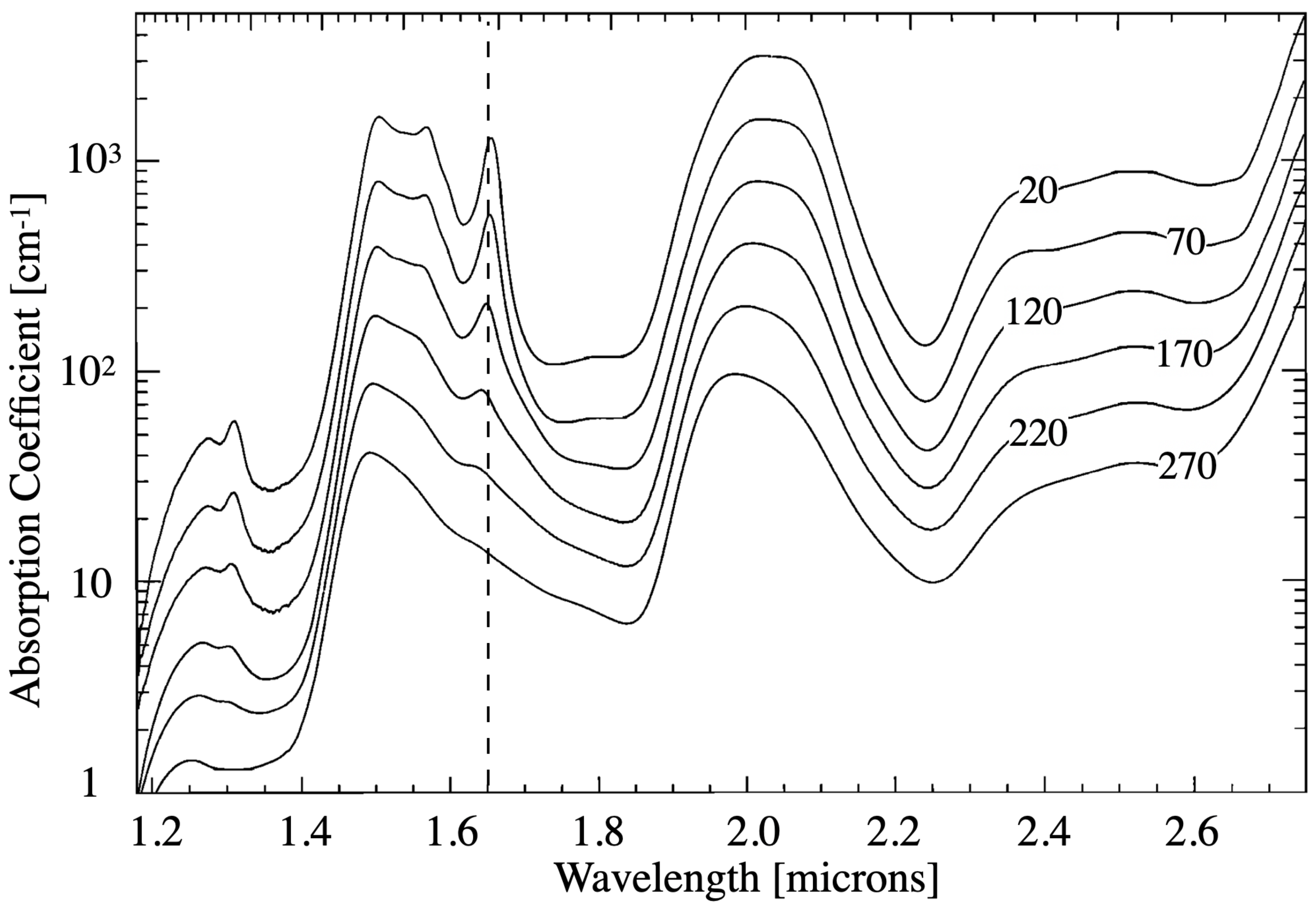}
\caption{Temperature dependence of the absorption coefficient in crystalline water ice from 20~K to 270~K, as labeled.  Note the strong changes in the 1.65~$\mu$m band, marked by a vertical dashed line.  The curves are vertically displaced for clarity.  Modified from \cite{Grundy98}.}
\label{grundy_schmitt}
\end{center}
\end{figure}

\begin{figure}[htb!]
\begin{center}
\includegraphics[width=7cm]{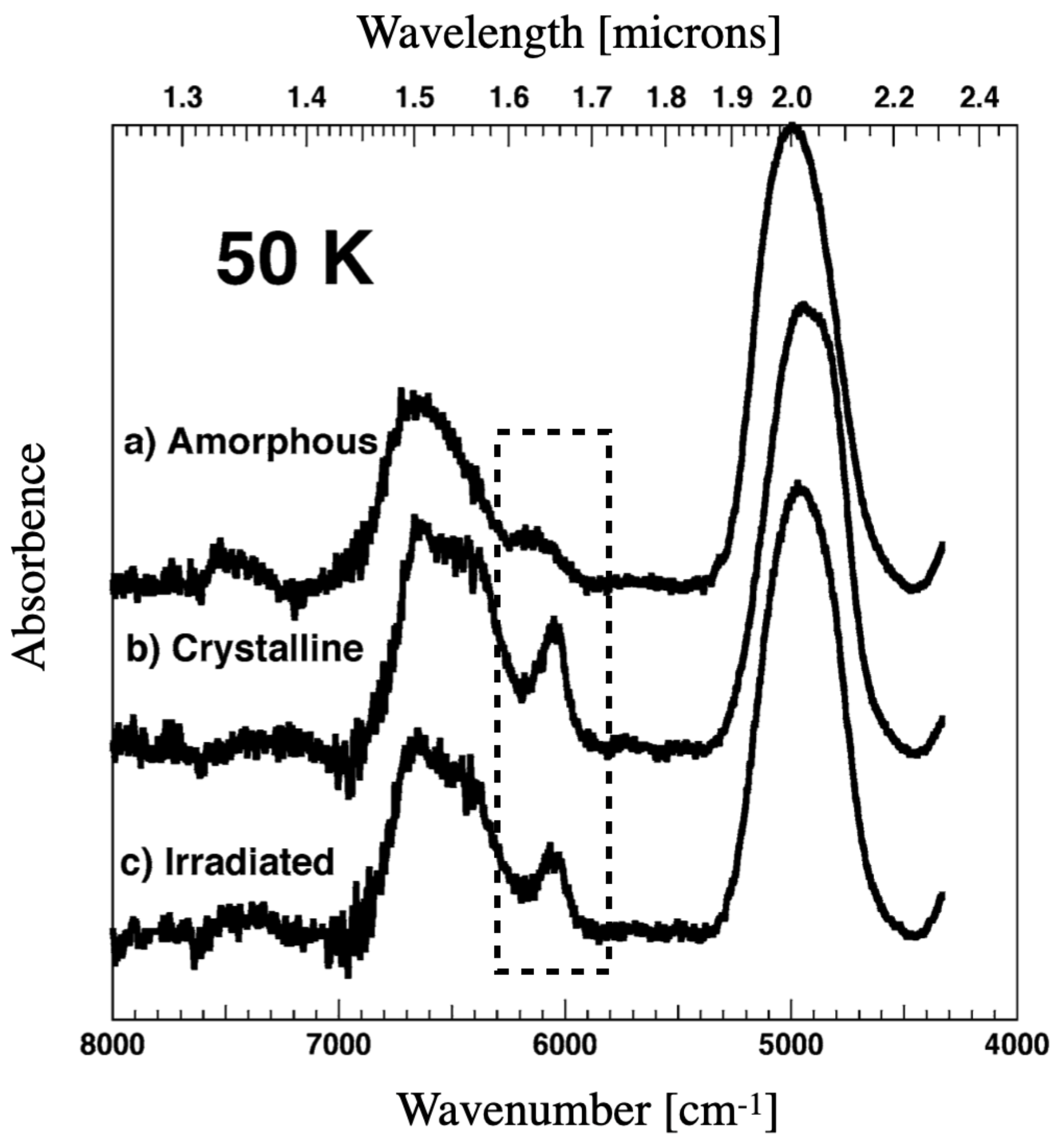}
\caption{Three near-IR spectra showing a) amorphous ice grown at 50~K b) crystalline ice obtained by heating the sample to 160~K and then cooling to 50 K and c) the crystallized sample after irradiation by 1~MeV protons to 16~eV per molecule.  The dotted box highlights variations in the 1.65~$\mu$m band. Modified from \cite{Mastrapa06}.}
\label{mastrapa}
\end{center}
\end{figure}

\begin{figure}[htb!]
\begin{center}
\includegraphics[width=7.25cm]{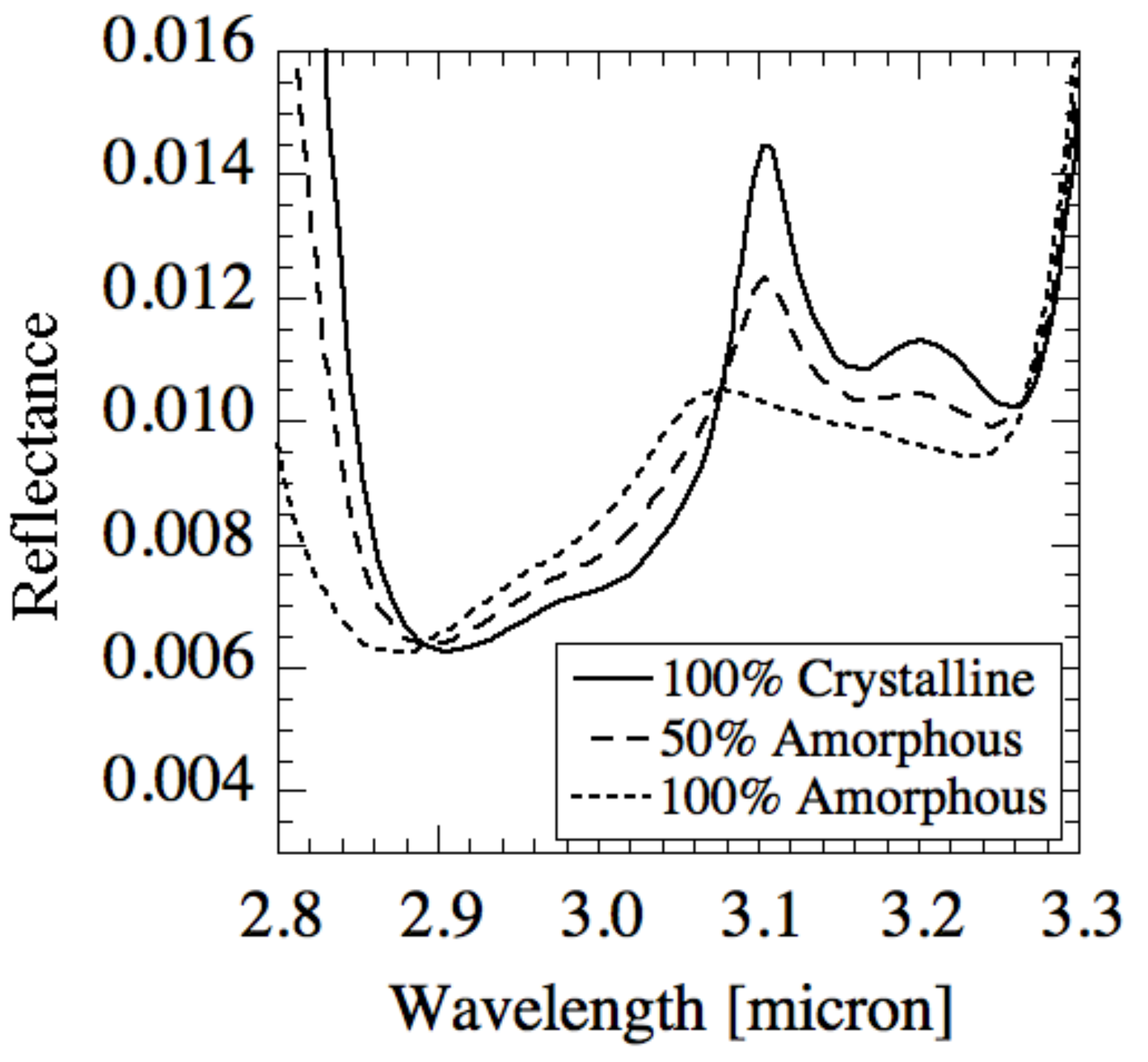}
\caption{Three micron spectra showing the sensitivity to amorphous vs.~crystalline structure in ice at 100 K, with a 20 $\mu$m grain size. Modified from \cite{Mastrapa13}.}
\label{mastrapa_3micron}
\end{center}
\end{figure}

Use of this feature is not without problems, however.  The band center, width and depth are all functions of the ice temperature as well as the crystallinity \citep{Fink75, Grundy98} and have, indeed, been used as a spectroscopic thermometer e.g.~\cite{Grundy99}.  The band  is half as deep at 100~K as it is at 20~K (Fig.~\ref{grundy_schmitt}) meaning that, without additional information, a weak 1.65~$\mu$m band could be due to the presence of amorphous ice or simply a result of higher temperatures.  Radiative transfer (scattering and absorption)  in granular ice can influence the band  parameters depending on the grain size \citep{Hansen04}, as can energetic particle irradiation \citep{Mastrapa06}.  Two of these effects are shown in Fig.~\ref{mastrapa}, where (a) shows the spectrum of amorphous ice deposited at 50~K, (b) shows the effect of heating the ice to 160~K (to crystallize it) then cooling back down to 50~K and (c) shows the additional effect on the same sample of irradiation by 1~MeV protons up to 16~eV/molecule.  The 1.65~$\mu$m band is clearly associated with the crystalline phase and diminished, but not erased, by bond destruction due to particle irradiation in spectrum (c).  The degree to which the 1.65~$\mu$m band is weakened by irradiation varies inversely with the ice temperature.  

Fig.~\ref{mastrapa_3micron} plots the reflectivity of granular ice in the  3 $\mu$m region \citep{Mastrapa13}.  This extremely deep absorption feature shows crystallinity-dependent band shifts and reflectivity differences (e.g.~especially near the 3.1 $\mu$m Fresnel peak) that will be of practical value once space-based observations at this wavelength are routine.

\subsection{Icy satellites}
\label{ssec:satellites}

The  Galilean satellites Europa, Ganymede and Callisto are  natural places to examine the effects of irradiation on ice crystal structure \citep{Hansen04}. All three have surfaces rich in water ice and orbit deep within Jupiter's magnetosphere, with orbital radii  9.4~$R_J$, 15.0~$R_J$ and 26.4~$R_J$, respectively, where $R_J = 7.14\times10^7$~m is the radius of Jupiter. The Jovian magnetosphere rotates with the planet at about 10 hours, which is short compared to the 3.6, 7.2 and 21.6 day orbital periods of the satellites. As a result, magnetospheric particles preferentially impact the trailing hemispheres of each satellite, leaving the leading hemispheres relatively unirradiated.  Furthermore, because the satellites rotate synchronously, this hemispherically asymmetric bombardment  is imprinted as a fixed longitudinal pattern on each satellite.

\begin{figure*}[htb!]
\begin{center}
\includegraphics[width=16.5cm]{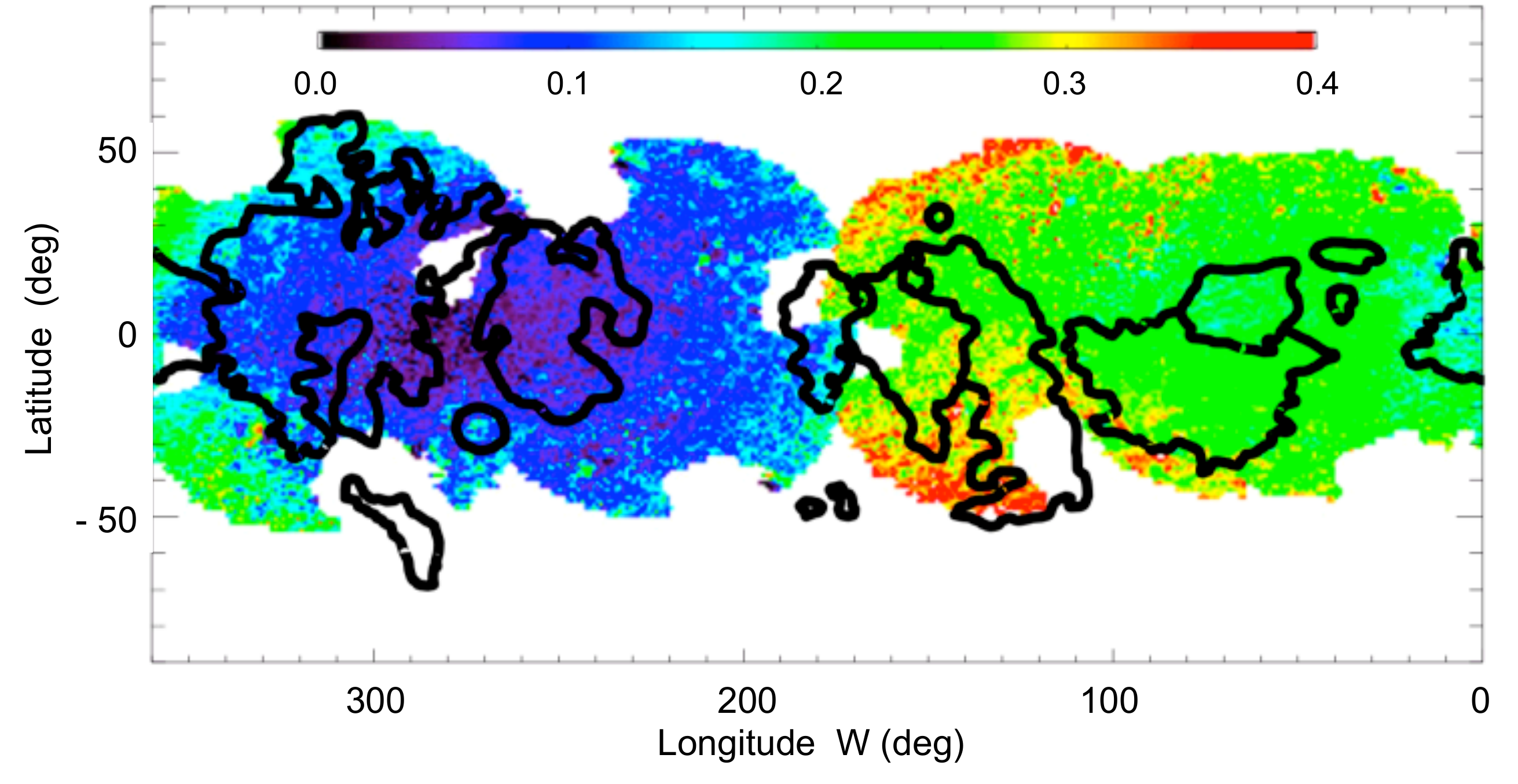}
\caption{Map of the water ice crystallinity on Europa, largely deduced from observations in the 1.65 $\mu$m band.  The amorphous fraction is color coded from black (0) to red (0.4); the ice on the trailing hemisphere (longitudes 0\degr~to 180\degr) is more amorphous than on the  leading hemisphere (180\degr~to 360\degr).  Irregular shapes outlined in black lines show geological surface units, with which the amorphous/crystalline pattern shows no simple correlation. Adapted from \cite{ligier16}.}
\label{ligier_europa2}
\end{center}
\end{figure*}

Fig.~\ref{ligier_europa2} shows the distribution of amorphous ice on the surface of Europa \citep{ligier16}; c.f.~\cite{Berdis20}.  The map shows a clear dichotomy, with the more amorphous ice (less deep 1.65~$\mu$m band) concentrated on the trailing hemisphere (longitudes centered near 90\degr) where the bombarding magnetospheric particle flux is highest.  This observation by itself provides solid evidence for the amorphization of ice by energetic particles in the natural environment.  According to these authors, temperatures on Europa are such that amorphous ice should crystallize on timescales $\sim$10 years, while the magnetospheric ion flux amorphizes crystalline ice in only $\sim$1 year.  

The magnetospheric particle flux also follows a strong  radial gradient with respect to distance from Jupiter,  decreasing by a factor of about 300 from Europa to Callisto.  It is slightly more complicated to compare ice spectra from the three satellites because, not only does the particle flux vary with orbital distance, but the temperatures of the satellites differ because of their different optical properties (principally, the Bond albedo, which controls the radiative temperature balance with sunlight and varies from 0.68 (Europa) to 0.43 (Ganymede) and 0.22 (Callisto).  The surface temperature differences are modest (few $\times$10~K) but, because the crystallization rate is exponentially dependent on the temperature, it can have a large effect.  Nevertheless, \cite{Hansen04} concluded that ice on  Europa, where the particle fluxes are highest is, on average, more amorphous than is ice on Callisto, while Ganymede is an intermediate case.  These authors also found evidence for a vertical crystallinity gradient in the top millimeter of the ice.  This inference derives from the observation that the 1.65~$\mu$m band, which has an absorption length $\sim$1~mm, gives a larger crystalline fraction than  the 3.1~$\mu$m band, which has an absorption length of only $\sim$10 $\mu$m.  A vertical crystallinity gradient is expected because of the steep energy spectrum; energetic particles able to penetrate deep into the ice are comparatively rare. For example, the stopping length in ice of 1~MeV electrons is $\sim$ 10~$\mu$m, while their flux is $\sim10^4$ times that of the 10~MeV electrons needed to reach $\sim$~1 mm depth \citep{Cooper01}.

The crystalline feature is also present in the spectra of other icy satellites including those of Uranus \citep{Grundy06, Guilbert09} and Pluto's satellite Charon \citep{Brown00}.

\subsection{Kuiper belt objects} 
\label{ssec:kuiper}

The Kuiper belt objects (KBOs) are cold enough for  amorphous ice to persist over the age of the solar system. Paradoxically, high signal-to-noise ratio astronomical spectra  instead reveal  the 1.65 $\mu$m absorption band characteristic of crystalline ice \citep{Jewitt04, Merlin07, Trujillo07, Barkume08}.   Table~\ref{terai} lists measurements  of the ``crystallinity factor'' (essentially, the fraction of crystalline ice) determined from narrowband photometry of the 1.65 $\mu$m feature \citep{Terai16}.  All of the objects show substantial crystallinity, regardless of their temperature over the 48~K to 82~K range.  The objects with the smallest (90482 Orcus) and the largest (315530 2008 AP129) crystallinity factors both have temperatures so low that amorphous ice, if present, would be indefinitely stable by Eq.~(\ref{lambda}).

\begin{table}[h]
\centering
\begin{tabular}{|c|c|c|c|c}
\hline
Object  & $r_H$ [AU] & $T$ [K] & CF \\
\hline
   (136108) Haumea & 51 & 48 & $0.77_{-0.05}^{+0.06}$ \\
   (90482) Orcus        & 48    & 49 &  $0.53_{-0.09}^{+0.08}$    \\
   (50000) Quaoar & 43 & 52 & $0.82_{-0.15}^{+0.15}$ \\
   (315530) 2008 AP129  & 38    & 56 & $1.00_{-0.47}^{+0.00}$ \\
   (38628) Huya & 29 & 66 & $0.92_{-0.93}^{+0.07}$ \\
     (42355) Typhon & 19 & 82 & $0.79_{-0.27}^{+0.21}$ \\
\hline
    \end{tabular}
    \caption{Crystallinity factors, CF, and infrared color temperatures, $T$, from \cite{Terai16}.  The color temperatures are slightly smaller than $T_{SS}$ from Eq.~(\ref{tlimits}). }
    \label{terai}
\end{table}

The detection of the crystalline ice feature is further puzzling because the surfaces of KBOs are exposed to cosmic ray bombardment, whose effect (just as with particle bombardment of the Galilean satellites) is to destroy the crystalline lattice and  convert the ice back to the amorphous form.   The timescale for this damage to occur is $\sim10^6$ year, very short compared to the age of the solar system.  At first, this observation was thought to indicate that the surfaces have been recently heated, perhaps through cryovolcanism or, more likely, through micrometeorite bombardment heating \citep{Porter10}.

\begin{figure}[h]
\begin{center}
\includegraphics[width=8cm]{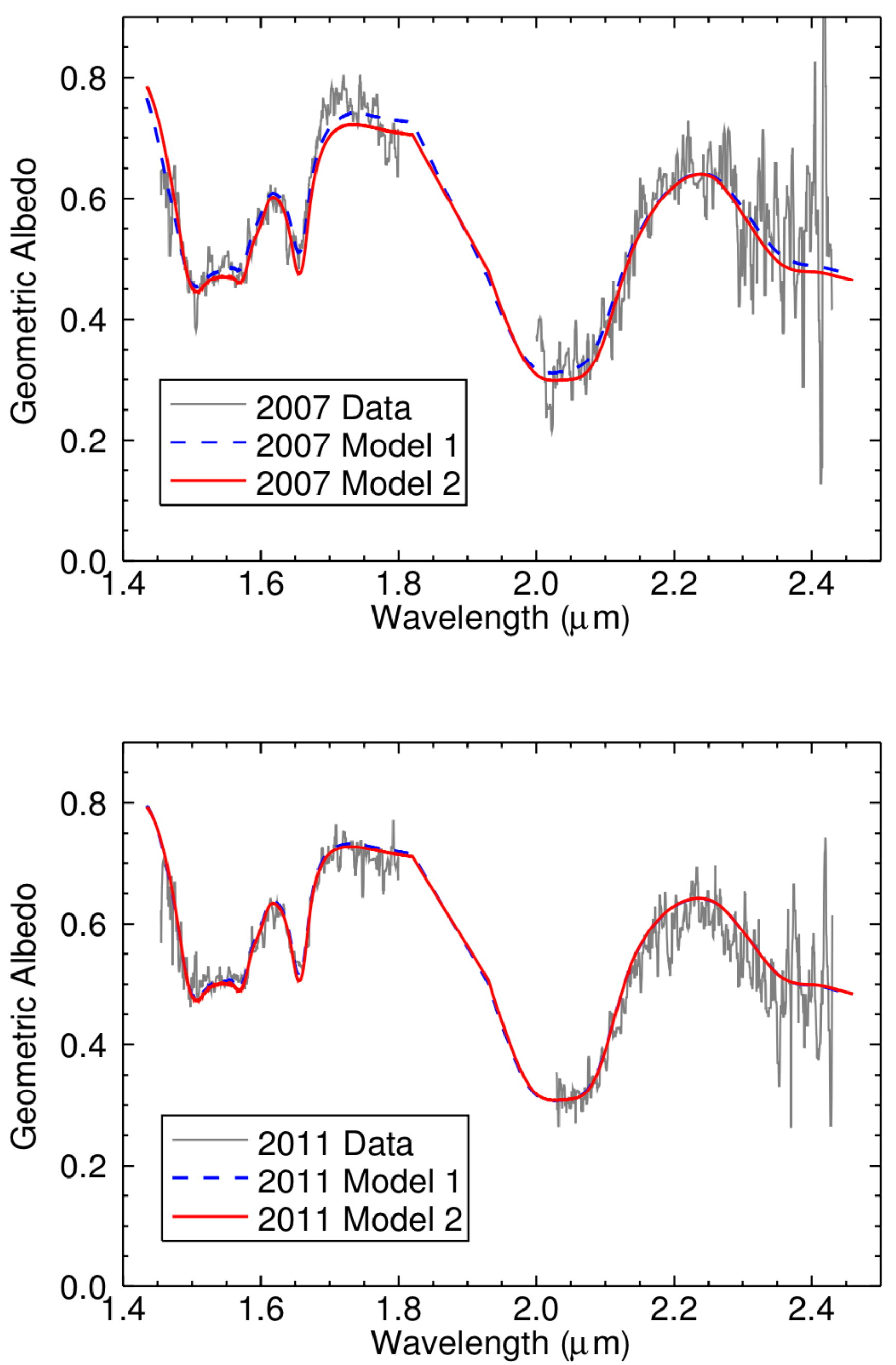}
\caption{Near infrared spectrum of KBO Haumea showing the 1.5, 1.65 and 2.0~$\mu$m bands of  water ice.  The dashed blue and solid red lines show models of pure crystalline ice and crystalline ice with a 10\% admixture of amorphous ice.  Only tiny differences between the models are apparent. From \cite{Gourgeot16}.}
\label{Haumea}
\end{center}
\end{figure}

One clue to the observational prevalence of crystalline ice comes from spectral models showing that amorphous ice can be difficult to detect if crystalline ice is present.  For example, Fig.~\ref{Haumea} shows that the addition of 10\% amorphous ice to an otherwise crystalline ice model creates only tiny differences in the reflectivity spectrum that require high signal-to-noise ratios to be detected.  For this reason, the absence of spectral evidence for amorphous ice is not necessarily evidence of its absence.

\begin{figure}[h]
\begin{center}
\includegraphics[width=8cm]{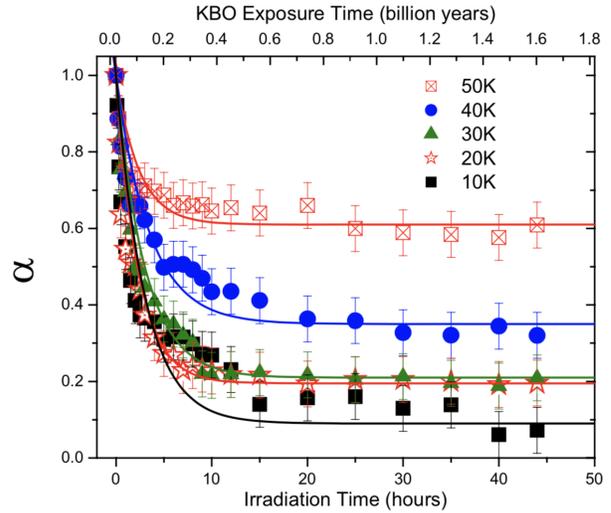}
\caption{1.65~$\mu$m band strength variation (plotted as the dimensionless parameter $\alpha$) as a function of irradiation dose and ice sample temperature.  The dose is related to the effective Kuiper belt exposure time on the upper y-axis, scaling from \cite{Cooper03}.  The 1.65~$\mu$m band is most strongly suppressed by a given dose at the lowest temperatures.  Figure from \cite{Zheng09}.}
\label{zheng}
\end{center}
\end{figure}

Laboratory experiments provide supporting evidence for this conclusion.  The degree to which irradiation can scramble the crystalline structure in ice is itself a function of temperature (see Fig.~\ref{zheng}).  At low temperatures the damage done to the crystal structure by energetic particles is evidently severe and long-lived while at higher temperatures  thermal vibrations result in partial resetting of the bonds, offsetting amorphization by the destructive ionizing particle flux \citep[][c.f.~Fig.~\ref{zheng}]{Zheng09}.  At 50 K, for example, these experiments show that the ratio of the surviving to the original 1.65 $\mu$m band area stabilizes at about $\alpha$ = 0.6 after $\sim$0.8 Gyr of equivalent cosmic ray exposure, while only at 10 K is the crystalline band  almost obliterated ($\alpha$ = 0.1). Although different experimenters \cite[c.f.][]{Mastrapa13} obtain  results for the dependence of $\alpha$ on $T$ that are different in detail from those in \cite{Zheng09}, the basic result that the degree of amorphization by cosmic rays depends on temperature is confirmed.  Therefore, the prevalence of the 1.65~$\mu$m band in KBOs is not evidence that amorphous ice is absent, only that  at least partially crystalline ice must be present.  Proper interpretation of the 1.65 $\mu$m band requires more detailed knowledge of KBO surface properties than we  normally possess.

\subsection{Centaurs}
\label{ssec:centaurs}

Centaurs are recently escaped KBOs, defined as having perihelia and semimajor axes in between the orbits of Jupiter and Neptune (excluding those trapped in mean-motion resonances with the giant planets, like the Jupiter and Neptune trojans).  These objects are precursors to the Jupiter family comets and are dynamically short-lived because of scattering by the giant planets. In order for their population to remain in steady-state, the Centaurs must be continually replenished.  The source region is thought to be the scattered disk component of the Kuiper belt.  These are high eccentricity objects with perihelia within $\lesssim$10~AU of the orbit of Neptune. The latter property renders them susceptible to scattering by Neptune, increasing their orbital eccentricities and giving a finite probability of being scattered into Neptune-crossing orbits that ultimately fall under the influence of Uranus and the gas giants.  The Centaurs are scattered amongst the giant planets until they either collide with a planet or, more frequently, are ejected from the planetary system or injected to Jupiter family comet orbits.  With perihelia beyond Neptune, the scattered KBOs also have isothermal blackbody temperatures $T_{BB} \le$ 50~K, providing indefinite stability for amorphous ice.

Observationally, we should expect that the physical properties of the Centaurs, for example the sizes and size distribution, the colors, albedos and rotational period distributions, should be identical to those of the KBOs in the source region from which they are derived.  The best-measured of these physical properties is the optical color.  For Centaurs with perihelion distances $q \gtrsim$ 10~AU the color distributions of Centaurs and scattered KBOs are indeed comparable.  In particular, the $q \gtrsim$ 10~AU Centaurs and KBOs are unique in the solar system in showing a substantial fraction of objects with ultra-red colors (color index B-R $>$1.6).  On the other hand, Centaurs with $q \lesssim$ 8 to 10~AU are depleted in ultrared surfaces, indicating a change in the nature of the surface material at isothermal blackbody temperatures $T_{BB} \sim$ 90~K to 100~K.  Instead, the optical colors of small-$q$ Centaurs are indistinguishable from those of short-period comets with orbits $q <$ 5~AU, in the water sublimation regime.  Coincidentally, mass loss from some Centaurs has been detected and is found preferentially in objects with $q \gtrsim$ 10~AU.   The similarity between the perihelion distance at which activity begins and the ultrared matter disappears suggests that activity and color change might have the same origin.  A likely mechanism is surface blanketing caused by the deposition of sub-orbital debris \citep{Jewitt15}.  

\begin{figure}[h]
\begin{center}
\includegraphics[width=7.5cm]{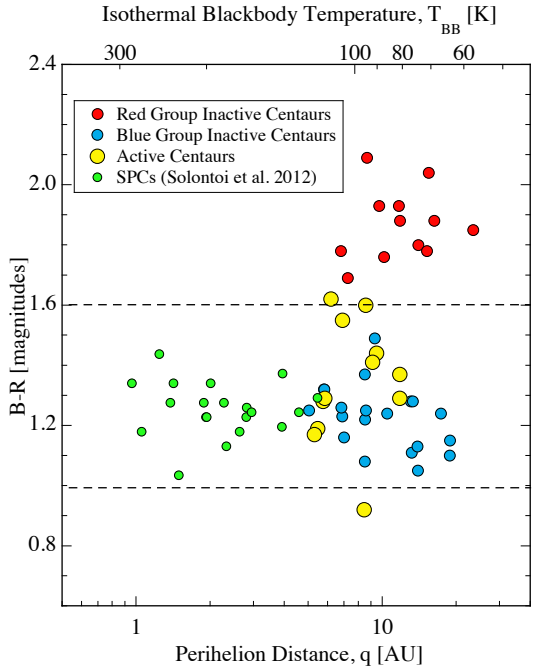}
\caption{Optical color vs.~perihelion distance for Centaurs and short-period comets, showing a bimodal distribution for $q \gtrsim$8 to 10 au and a unimodal distribution at smaller $q$.  Dashed horizontal lines show the color of the Sun at B-R = 0.99 and the onset of ultrared colors at B-R = 1.60. Adapted from \cite{Jewitt15}.} 
\label{fig:depres}
\end{center}
\end{figure}

Crystallization of exposed amorphous ice occurs on timescales comparable to the orbit period at about 8 to 15~AU \cite{Guilbert12} and therefore offers a plausible, but non-unique, explanation of the dichotomy in the Centaur observations.  Water ice is involatile at 10~AU and cannot drive the observed mass loss from Centaurs.  Exposed supervolatile ices (e.g.~CO, N$_2$ and even CO$_2$) are too volatile in that they sublimate strongly even beyond 10~AU, where no activity is detected in even the most sensitive data \citep{Li20}.  They provide no natural explanation for the observed changes at 10~AU.  Sublimation could be suppressed and delayed by burial of supervolatiles under an inert mantle but, again, there is no natural explanation for the observed critical distance of 8 to 10~AU.  Other species, less volatile than CO but more volatile than water have been suggested, including H$_2$S, but the match to the critical distance is poor and spectral confirmation of these substances has so far failed \citep{Wong19}.

\subsection{Comets} 
\label{comets}
Most comets become bright only when close to the Sun and therefore remain unobserved  at large distances.  However, near the Sun, cometary ice  is too warm to exist in the amorphous state, whether or not it is amorphous inside the cometary nucleus.  For this reason,  attempts to spectroscopically identify amorphous ice  must necessarily be focused on distant, intrinsically faint and  observationally unattractive targets.  Few such observations exist.

\begin{figure}[h]
\begin{center}
\includegraphics[width=7.5cm]{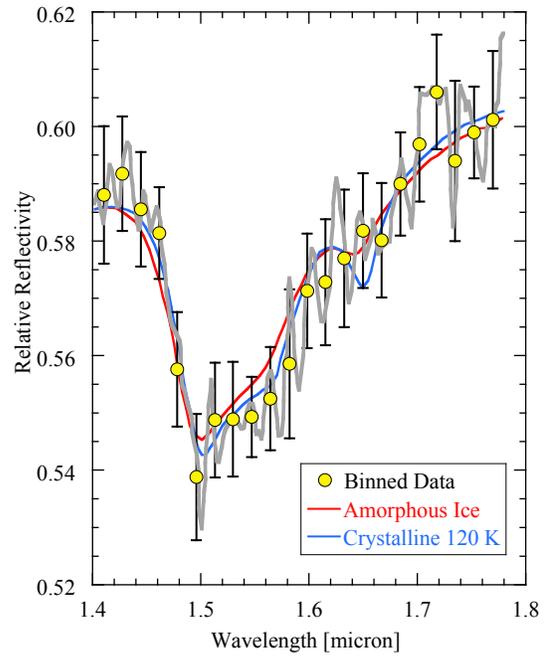}
\caption{Near infrared spectrum of the coma of comet C/2002 T7 at 3.52 au showing the absence of a 1.65 $\mu$m band. The raw data are shown as a thick grey line, the yellow circles with error bars show the data binned to resolution 0.016 $\mu$m. The spectrum is compared with a model of crystalline ice at 120 K (blue line) and one of  amorphous ice at 100 K (red line).  Within the noise in the data the model fits are indistinguishable.  Adapted from \cite{Kawakita06}.} 
\label{kawakita}
\end{center}
\end{figure}

To see where amorphous ice might be detected, if it is present, consider that the best  ground-based resolution on comets  is limited to a distance $L \sim 10^3$~km, and that small cometary dust grains are well-coupled to the gas flow and so leave the nucleus at a speed comparable to the speed of sound, $V \sim 1$~km~s$^{-1}$.  Then, we set $\tau_{CR} = L/V $ and solve for $T$ using Eq.~(\ref{lambda}).  This gives $T \sim 145$~K, corresponding to the isothermal  temperature at $r_H \sim 3.7 (1-A)^{1/2}$ au, where $A$ is the Bond albedo.  Cometary nuclei have $A \sim$ 0.01. Amorphous ice  grains  with this low albedo in comets with $r_H \sim$ 3.7~AU would sublimate before traveling a single spatial resolution element in ground-based spectra and so could not be detected.   \cite{Zubko17} estimates the cometary dust albedo $A \sim$ 0.1 to 0.2; amorphous ice grains might then be detected down to $r_H \sim$ 3.3~AU.  Pure ice grains could, in principle, have  larger $A$ and survive to smaller distances.

A few relevant spectroscopic observations have been reported. \cite{Davies97} failed to detect the 1.65 $\mu$m band in comet C/Hale-Bopp when at $r_H$ = 7~AU and speculated that the ice might be amorphous.  However, the 1.5~$\mu$m and 2.0~$\mu$m bands in their spectra are also very weak, and it is not clear that a significant limit on the 1.65~$\mu$m band, or on amorphous ice, can be placed. \cite{Kawakita04} reported that the 1.65~$\mu$m crystalline band was missing in the spectrum of C/2002 T7 (LINEAR) when at $r_H$ = 3.52~AU.  They at first speculated that this might mean that the ice is amorphous but later determined that crystalline ice can fit the data equally well, within the observational uncertainties (c.f.~Fig.~\ref{kawakita}).   The  isothermal blackbody temperature at 3.52~AU is 148~K, giving a crystallization time $\tau_{CR} \sim$ 10 minutes by Eq.(\ref{lambda}).  Outbursting comet P/2010 H2 (Vales) at $r_H$ = 3.1~AU showed a strong 1.65~$\mu$m band associated with crystalline ice, not surprising given that the spherical blackbody temperature at this distance is $T_{BB}$ = 158 K and the crystallization time (Eq.~(\ref{lambda}))  is only $\tau_{CR} \sim$ 55 s \citep{Yang10}. 

\section{\textbf{CRYSTALLIZATION AS THE DRIVER OF \\ COMETARY ACTIVITY}} 
\label{sec:outbursts}

\subsection{Evolutionary considerations}
\label{ssec:evolution}
Cometary behavior is largely determined by the orbit, and hence activity patterns vary widely
among comets; nevertheless, it is possible to assess the effect of crystallization in comets in a general manner,
regardless of orbit \citep{Prialnik1993}. A heat wave generated by a crystallization front propagating from the surface of the nucleus inwards, into a cold 
isothermal  medium of porous ice, and generating
sublimation, causes a sharp temperature gradient to arise between the regions lying in front and
behind it, where the temperatures are almost uniform.
The uniform low temperature ahead of the front
is maintained because the very {\it low} thermal
conductivity of porous amorphous ice
prevents heat dissipation. The uniform, relatively high temperature behind it is due to the stabilizing effect of the ice-vapor mixture.
The thickness $\ell$ of the crystallization front may be estimated according to the thermal diffusivity $\kappa=K/\rho c$ of the amorphous ice medium into which the front propagates and the time constant $\tau_{CR}$ of the reaction, both functions of temperature; defining $\lambda=\tau_{CR}^{-1}$, the velocity of the front $v=\ell/\tau$ is thus obtained:
\begin{equation}
    \ell(T) = \sqrt{\kappa(T)\tau_{CR}(T)}\ \ {\rm and}\ \  v(T)=\sqrt{\kappa(T)\lambda(T)}.
\end{equation}
It ranges between a few cm/year at 100~K and a few km/year around 150~K. Hence, no amorphous ice will be found in any part of a cometary nucleus that has been heated at some point in time to $\sim$140~K. At the surface, this corresponds to the isothermal blackbody temperature at $\sim$4~AU. 

\begin{figure}[h]
\begin{center}
\includegraphics[width=7cm]{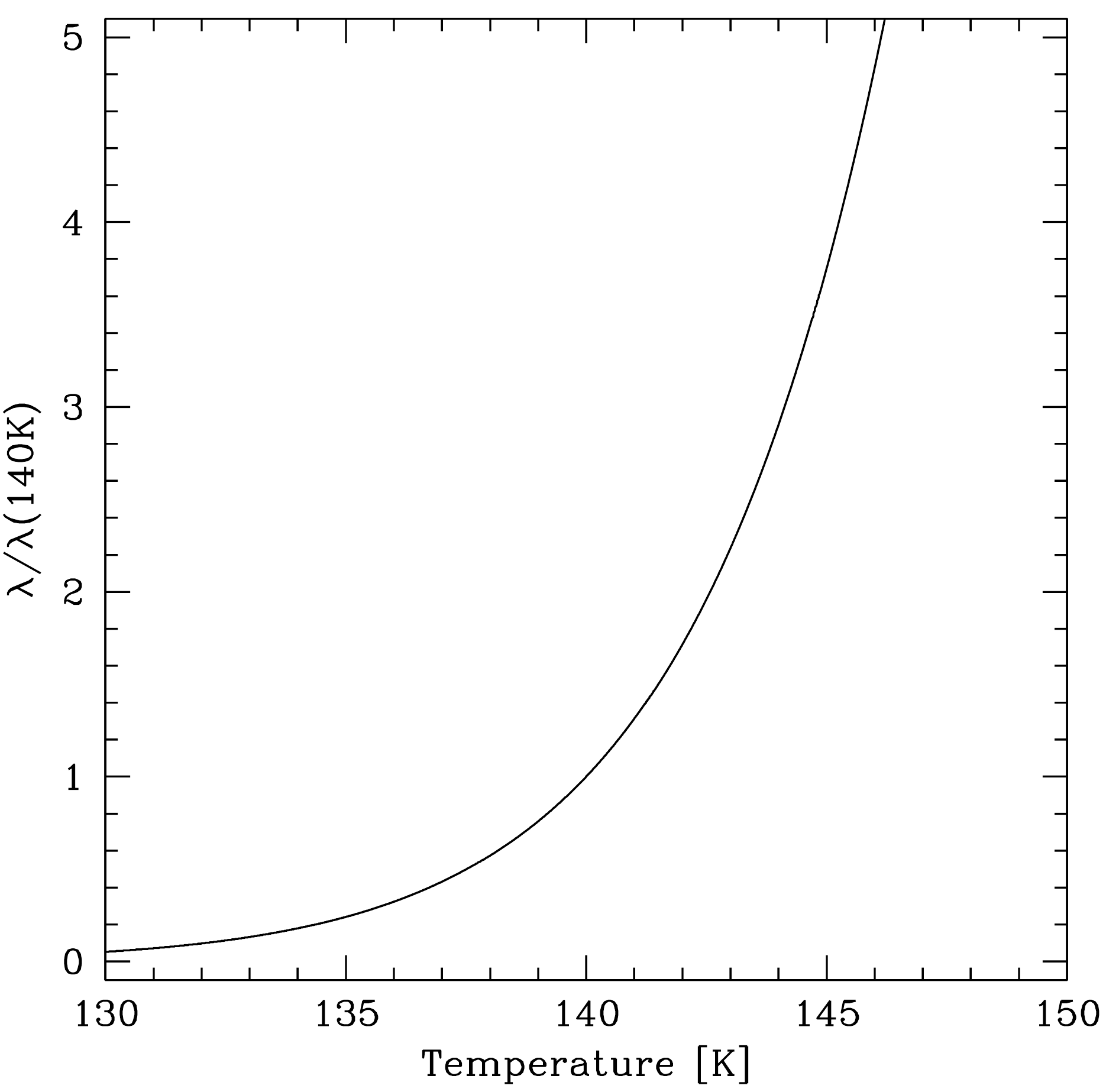}
\caption{Rate of crystallization, $\tau_{CR}(T)^{-1}$ as a function of temperature, normalized by the rate at $T=140$~K ($2.26\times 10^{-4}$s$^{-1}$).} 
\label{taut}
\end{center}
\end{figure}

Given the mass fraction of amorphous ice $X_a$ and the density $\rho$, the resulting flux of volatiles, for example CO, released from the amorphous ice may be estimated by 
\begin{equation}
 J_{\rm CO}=vf_{{\rm CO},r}X_a\rho/m_{\rm CO},   
\end{equation}
where $m_{\rm CO}$ is the molecular mass. At $T=140$~K, $J_{\rm CO}\sim10^{19}$~molecules~s$^{-1}$~m$^{-2}$ within an order of magnitude, depending on the thermophysical properties assumed and on composition.
Although the crystallization rate is a continuous function of temperature, laboratory experiments \citep{BarNun1987}, analytical considerations \citep{Prialnik1993} and numerical simulations \citep{Tancredi1994}, all show that the process actually occurs within a narrow range of temperatures around 140~K. The reason is that the steep rise of $\lambda$ with temperature in this range, as shown in Fig.~\ref{taut}, results in depletion of the amorphous ice content before the temperature rises any further. Consequently, one may speak of a crystallization temperature $T_c$, with an uncertainty of less than 10\%.

\begin{figure}[h]
\begin{center}
\includegraphics[width=7.0cm]{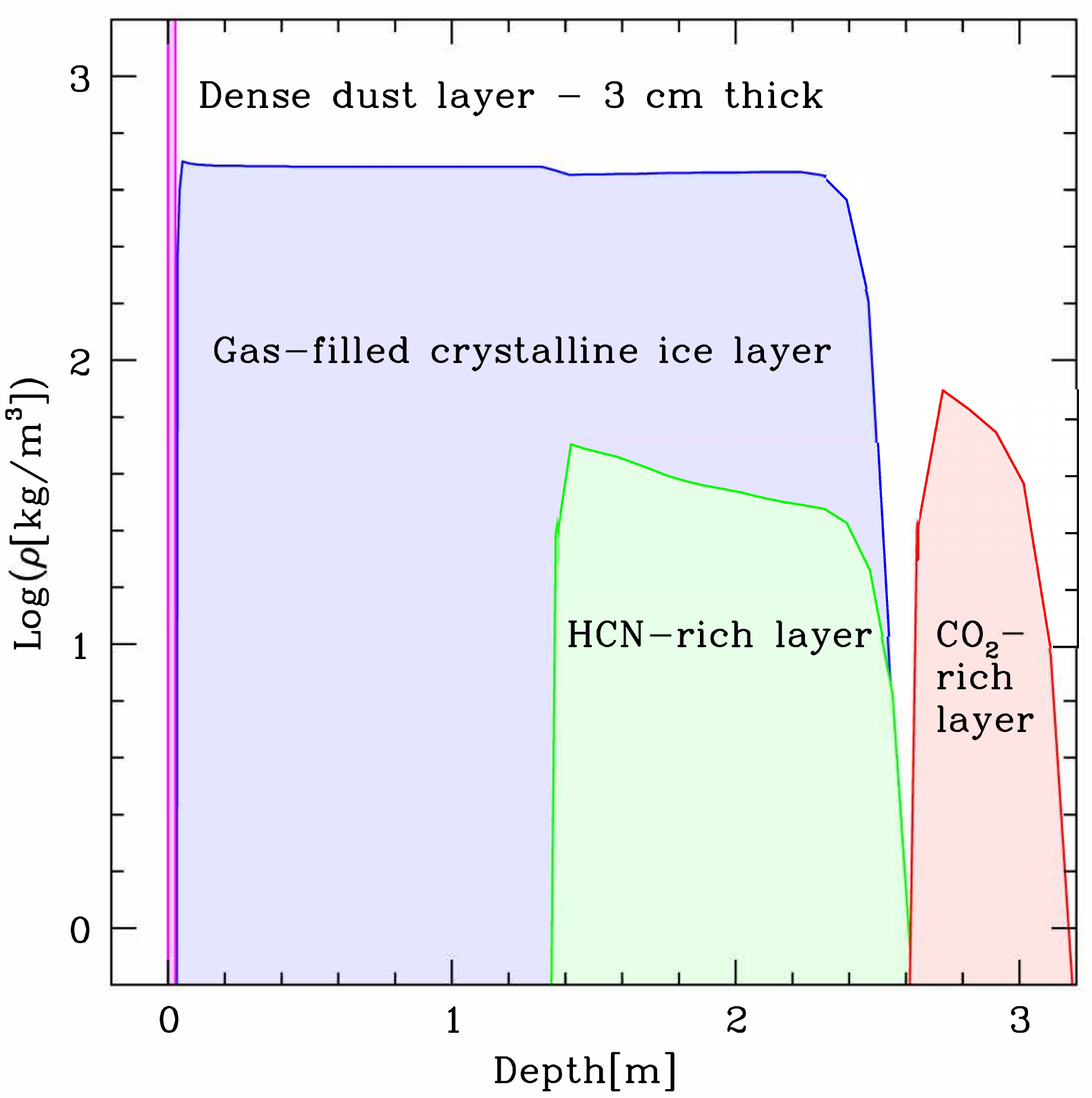}
\caption{Densities of volatile ices in the outer layer of a model nucleus following crystallization; the outer dust layer (marked) is depleted of volatiles. Adapted from \cite{Prialnik2008}.}
\label{fig:volat}
\end{center}
\end{figure}

The trapped gas released upon crystallization flows both outwards to the surface and inwards, and if it encounters sufficiently cold regions, it may refreeze, creating a volatile ice enriched layer. Upon subsequent heating, this layer will evaporate independently of crystallization. Hence the composition of volatile ices mixed with water ice should not necessarily be pristine. The gas flowing outwards may be impeded from escaping by the presence of a dense and less permeable outer crust, the result being a high pressure gas pocket. A model of such a configuration is shown in Fig.~\ref{fig:volat}.

\subsection{Spurts of crystallization and runaway}
\label{ssec:runaway}

As noted in Section \ref{ssec:latent}, the maximum energy released by the crystallization of pure amorphous water ice is $H_a \sim 10^5$ J kg$^{-1}$. This is only $\sim$1/25 of the latent heat of sublimation, $2.5\times10^6$ J kg$^{-1}$ and, for this reason,  crystallization  is not capable of driving substantial sublimation by itself.  However, the warming of the ice by $H_a$ is capable of  driving a crystallization runaway, potentially leading to the explosive release of trapped gases, as we now describe.

The progress of crystallization depends on the ambient temperature $T_0$ of the medium into which it propagates.
If a mass element $\Delta m_1$ has just crystallized, liberating an amount
of heat $H_a\Delta m_1$, of which a fraction $\eta$ is absorbed by an adjacent mass
element $\Delta m_2$, raising its temperature up to $T_c$ (the rest being dissipated over a more extended region), then $\Delta m_2$, in turn, crystallizes. 
Thus
\begin{equation}
\eta H_a\Delta m_1=[u(T_c)-u(T_0)]\Delta m_2,
\label{eq:8c1}
\end{equation}
and heat is again released in an amount $H_a\Delta m_2$, of which the absorption of a fraction $\eta$ causes another mass element to crystallize, and so forth. 
The total amount of mass $\Delta m$ that will ultimately crystallize (starting spontaneously with the crystallization 
of $\Delta m_1$) is then given by the sum of a geometric series with the factor $q={\eta H_a/([u(T_c)-u(T_0)]}$.
If $q\ge1$, the sum diverges, meaning that crystallization will continue indefinitely. However, if $q<1$, the sum converges to $\Delta m_1/(1-q)$, meaning that
it will stop after a while. A critical initial temperature may be defined, $T_{\rm crit}$,
corresponding to $q=1$. Accordingly, if $T_0\ge T_{\rm crit}$,
crystallization will proceed continuously in a runaway process; otherwise ($T_0<T_{\rm crit}$), it will stop and will need to be 
triggered again. For $\eta=0.5$, $T_{\rm crit}\approx100$~K, whereas typically, $T_0\aplt50$~K, as in the Kuiper belt. Therefore, spurts of crystallization should be more common than runaway.

\begin{figure}[h!]
\begin{center}
\includegraphics[width=7cm]{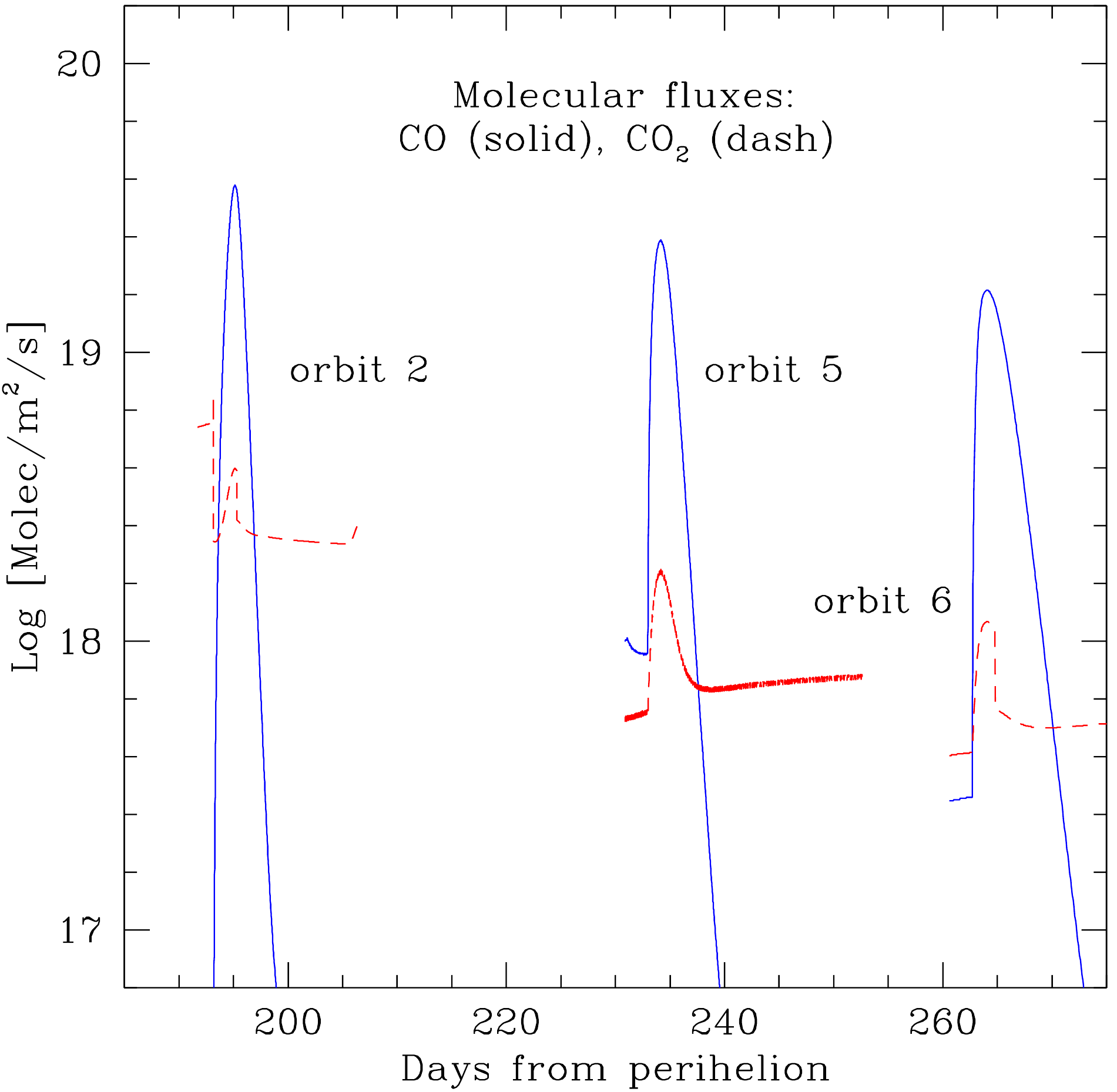}
\caption{Examples of post-perihelion outbursts on different orbits for orbital parameters of comet 67P/Churyumov-Gerasimenko, illustrated by production rates of CO and CO$_2$ - from evolutionary calculation. Adapted from \cite{Prialnik2017}.} 
\label{fig:CGflux}
\end{center}
\end{figure}

A burst of crystallization may be initiated by the heat wave propagating inwards from the insolated comet surface to the
crystalline-amorphous ice boundary, provided that reaching this boundary, it still carries sufficient energy for raising 
the local temperature 
to $T_c$. However, once this has occurred and the boundary has moved deeper into the nucleus, later heat waves originating
at the surface will be too weak when reaching the boundary to rekindle crystallization. A quiescent period thus ensues,
until the surface recedes (by sublimation) to a sufficiently short distance from the crystalline-amorphous ice boundary for
a new spurt of crystallization to take place. Since in the meantime the interior ice temperature has risen to some extent, crystallization advances deeper into the nucleus than at the previous spurt. This, in turn, affects the time span to the next spurt of crystallization. Evolutionary simulations through repeated outbursts triggered by crystallization illustrate this behavior trend \citep{Prialnik1987,Tancredi1994,Gonzalez2008}. An example is shown in Fig.~\ref{fig:CGflux}.

By the time the heat wave reaches the amorphous ice boundary and triggers a new spurt of crystallization, the comet may have moved far away from the Sun. This is how crystallization can explain the distant activity of comets.
In new comets, which may have amorphous ice close to the surface, crystallization sets in when the surface temperature approaches $T_c$, which occurs far beyond the distance where ice sublimation controls cometary activity. New comet C/2003 A2 (Gleason), for example, was observed to be active at 11.49 AU pre-perihelion and \cite{Meech2009} suggested crystallization as the cause. Nevertheless,  
inbound comets  C/2017 K2 (PANSTARRS) \citep{Jewitt17,Meech2017} and C/2014 UN271 (Bernardinelli-Bernstein) \citep{Farnham21} were both observed to be active at 23 AU, where temperatures are too low for triggering crystallization.

Perhaps counterintuitively, runaway crystallization is favored by a low conductivity of the amorphous ice, in which case most of the heat released is absorbed near the crystallization front, raising the temperature sufficiently to keep it going. A high conductivity results in spread out and waste of heat.
Indeed, \cite{Haruyama1993} studied the thermal history of comet nuclei heated by radioactive decay during their residence in the Oort cloud and found that for a very low $K$, almost all the ice crystallizes, while for a sufficiently large $K$, the initial amorphous ice is almost completely preserved. The refractory component of the nucleus, presumed to be intimately mixed with the ice, also affects the possibility of runaway crystallization.  A large refractory/ice ratio inhibits runaway by effectively reducing $\eta$.  The likelihood of runaway 
in any given comet 
is influenced by many factors.

\subsection{Non-uniform crystallization}
\label{ssec:nuniform}

\begin{figure}[h!]
\begin{center}
\includegraphics[width=7.75cm]{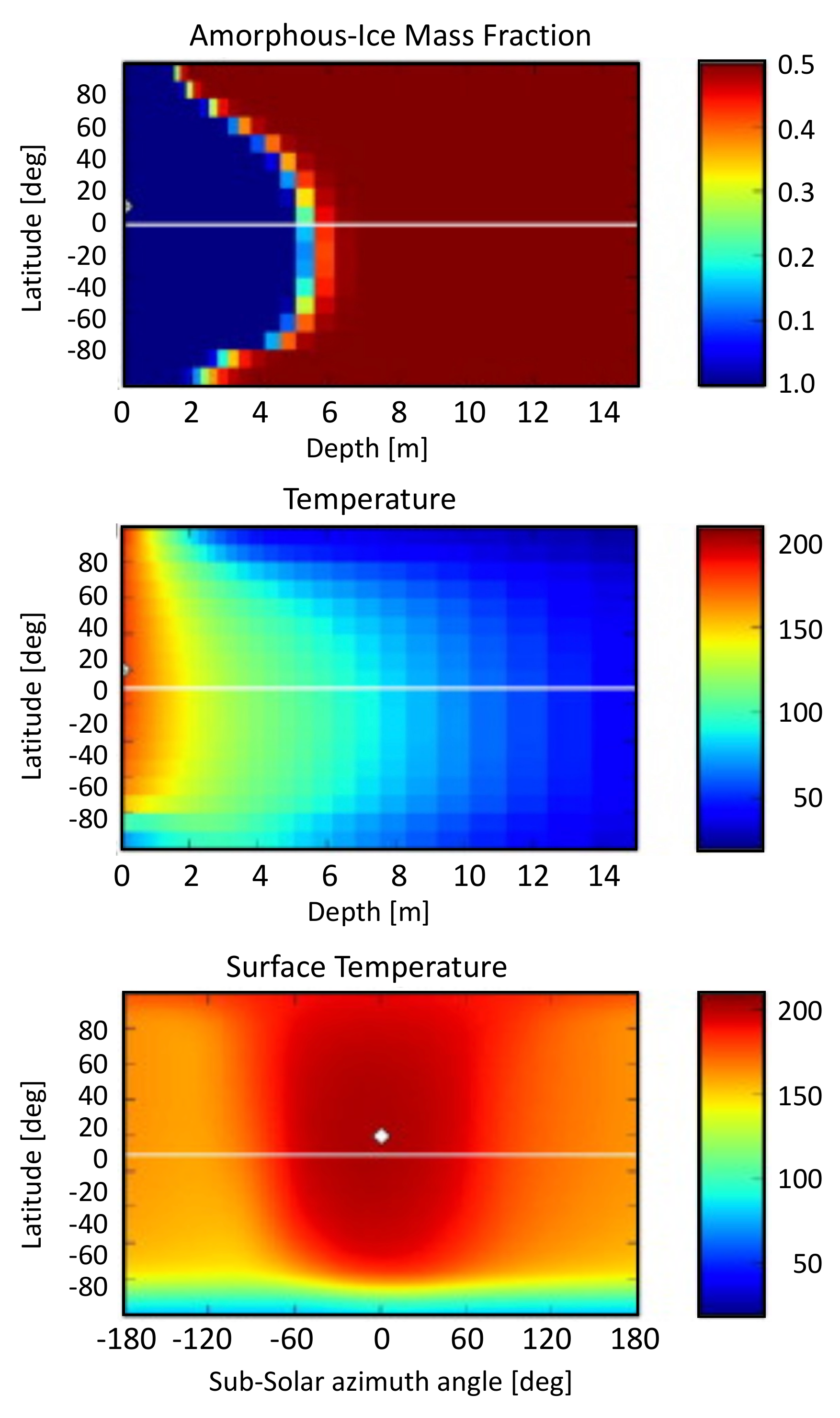}
\caption{Advance of the crystallization front in a 3D-simulation of a model with a spin period of 12.6 hr, an axis tilt angle of 15$\degr$, illustrating the effect of latitude. Adapted from \cite{Rosenberg2007}.} 
\label{fig:3D}
\end{center}
\end{figure}

As the comet nucleus surface is not uniformly heated by solar radiation, and since the thermal conductivity is low, we expect the depth of the crystallization front to vary with latitude and be affected by spin axis orientation and surface inhomogeneities. These effects have been studied by means of multi-dimensional evolution codes \citep[e.g.,][]{Guilbert2016}. The dependence on latitude is illustrated in Fig.~\ref{fig:3D}.

\begin{figure}[h]
\begin{center}
\includegraphics[width=7.5cm]{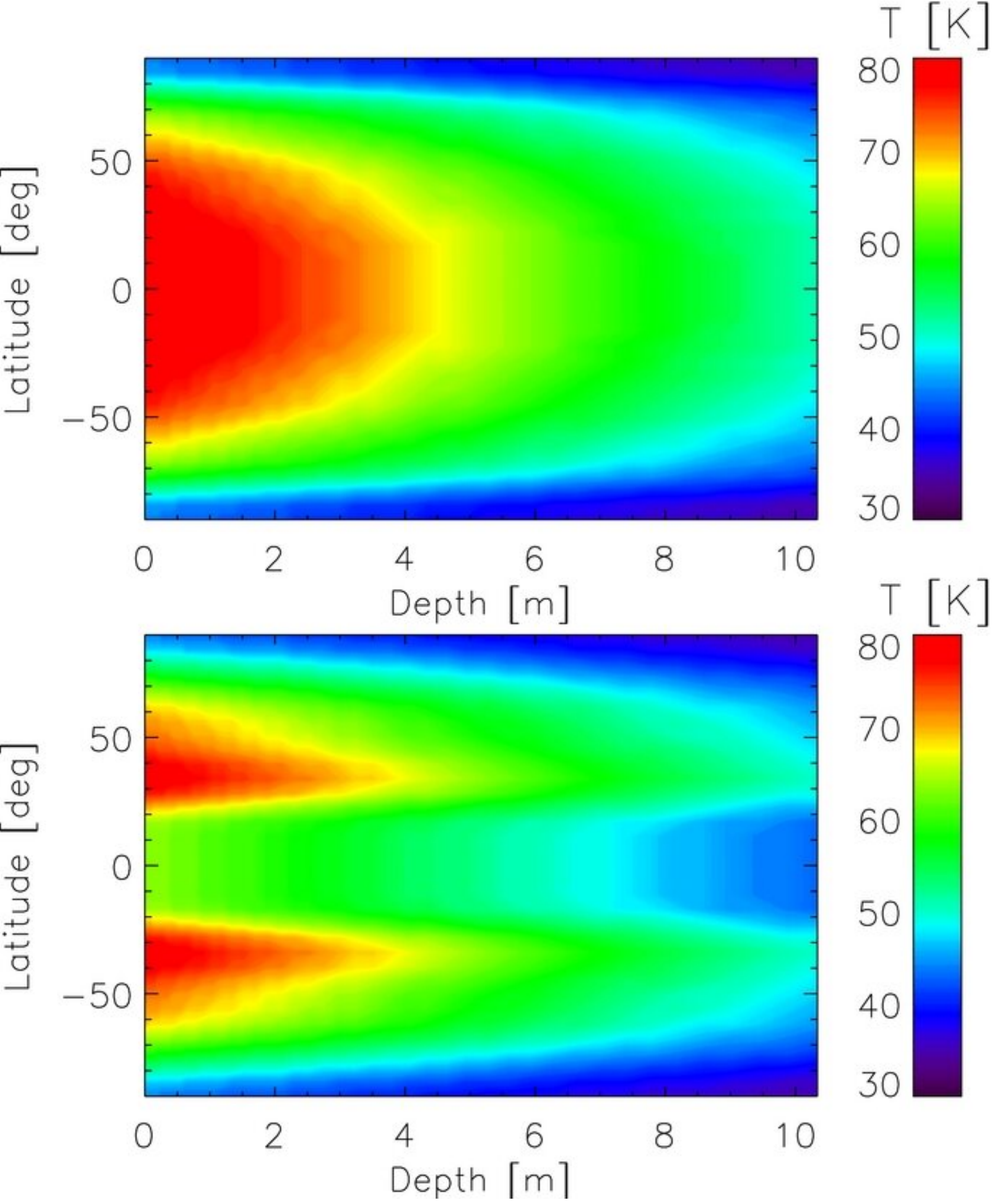}
\caption{Sub-surface temperature shadows resulting from surface albedo variations on a cometary nucleus. From \cite{Guilbert2011}.} 
\label{fig:GJ3D}
\end{center}
\end{figure}

The effect of non-uniform surface heating, modeled as a patch of high albedo, was shown by \cite{Guilbert2011} to induce varying temperature gradients in the subsurface layers, as illustrated in Fig.~\ref{fig:GJ3D}. When coupled with rotational and seasonal effects, it is clear that a spatially complex and time-varying subsurface thermal structure can be expected, resulting in local crystallization spurts at various heliocentric distances. The frequent, localized mini-outbursts detected by {\it Rosetta} on comet 67P/Churyumov-Gerasimenko may have resulted from  such behavior. 

\subsection{Crystallization during early evolution}
\label{ssec:early}

Cometary nuclei formed beyond the snowline of the solar nebula, where radiogenic heating was stronger than insolation.  This was especially true during the first few Myr, when the strongest heat source, $^{26}$Al (half-life 0.7 Myr), was abundant.  The length scale over which conduction can transport heat in time $\tau_c$ is $\ell \sim (\kappa \tau_c)^{1/2}$, where $\kappa$ is the thermal diffusivity.  We take $\kappa$ = 10$^{-8}$ m$^2$ s$^{-1}$ as representative.  With $\tau_c$ = 7 Myr (i.e.~10 half-lives of $^{26}$Al), we find $\ell \sim$ 1.5 km.  Nuclei larger than 1.5 km should capture the energy released by the decay of $^{26}$Al below a 1.5~km thick outer mantle, and so the core would suffer strong heating, comparable to that of the main-belt asteroids, which are known to have formed quickly and to be highly metamorphosed. Setting $\tau_c$ = 4.5 Gyr, we estimate $\ell \sim$ 40 km;  bodies larger than $\sim$40 km would capture much of the energy released by the decay of longer-lived radiogenic elements (K, U, Th) and be similarly metamorphosed.  These estimates are obviously simplistic (heat can also be transported by radiation, by gas and, if the liquid phase is generated, by convection) but they serve to illustrate the potential importance of radiogenic decay heating in comets.

The extent of nucleus heating and the question of the long-term survival of amorphous ice are difficult to address with confidence because of the lack of knowledge of key nucleus physical parameters. First, the low density ($\sim$500 kg m$^{-3}$,\cite{Jorda2016}) of the nucleus of 67P/Churyumov-Gerasimenko suggests substantial porosity, favoring lower values of the diffusivity, possibly by one or more orders of magnitude. At the same time, a high porosity increases the permeability to gas flow and hence the efficiency of heat transport by advection.  
Secondly, the amount of $^{26}$Al and the radiogenic power production in a given nucleus are proportional to the dust/ice ratio, $\mathcal{R}$. Modern astronomical data suggest large dust/ice ratios, with $\mathcal{R} > $1 for 67P/Churyumov-Gerasimenko \citep{Choukroun20}, $\mathcal{R} \sim$ 3 for Kuiper belt objects \citep{Fulle19}, and published values reaching $\mathcal{R}$ = 10 to 30 for 2P/Encke \citep{Reach00}. These large values indicate that radiogenic heating of cometary ices might have been substantial.  
  Lastly, we consider the accretion mechanism itself: recently, popular models of streaming-instability/pebble accretion \citep[c.f.][]{Davidsson2021} invoke rapid nucleus formation scenarios in which short-lived isotopes would be efficiently trapped, leading to high initial abundances of $^{26}$Al, but it is still possible that the growth of comets was slow \citep{Golabek21}, allowing $^{26}$Al to decay before its incorporation into the nuclei.

Against the indications of the likely importance of radioactive decay heating, stand  spectroscopic observations showing that comets preserve supervolatiles like CO and CO$_2$  in substantial abundance. A plausible resolution of this conflict is that the supervolatiles were  trapped in  amorphous ice, some of which was destroyed in the epoch of $^{26}$Al decay, but part of which survived in the outer, low-temperature regions of radiogenically heated nuclei. This scenario is indeed supported by detailed thermal modeling.

\begin{figure}[h]
\begin{center}
\includegraphics[width=8.5cm]{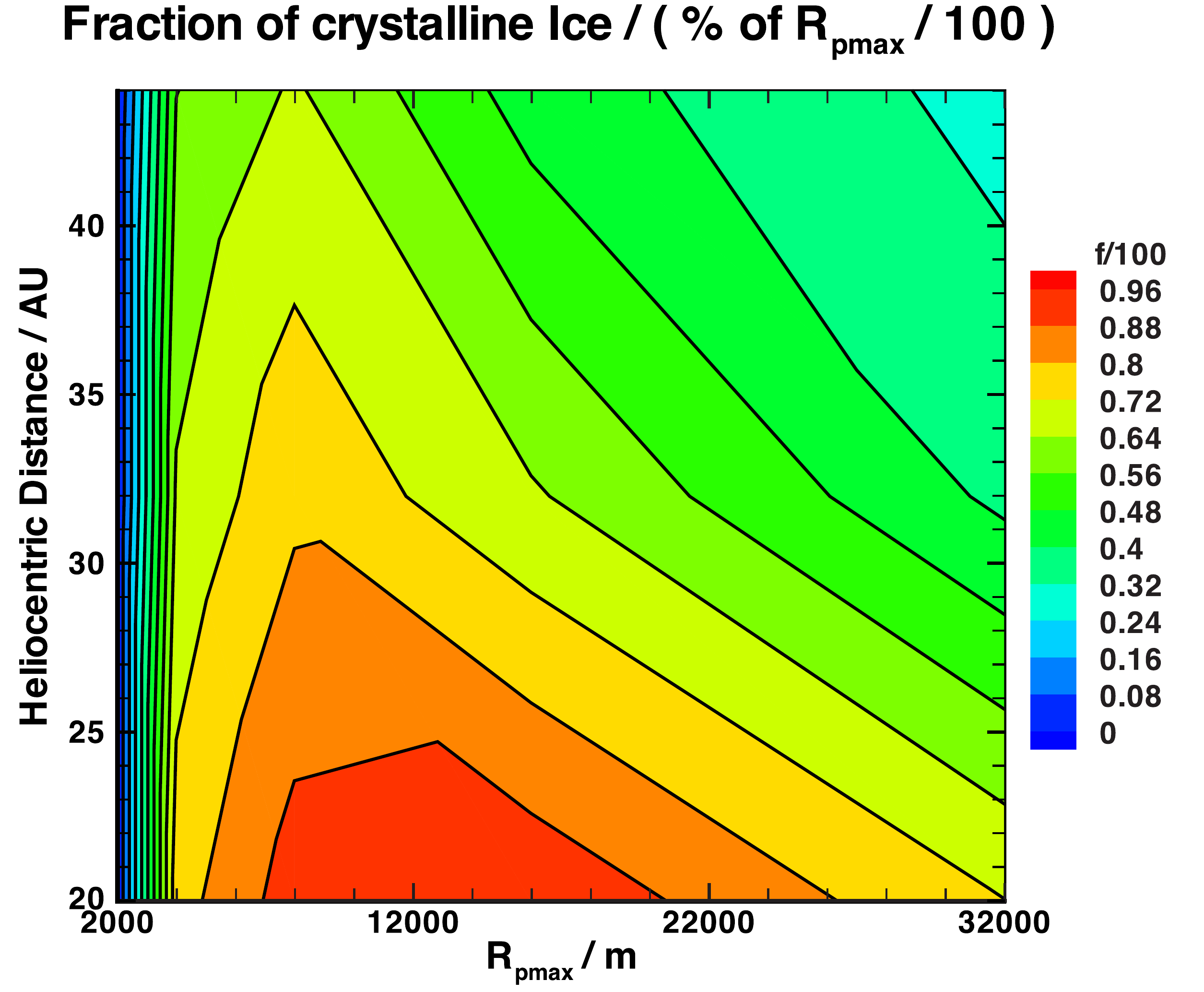}
\caption{Fractional size of the crystalline core for ice-rich objects as function of final radius and heliocentric distance. Note that half the total radius means only 12.5\% of the volume. Adapted from \cite{Merk2006}.}
\label{fig:radio}
\end{center}
\end{figure}

The first to have modeled radiogenic heating of comets by $^{26}$Al was \cite{Wallis1980}, using a simple analytic approach; the aim, however, was melting rather than crystallization. \cite{Prialnik1995} simulated numerically the early evolution of icy bodies in the Kuiper belt region, including long-lived radioisotopes and $^{26}$Al, and found that, depending on size, physical parameters---such as porosity and thermal conductivity---and the initial $^{26}$Al abundance, KBOs may emerge from the long-term evolution in three different configurations: (a) preserving their pristine structure throughout; (b) having a large crystallized core; (c) having a crystallized core, a layer of frozen gas (originally occluded in the amorphous ice) and an outer layer of unaltered pristine material. \cite{Merk2003, Merk2006} refined the calculations by considering concomitant growing of KBOs by accretion and thermal evolution due to radiogenic heating, but neglecting gas flow through the porous nucleus. An example of the results is given in Fig.\ref{fig:radio}. The occurrence of the peak in the extent of the crystalline ice core is due to the competition between the advance of the crystallization front and the accretion front and its location strongly depends on assumed parameters.

 Finally, \cite{Yabushita1993} showed that in large objects ($R\ge40$~km), crystallization in the core may occur by radioactive heating even due to long-lived radioisotopes alone. However, in all cases a sufficiently extended outer layer of amorphous ice would be preserved, to become active when these objects eventually become comets.

\subsection{Observed outbursts modelled by crystallization}
\label{ssec:models}

\begin{figure}[h]
\begin{center}
\includegraphics[width=6cm]{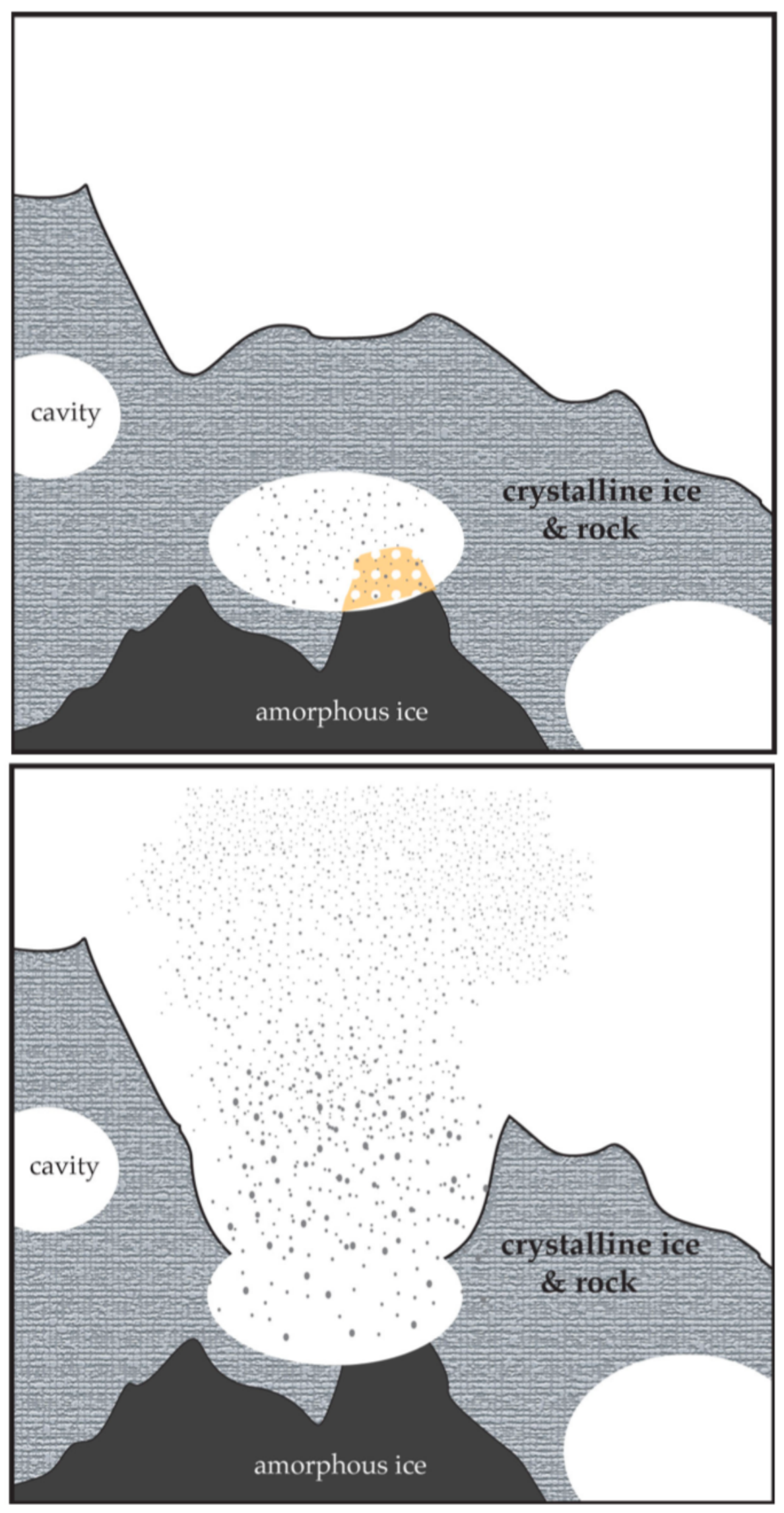}
\caption{Schematic diagram showing a cometary outburst produced by pressure build-up in a buried cavity within a nucleus whose surface layers have already crystallized. The yellow patch represents amorphous ice undergoing crystallization and releasing gas into a cavity. Adapted from \cite{Reach10}.}
\label{Reach_schematic}
\end{center}
\end{figure}
Crystallization of amorphous ice and release of trapped gases---accompanied by drag of dust particles---has been invoked to explain many observed cometary outbursts, as we describe in this section. Other interpretations are possible, generally involving the build-up of internal pressure due to sublimation of pure supervolatiles within confined spaces in cometary nuclei that maintain significant cohesive strength.  Of course, crystallization and gas confinement can occur together, leading to outburst \citep{Samarasinha01}. 

The occurrence of outbursts triggered by crystallization depends on the depth of the amorphous-crystalline ice boundary. A schematic view of such an outburst is shown in Fig.~\ref{Reach_schematic}. For a new comet, where the boundary is close to the surface, outbursts are expected on the inbound leg of the orbit, at the distance where the surface temperature is close to the crystallization temperature, from Eq.~(\ref{tlimits}) beyond 8~AU.  

\begin{figure}[h]
\begin{center}
\includegraphics[width=8cm]{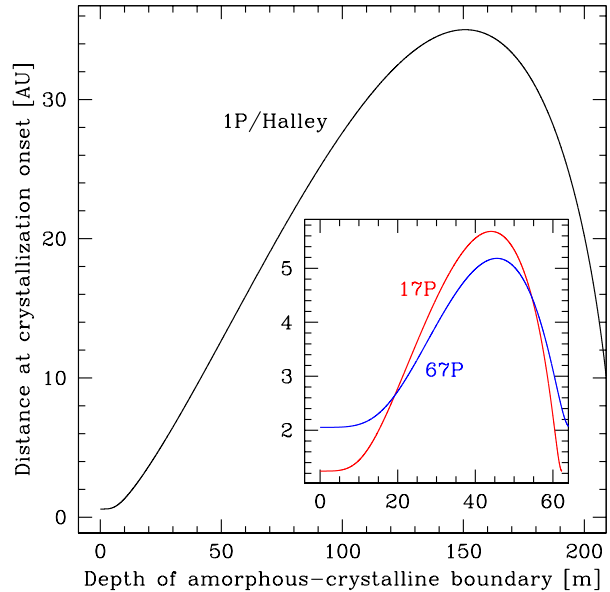}
\caption{Illustration of the time lag of crystallization, for orbital parameters of different comets. Large depths may be reached by the heat wave after aphelion, keeping in mind that the amplitude decreases with depth.} 
\label{fig:rdepth}
\end{center}
\end{figure}

For an old comet, where the boundary lies at a depth below the surface, the heat wave generated by absorption of solar radiation close to perihelion penetrates to deeper layers with a time lag determined by the thermal diffusivity and crystallization sets in when it reaches the amorphous ice boundary, which may occur at large heliocentric distances post-perihelion 
(see Section~\ref{ssec:runaway}). 
A simple estimate is shown in Fig.~\ref{fig:rdepth}. If the boundary is found very deep, a burst of crystallization may occur as late as the next inbound leg of the orbit. Indeed, according to the survey of the distant activity of SPC by \cite{Mazzota2009}, outbursts occur predominantly post-perihelion and only on rare occasions, on the inbound leg of the orbit.

{\bf Comet 1P/Halley:} In Feb. 1991, when outbound at 14.3 AU, the comet was found to have undergone a major outburst that lasted for about three months, with a 300-fold brightness increase and the development of a tail \citep{West1991}, the first comet outburst detected at such a large distance. Crystallization of amorphous ice was immediately invoked as the most plausible mechanism to explain the outburst \citep{Schmitt1991,Weissman1991}. \cite{Prialnik1992} showed by numerical simulations that the outburst features---heliocentric distance, characteristic time, and outburst amplitude---are indeed obtained for a model of low density, where the outburst is initiated by crystallization and CO release at a depth of a few tens of meters.

{\bf Comet C/Hale-Bopp:} The comet was discovered in July 1995, being characterized by an unusually
bright dust coma at a distance of about 7 AU. In late 1996, \cite{Jewitt1996} detected a very large flux
of CO molecules, which increased rapidly. Such rapid
brightening is unlikely to have resulted from surface (or
subsurface) sublimation of CO ice in response to insolation.
Moreover, CO ice should have been depleted much
earlier on the orbit, since at 7 AU the surface temperature is
already above 100~K, far above the sublimation temperature of CO. 

\cite{Prialnik1997} modelled the thermal evolution of comet Hale-Bopp and showed that a sharp rise in the activity of the nucleus  at  7 AU pre-perihelion, marked by an increase in the CO flux and in the rate of dust emission by several
orders of magnitude, is consistent with runaway crystallization taking place a few meters below the surface,
accompanied by the release of the trapped CO. The runaway was eventually quenched, and a period of sustained, but variable, activity ensued, in agreement with observations. Similar results were obtained by \cite{Enzian1998}, using a quasi-3D model of the nucleus. They showed that crystalline ice cannot account for the observed activity. \cite{Capria2000} showed that to account for the observed CO production, one must invoke gas trapped in amorphous ice, a fraction of which remains trapped until the ice sublimates.

\begin{figure}[h!]
\begin{center}
\includegraphics[width=8.5cm]{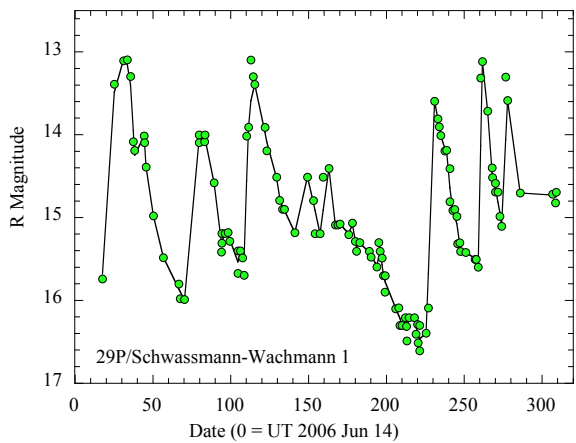}
\caption{Approximately one year of photometry of 29P/Schwassmann-Wachmann 1 (circles). The solid line is a running mean added to guide the eye.  Adapted from \cite{Trigo08}.} 
\label{29P}
\end{center}
\end{figure}

{\bf Comet 29P/Schwassmann-Wachmann 1:} 
The orbit of 29P  is confined to a small range of heliocentric distances around 6 AU, just in the temperature range where crystallization occurs on orbital time scales (see Fig.~\ref{fig:tau}). Despite the large distance from the Sun, the comet is always active \citep{Jewitt90} with superimposed outbursts  by factors up to several hundreds (see Fig.~\ref{29P}).   Production of CO occurs at about 10$^3$ kg s$^{-1}$ but, unlike the continuum, has been observed to vary by only a factor of $\sim$2 since the first detection nearly three decades ago \citep{Senay94, Festou01, Gunnarsson08, Wierzchos20}. The nucleus is estimated to be about 32$\pm$3 km in radius \citep{Schambeau21}. The total  mass of CO lost  since 1994  is $\Delta M_{CO} \sim 10^{12}$ kg, corresponding to a layer $\sim$ 0.15 m deep over the whole nucleus or, more likely, to deeper excavation, or progressive cliff collapse, over a smaller portion of the surface.  

\cite{Cabot1996} analysed the activity of the comet and suggested that it  
shows a maximum after each perihelion passage and that the delay between this peak of activity and perihelion tends to increase from one revolution to the next: less than one year in 1940, more than two years in 1957 and almost five years in 1973. 
They also noted that the average level of activity increased after the orbit change in the 1980s. They further showed that this erratic activity can be explained by the crystallization of amorphous ice, triggered by the propagation of a heat wave into the nucleus. The modelled activity
is correlated with heliocentric distance: the maximum activity is delayed with respect to perihelion and the delay seems to increase between successive revolutions (see Section~\ref{ssec:runaway}). \cite{Enzian1997} applied to this comet a 2.5-D model (that considers radial and latitudinal effects, summing over meridians for a spin period), and showed that crystallization of amorphous ice and release of trapped gas are crucial and uniquely successful for explaining the activity features of this comet.   

\begin{figure}[h]
\begin{center}
\includegraphics[width=8cm]{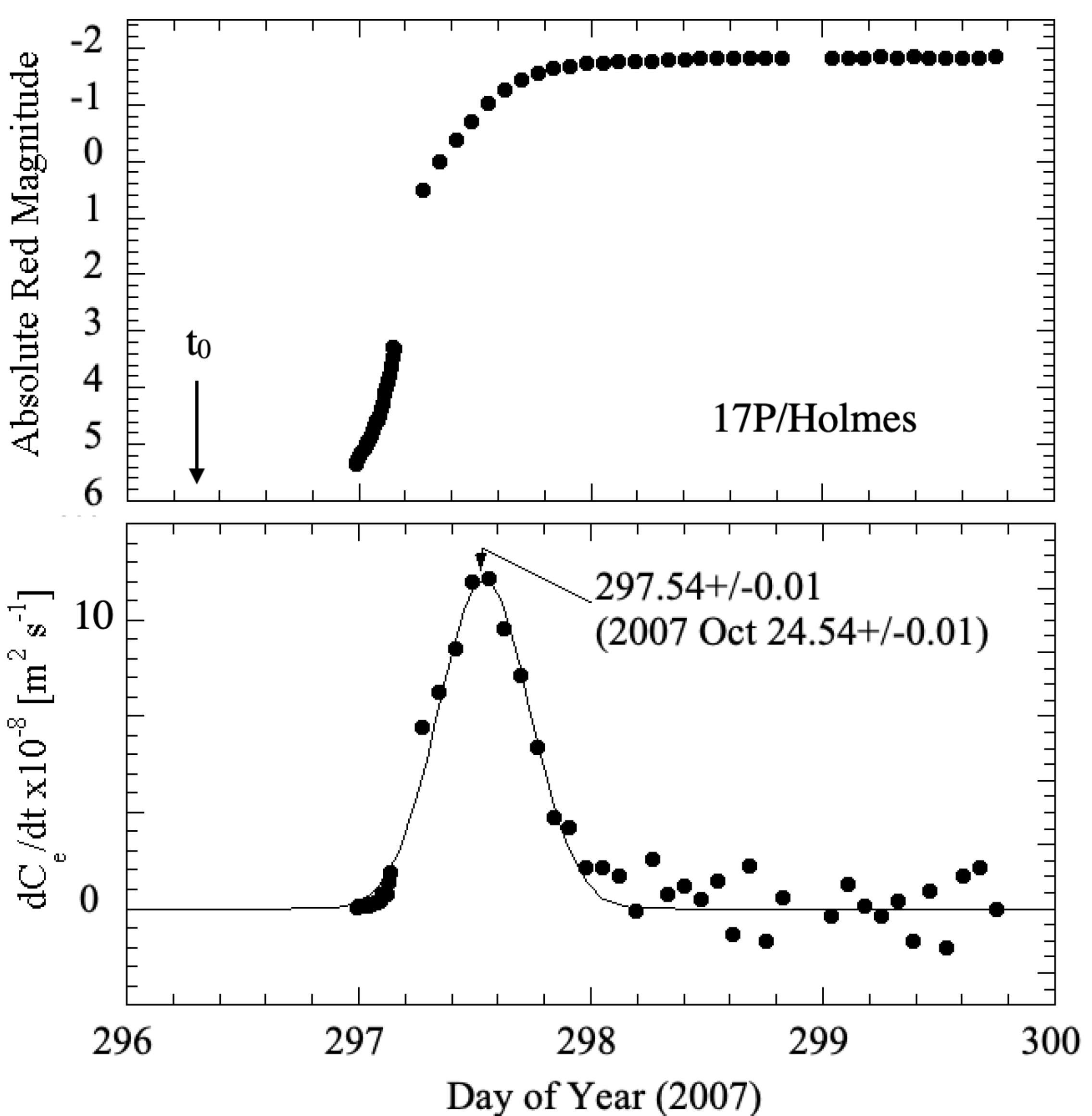}
\caption{Upper panel: Integrated photometry of 17P/Holmes showing outburst brightening which began at $t_0$ = 296.3 (UT 2007 October 23.3). Lower panel: Rate of change of dust scattering cross-section revealing a single release of dust.  The peak rate of brightening corresponds to 1000 km$^2$ s$^{-1}$ and mass loss rate $3\times10^5$ kg s$^{-1}$. Adapted from \cite{Li11}.} 
\label{17P}
\end{center}
\end{figure}

{\bf Comet 17P/Holmes:}
\begin{figure}[h!]
\begin{center}
\includegraphics[width=5.5cm]{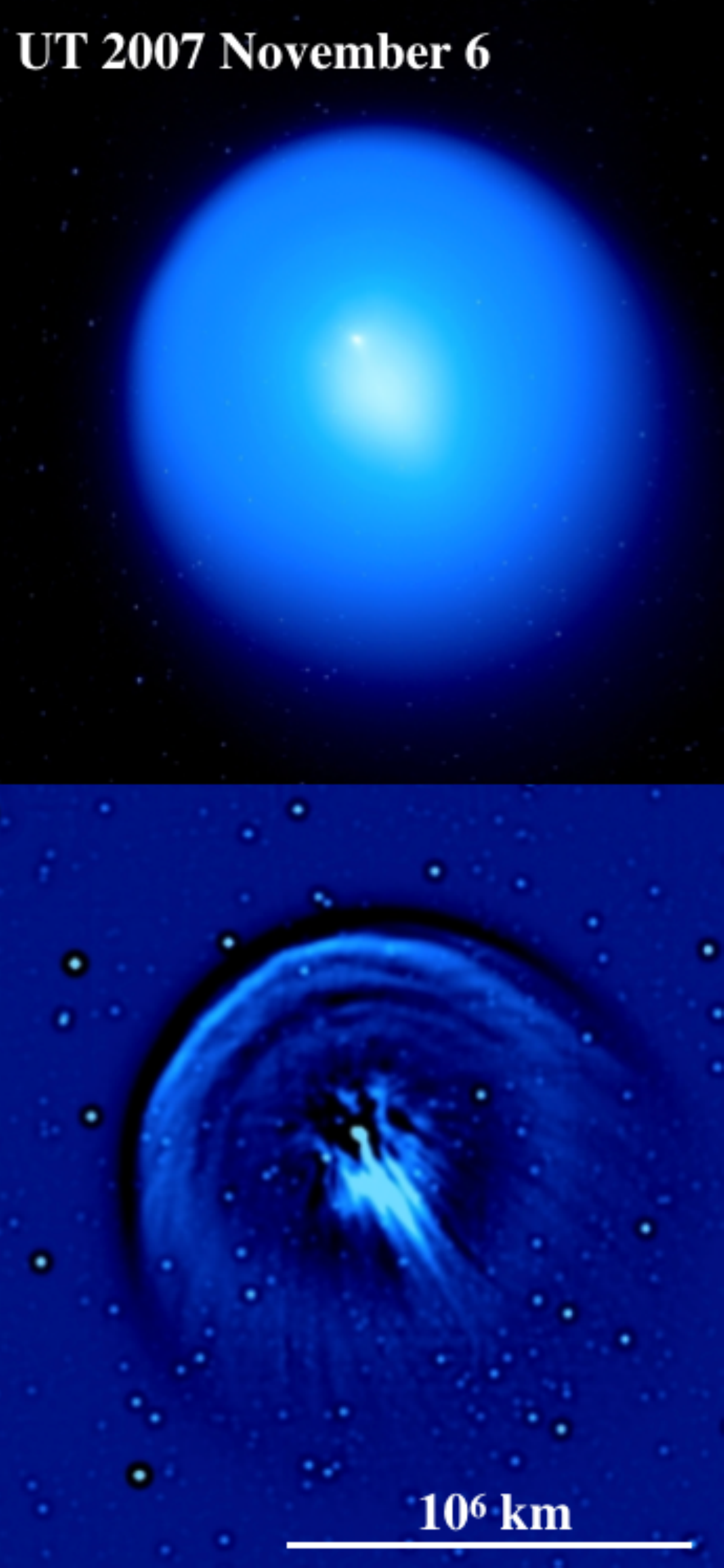}
\caption{Outburst of 17P/Holmes on UT 2007 November 6 shown (top) as a raw image, revealing the off-center nucleus and the down-tail debris cloud and (bottom) spatially filtered to reveal structure in the coma. About 14 days after the start of the outburst, the $\sim10^6$ km diameter of the coma indicates an expansion speed $\sim$400 m s$^{-1}$, a large fraction of the local gas thermal speed. Adapted from \cite{Stevenson10}.} 
\label{Holmes_vertical}
\end{center}
\end{figure}
This short-period comet displayed two large post-perihelion outbursts in November 1892 and January 1893. Sixteen uneventful orbits later, an even greater and much more closely observed post-perihelion outburst occurred in October of 2007 \citep{Hsieh10, Reach10, Stevenson10, Li11}.  The total brightening was about 19 magnitudes, from likely quiescent magnitude $V\sim17$ to a peak near -2, \citep{Li11} corresponding to a factor of about 40 million (see Fig.~\ref{Holmes_vertical}).  Although the effects were measurable for months, the mass loss rate was sharply peaked with a full-width at half maximum of 0.4 day, centered about 1 day after the start of activity (see Fig.~\ref{17P}), and reaching a maximum mass ejection rate $\sim3\times10^5$~kg~s$^{-1}$, with dust ejection speeds $\sim$400~m~s$^{-1}$.  Using a quasi-3D model, \cite{Hillman2012} were able to simulate the observed behavior, resulting from crystallization of amorphous ice and release of occluded gas species, in particular the interval between outbursts, the close post-perihelion occurrence and the production rates of CO, CO$_2$ and NH$_3$.

\cite{Kossacki2010} applied a thermal model to comet 17P, taking into account  amorphous water ice and dust, but no other volatiles. They showed that in this case crystallization is unable to account for the observed, powerful outburst. This result, in fact, strengthens the conclusion that supervolatile gases expelled from the ice upon crystallization are crucial to the outburst mechanism, since the heat released is only a few percent of the heat required to sublimate water ice (see Section~\ref{ssec:runaway}).

{\bf 2060 Chiron:} First classified as an asteroid, Chiron is the first object identified as a Centaur, exhibiting cometary activity: developing a coma and undergoing outbursts. Its distance from the Sun is too large for water sublimation to drive this activity. Both \cite{Prialnik95} and \cite{Capria2000AJ} have modeled the activity of this object and showed that the observed characteristics may be accounted for by a composition of porous gas-laden amorphous ice and dust.

{\bf Comet 67P/Churyumov-Gerasimenko:}  \cite{Mousis2015} have suggested that pits observed by the {\textit{Rosetta}} mission on the surface of comet 67P may result from disrupting outgassing due to increased gas pressure that could be caused by clathrate destabilization and amorphous ice crystallization. They discarded the possibility of impacts because of inconsistency between the size distributions of pits and impactors, and also the possibility of erosion due to insolation. \cite{Prialnik2017} explored a possible mechanism that may explain sudden depressions of surface areas on a comet nucleus, as those observed on this comet. Assuming the area is covered by a thin, compact dust layer of low permeability to gas flow compared to deeper, porous layers, gas can accumulate below the surface when a surge of gas release from amorphous ice occurs upon crystallization, as we have shown in Fig.~\ref{fig:volat}. The gas pressure is found to exceed the hydrostatic pressure down to a depth of a few metres. The rapid build-up of pressure may weaken the already fragile, highly porous structure. Eventually, the high pressure gradient that arises drives the gas out and the pressure falls well below the hydrostatic pressure. The rapid pressure drop results in collapse. Since the crystallization front lies at some depth below the surface, the location on the orbit when this phenomenon occurs is determined by the thermal lag (see Fig.~\ref{fig:rdepth}).

\begin{figure*}[h!]
\begin{center}
\includegraphics[width=17cm]{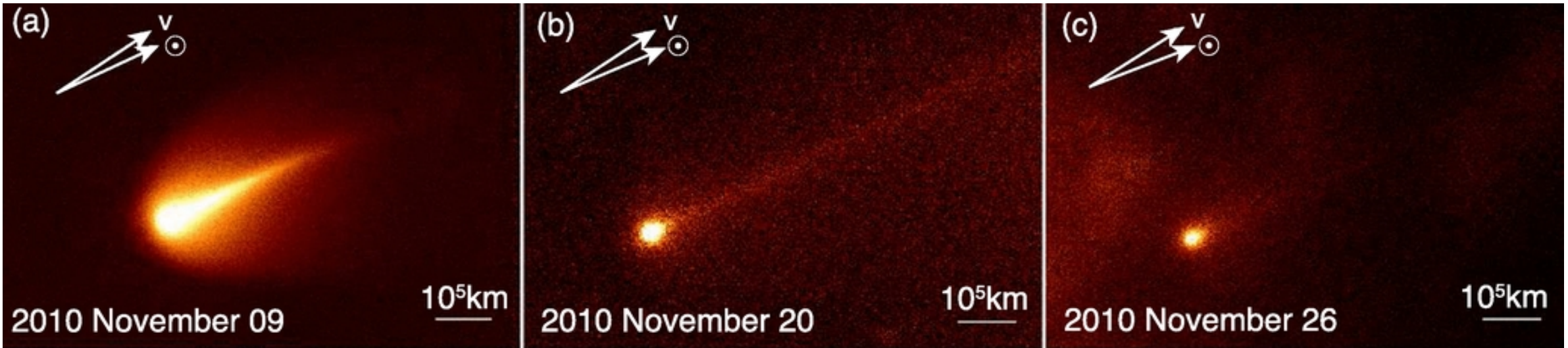}
\caption{Outbursting Comet P/2010 V1: time-series RC-band images. The projected anti-solar direction ($\odot$) and the projected negative heliocentric velocity vector (V) are marked. Adapted from \cite{Ishiguro14}.}
\label{fig:IM}
\end{center}
\end{figure*}

{\bf Comet 332P/2010 V1 (Ikeya-Murakami):} Similarly to Comet 17P/Holmes, the comet was discovered due to a large-scale explosion it underwent between UT 2010 October 31 and November 3 (see Fig.~\ref{fig:IM}). In both cases the outbursts occurred post-perihelion---20 days for 332P and 172 days for 17P/Holmes---and the decline times were very similar, 70 and 50 days, respectively. Although the mass and energy were orders of magnitude smaller for 332P than for 17P, the energy per unit mass was similar, $\sim 10^4$~J kg$^{-1}$. The sudden ejection and the derived energy per unit mass of the ejecta were found consistent with runaway crystallization of
buried amorphous ice as the source of energy to drive the outburst \citep{Ishiguro14}.  The main nucleus of 332P was later determined to be $\le$275 m in radius (0.04 albedo assumed), and underwent partial fragmentation in 2015 \citep{Jewitt16}.  The relationship between the outburst in 2010 and the subsequent fragmentation is not known.
But stresses induced by crystallization of amorphous ice might lead to fracture in a cometary nucleus \citep{Prialnik1993a}, especially in a small one.

{\bf Comet P/2010 H2 (Vales):} This quasi-Hilda asteroid underwent a spectacular photometric outburst by $\ge7.5$~mag (factor of $\apgt 10^3$) in 2010, 37 days post-perihelion, at a distance of 3.1~AU. While the rising phase of the outburst was very steep (brightness doubling time of hours), subsequent fading occurred slowly (fading timescales increasing from weeks to months). The specific energy of the ejecta was estimated at 220~J~kg$^{-1}$ \citep{Jewitt20}. Low-energy processes known to drive mass loss in active asteroids, including rotational disruption; thermal and desiccation stress cracking; and electrostatic repulsion, cannot generate the high particle speeds measured in P/Vales, and were discounted. Impact origin was deemed unlikely, given the short dynamical lifetimes of the quasi-Hildas and the low collision probabilities of these objects. \cite{Jewitt20} conclude that comet Vales is most likely a temporarily captured comet in which conductive heating of subsurface ice has triggered
an outburst, probably through exothermic crystallization
of amorphous ice. To account for the outburst energy, an ice volume about
0.4~km$^2$ in areal extent ($\sim$1\% of the nucleus surface)
and $\apgt 5$~m thick, and buried beneath a refractory layer a
few meters thick, is inferred.

{\bf Comet 9P/Tempel 1:} Mini-outbursts observed by the {\textit{Deep Impact}} mission on comet 9P/Tempel 1 \citep{AHearn2005} were found to occur in regions of lowest effective surface gravity; such regions would be the first to yield to internal stresses resulting from gas pressure build-up due to gas release by amorphous ice crystallization \citep{Belton2008}. Furthermore, \cite{Belton2009} propose that the smooth terrains observed on the nucleus surface are produced by spurts of CO  release at the crystallization front, at a depth of several tens of meters, which result in fluidization of the dust and ice layer above it and extrude fluidized material onto the surface at low velocity (cometary cryo-volcanism). The inferred CO loss rate of $\sim4\times10^{18}$~molecules/s/m$^2$ \citep{Feldman2006,Belton2009} is consistent with gas release upon crystallization.

There are more examples of cometary outbursts explained by the amorphous ice crystallization, but those listed are so different from each other and yet accounted for by the same basic mechanism, that they strengthen the hypothesis which links cometary activity to amorphous ice. 

The most plausible picture that emerges is the following. Comets formed out of amorphous ice with large amounts of occluded gases, in environments which were too hot for extremely volatile species to be wholly frozen as bulk ice, but cold enough to preclude clathrate formation. Radiogenic heating, even by powerful short-lived radionuclides, did not affect this composition over most of the comet nucleus. When comets approach the sun, runaway but transient crystallization of the amorphous ice manifests itself as outbursts in which a fraction of the occluded gas is expelled. The remaining gas is trapped in part in the crystalline ice to form clathrates and in part diffuses toward colder regions of the nucleus, where gases refreeze down the temperature gradient in order of volatility, forming a stratified pattern. Later on and in subsequent orbits, cometary activity is bound to be complex, as there are now various spatially and compositionally distinct sources of volatiles. The decomposition of newly formed clathrate and the sublimation of newly frozen volatiles is induced by solar energy absorption, but the resulting production rates and the ratios between them, as well as between them and water production, will differ among comets and epochs, depending on size, shape, nucleus spin and spin axis orientation, and on the dynamical history of the comet. Heat propagating from the surface will trigger further bursts of crystallization of amorphous ice leading to outbursts of various strengths, depending again on the nucleus characteristics. Driven by conducted heat, outbursts may occur at any distance from the sun, including large distances. As a by-product, fresh clathrates and ices will form. All this accounts for the enormous complexity and variability of cometary activity, despite the fact that the initial physical and compositional state of the ice is common and simple. The long-standing controversy over whether cometary ice is amorphous or crystalline \citep[e.g.,][]{Luspay16} is thus a false dichotomy.

\section{\textbf{THE FUTURE }}
\label{sec:conclusions}

Despite  hundreds of observations, laboratory studies, results from space missions, and sophisticated numerical simulations, evidence for the role of amorphous ice in comets remains indirect, although this is still the most plausible form of cometary ice. One reason is that, by its very nature, amorphous ice is only long-lived in the more frigid, distant regions of the solar system, precisely where observations are the most difficult to obtain.  Another  is that, in evolved comets, amorphous ice is expected to be buried below the surface, being inaccessible to direct observation and manifesting itself only indirectly.  Moreover, the accessible indicators of amorphous ice, principally the strength of the band at 1.65 $\mu$m,  provide an imperfect measure of the fraction of amorphous ice  because of the confounding influences of ice temperature, irradiation history, grain size and other variables.    More laboratory work is needed to determine the thermophysical properties of amorphous and crystalline ices, especially when loaded with other volatiles.  

Observational improvements in characterization of the crystalline state are expected in the era of the JWST, driven by its greater near infrared sensitivity and by access to the much stronger 3 $\mu$m water ice band (c.f.~Figure \ref{mastrapa_3micron}). Ideally, of course, we would like to sample and return cometary ices to Earth while preserving their physical and chemical properties \citep{Westphal20, Vernazza21}.  A cryogenic sampling mission to excavate nucleus material from depth and bring it back to the Earth without causing it to crystallize, would be one of the most scientifically interesting, but technically challenging, endeavors ever attempted.  

\vskip .3in
\noindent \textbf{Acknowledgments} 
\vskip .1in
\noindent We thank Murthy Gudipati for comments on a draft of the manuscript. D.P. gratefully acknowledges support from the Israeli Science Foundation grant 566/17.  

\bibliographystyle{sss-full.bst}
\bibliography{AmorphousIce.bib}

\begin{thebibliography}{139}
\providecommand{\natexlab}[1]{#1}
\parskip=0pt \itemsep=0pt \small \baselineskip=11pt

\bibitem[{\emph{{A'Hearn} et~al.}(2008)\emph{{A'Hearn}, {Belton}, {Collins},
  {Farnham}, {Feaga}, {Groussin}, {Lisse}, {Meech}, {Schultz}, and
  {Sunshine}}}]{AHearn2008}
{A'Hearn} M.~F., {Belton} M. J.~S., {Collins} S.~M., {Farnham} T.~L., {Feaga}
  L.~M., {Groussin} O., {Lisse} C.~M., {Meech} K.~J., {Schultz} P.~H., and
  {Sunshine} J.~M. (2008) \emph{{Deep Impact and sample return}}, \emph{Earth,
  Planets and Space}, \emph{60}, 61--66.

\bibitem[{\emph{{A'Hearn} et~al.}(2005)\emph{{A'Hearn}, {Belton}, and
  {Delamere}}}]{AHearn2005}
{A'Hearn} M.~F., {Belton} M.~J.~S., and {Delamere} W.~A. (2005) \emph{{Deep
  Impact: Excavating Comet Tempel 1}}, \emph{Science}, \emph{310}, 258--264.

\bibitem[{\emph{{Andersson} and {Suga}}(1994)}]{Andersson1994}
{Andersson} O. and {Suga} H. (1994) \emph{{Thermal conductivity of low-density
  amorphous ice}}, \emph{Solid State Communications}, \emph{91}, 985--988.

\bibitem[{\emph{{Andersson} and {Suga}}(2002)}]{Andersson2002}
{Andersson} O. and {Suga} H. (2002) \emph{{Thermal conductivity of amorphous
  ices}}, \emph{\prb}, \emph{65}, 140201.

\bibitem[{\emph{{Bar-Nun} et~al.}(1987)\emph{{Bar-Nun}, {Dror}, {Kochavi}, and
  {Laufer}}}]{BarNun1987}
{Bar-Nun} A., {Dror} J., {Kochavi} E., and {Laufer} D. (1987) \emph{{Amorphous
  water ice and its ability to trap gases}}, \emph{\prb}, \emph{35},
  2427--2435.

\bibitem[{\emph{{Bar-Nun} et~al.}(1985)\emph{{Bar-Nun}, {Herman}, {Laufer}, and
  {Rappaport}}}]{BarNun1985}
{Bar-Nun} A., {Herman} G., {Laufer} D., and {Rappaport} M.~L. (1985)
  \emph{{Trapping and release of gases by water ice and implications for icy
  bodies}}, \emph{\icarus}, \emph{63}, 317--332.

\bibitem[{\emph{{Bar-Nun} and {Kleinfeld}}(1989)}]{BarNun1989}
{Bar-Nun} A. and {Kleinfeld} I. (1989) \emph{{On the temperature and gas
  composition in the region of comet formation}}, \emph{\icarus}, \emph{80},
  243--253.

\bibitem[{\emph{{Bar-Nun} et~al.}(1988)\emph{{Bar-Nun}, {Kleinfeld}, and
  {Kochavi}}}]{BarNun1988}
{Bar-Nun} A., {Kleinfeld} I., and {Kochavi} E. (1988) \emph{{Trapping of gas
  mixtures by amorphous water ice}}, \emph{\prb}, \emph{38}, 7749--7754.

\bibitem[{\emph{{Bar-Nun} and {Laufer}}(2003)}]{BarNun2003}
{Bar-Nun} A. and {Laufer} D. (2003) \emph{{First experimental studies of large
  samples of gas-laden amorphous ``cometary'' ices}}, \emph{\icarus},
  \emph{161}, 157--163.

\bibitem[{\emph{{Bar-Nun} et~al.}(2007)\emph{{Bar-Nun}, {Notesco}, and
  {Owen}}}]{BarNun2007}
{Bar-Nun} A., {Notesco} G., and {Owen} T. (2007) \emph{{Trapping of N $_{2}$,
  CO and Ar in amorphous ice{\textemdash}Application to comets}},
  \emph{\icarus}, \emph{190}, 655--659.

\bibitem[{\emph{{Barkume} et~al.}(2008)\emph{{Barkume}, {Brown}, and
  {Schaller}}}]{Barkume08}
{Barkume} K.~M., {Brown} M.~E., and {Schaller} E.~L. (2008)
  \emph{{Near-Infrared Spectra of Centaurs and Kuiper Belt Objects}},
  \emph{\aj}, \emph{135}, 55--67.

\bibitem[{\emph{{Belton} et~al.}(2008)\emph{{Belton}, {Feldman}, {A'Hearn}, and
  {Carcich}}}]{Belton2008}
{Belton} M. J.~S., {Feldman} P.~D., {A'Hearn} M.~F., and {Carcich} B. (2008)
  \emph{{Cometary cryo-volcanism: Source regions and a model for the UT 2005
  June 14 and other mini-outbursts on Comet 9P/Tempel 1}}, \emph{\icarus},
  \emph{198}, 189--207.

\bibitem[{\emph{{Belton} and {Melosh}}(2009)}]{Belton2009}
{Belton} M. J.~S. and {Melosh} J. (2009) \emph{{Fluidization and multiphase
  transport of particulate cometary material as an explanation of the smooth
  terrains and repetitive outbursts on 9P/Tempel 1}}, \emph{\icarus},
  \emph{200}, 280--291.

\bibitem[{\emph{{Berdis} et~al.}(2020)\emph{{Berdis}, {Gudipati}, {Murphy}, and
  {Chanover}}}]{Berdis20}
{Berdis} J.~R., {Gudipati} M.~S., {Murphy} J.~R., and {Chanover} N.~J. (2020)
  \emph{{Europa's surface water ice crystallinity: Discrepancy between
  observations and thermophysical and particle flux modeling}}, \emph{\icarus},
  \emph{341}, 113660.

\bibitem[{\emph{{Blake} et~al.}(1991)\emph{{Blake}, {Allamandola}, {Sandford},
  {Hudgins}, and {Freund}}}]{Blake1991}
{Blake} D., {Allamandola} L., {Sandford} S., {Hudgins} D., and {Freund} F.
  (1991) \emph{{Clathrate Hydrate Formation in Amorphous Cometary Ice Analogs
  in Vacuo}}, \emph{Science}, \emph{254}, 548--551.

\bibitem[{\emph{{Brown} and {Calvin}}(2000)}]{Brown00}
{Brown} M.~E. and {Calvin} W.~M. (2000) \emph{{Evidence for Crystalline Water
  and Ammonia Ices on Pluto's Satellite Charon}}, \emph{Science}, \emph{287},
  107--109.

\bibitem[{\emph{{Cabot} et~al.}(1996)\emph{{Cabot}, {Enzian}, {Klinger}, and
  {Majolet}}}]{Cabot1996}
{Cabot} H., {Enzian} A., {Klinger} J., and {Majolet} S. (1996)
  \emph{{Complementary studies on the unexpected activity of comet
  Schwassmann-Wachmann 1}}, \emph{\planss}, \emph{44}, 1015--1020.

\bibitem[{\emph{{Capria} et~al.}(2000{\natexlab{a}})\emph{{Capria}, {Coradini},
  {De Sanctis}, and {Orosei}}}]{Capria2000AJ}
{Capria} M.~T., {Coradini} A., {De Sanctis} M.~C., and {Orosei} R.
  (2000{\natexlab{a}}) \emph{{Chiron Activity and Thermal Evolution}},
  \emph{\aj}, \emph{119}, 3112--3118.

\bibitem[{\emph{{Capria} et~al.}(2000{\natexlab{b}})\emph{{Capria}, {Coradini},
  {De Sanctis}, and {Orosei}}}]{Capria2000}
{Capria} M.~T., {Coradini} A., {De Sanctis} M.~C., and {Orosei} R.
  (2000{\natexlab{b}}) \emph{{CO emission mechanisms in C/1995 O1
  (Hale-Bopp)}}, \emph{\aap}, \emph{357}, 359--366.

\bibitem[{\emph{{Choukroun} et~al.}(2020)\emph{{Choukroun}, {Altwegg},
  {K{\"u}hrt}, {Biver}, {Bockel{\'e}e-Morvan}, {Dra{\.z}kowska}, {H{\'e}rique},
  {Hilchenbach}, {Marschall}, {P{\"a}tzold}, {Taylor}, and
  {Thomas}}}]{Choukroun20}
{Choukroun} M., {Altwegg} K., {K{\"u}hrt} E., {Biver} N., {Bockel{\'e}e-Morvan}
  D., {Dra{\.z}kowska} J., {H{\'e}rique} A., {Hilchenbach} M., {Marschall} R.,
  {P{\"a}tzold} M., {Taylor} M. G.~G.~T., and {Thomas} N. (2020)
  \emph{{Dust-to-Gas and Refractory-to-Ice Mass Ratios of Comet
  67P/Churyumov-Gerasimenko from Rosetta Observations}}, \emph{\ssr},
  \emph{216}, 44.

\bibitem[{\emph{{Ciesla}}(2014)}]{Ciesla14}
{Ciesla} F.~J. (2014) \emph{{The Phases of Water Ice in the Solar Nebula}},
  \emph{\apjl}, \emph{784}, L1.

\bibitem[{\emph{{Ciesla} et~al.}(2018)\emph{{Ciesla}, {Krijt}, {Yokochi}, and
  {Sandford}}}]{Ciesla18}
{Ciesla} F.~J., {Krijt} S., {Yokochi} R., and {Sandford} S. (2018) \emph{{The
  Efficiency of Noble Gas Trapping in Astrophysical Environments}},
  \emph{\apj}, \emph{867}, 146.

\bibitem[{\emph{{Cooper} et~al.}(2003)\emph{{Cooper}, {Christian},
  {Richardson}, and {Wang}}}]{Cooper03}
{Cooper} J.~F., {Christian} E.~R., {Richardson} J.~D., and {Wang} C. (2003)
  \emph{{Proton Irradiation of Centaur, Kuiper Belt, and Oort Cloud Objects at
  Plasma to Cosmic Ray Energy}}, \emph{Earth Moon and Planets}, \emph{92},
  261--277.

\bibitem[{\emph{{Cooper} et~al.}(2001)\emph{{Cooper}, {Johnson}, {Mauk},
  {Garrett}, and {Gehrels}}}]{Cooper01}
{Cooper} J.~F., {Johnson} R.~E., {Mauk} B.~H., {Garrett} H.~B., and {Gehrels}
  N. (2001) \emph{{Energetic Ion and Electron Irradiation of the Icy Galilean
  Satellites}}, \emph{\icarus}, \emph{149}, 133--159.

\bibitem[{\emph{{Cuppen} and {Herbst}}(2007)}]{Cuppen2007}
{Cuppen} H.~M. and {Herbst} E. (2007) \emph{{Simulation of the Formation and
  Morphology of Ice Mantles on Interstellar Grains}}, \emph{\apj}, \emph{668},
  294--309.

\bibitem[{\emph{{Davidsson}}(2021)}]{Davidsson2021}
{Davidsson} B. J.~R. (2021) \emph{{Thermophysical evolution of planetesimals in
  the primordial disc}}, \emph{\mnras}, \emph{505}, 5654--5685.

\bibitem[{\emph{{Davies} et~al.}(1997)\emph{{Davies}, {Roush}, {Cruikshank},
  {Bartholomew}, {Geballe}, {Owen}, and {de Bergh}}}]{Davies97}
{Davies} J.~K., {Roush} T.~L., {Cruikshank} D.~P., {Bartholomew} M.~J.,
  {Geballe} T.~R., {Owen} T., and {de Bergh} C. (1997) \emph{{The Detection of
  Water Ice in Comet Hale-Bopp}}, \emph{\icarus}, \emph{127}, 238--245.

\bibitem[{\emph{{Enzian} et~al.}(1997)\emph{{Enzian}, {Cabot}, and
  {Klinger}}}]{Enzian1997}
{Enzian} A., {Cabot} H., and {Klinger} J. (1997) \emph{{A 2 1/2 D thermodynamic
  model of cometary nuclei. I. Application to the activity of comet
  29P/Schwassmann-Wachmann 1.}}, \emph{\aap}, \emph{319}, 995--1006.

\bibitem[{\emph{{Enzian} et~al.}(1998)\emph{{Enzian}, {Cabot}, and
  {Klinger}}}]{Enzian1998}
{Enzian} A., {Cabot} H., and {Klinger} J. (1998) \emph{{Simulation of the water
  and carbon monoxide production rates of comet Hale-Bopp using a quasi 3-D
  nucleus model}}, \emph{\planss}, \emph{46}, 851--858.

\bibitem[{\emph{{Espinasse} et~al.}(1991)\emph{{Espinasse}, {Klinger}, {Ritz},
  and {Schmitt}}}]{Espinasse1991}
{Espinasse} S., {Klinger} J., {Ritz} C., and {Schmitt} B. (1991)
  \emph{{Modeling of the thermal behavior and of the chemical differentiation
  of cometary nuclei}}, \emph{\icarus}, \emph{92}, 350--365.

\bibitem[{\emph{{Farnham} et~al.}(2021)\emph{{Farnham}, {Kelley}, and
  {Bauer}}}]{Farnham21}
{Farnham} T.~L., {Kelley} M. S.~P., and {Bauer} J.~M. (2021) \emph{{Early
  Activity in Comet C/2014 UN271 Bernardinelli-Bernstein as Observed by TESS}},
  \emph{PSJ}, \emph{2}, 236.

\bibitem[{\emph{{Feldman} et~al.}(2006)\emph{{Feldman}, {Lupu}, {McCandliss},
  {Weaver}, {A'Hearn}, {Belton}, and {Meech}}}]{Feldman2006}
{Feldman} P.~D., {Lupu} R.~E., {McCandliss} S.~R., {Weaver} H.~A., {A'Hearn}
  M.~F., {Belton} M. J.~S., and {Meech} K.~J. (2006) \emph{{Carbon Monoxide in
  Comet 9P/Tempel 1 before and after the Deep Impact Encounter}}, \emph{\apjl},
  \emph{647}, L61--L64.

\bibitem[{\emph{{Festou} et~al.}(2001)\emph{{Festou}, {Gunnarsson}, {Rickman},
  {Winnberg}, and {Tancredi}}}]{Festou01}
{Festou} M.~C., {Gunnarsson} M., {Rickman} H., {Winnberg} A., and {Tancredi} G.
  (2001) \emph{{The Activity of Comet 29P/Schwassmann-Wachmann 1 Monitored
  through Its CO J=2-->1 Radio Line}}, \emph{\icarus}, \emph{150}, 140--150.

\bibitem[{\emph{{Fink} and {Larson}}(1975)}]{Fink75}
{Fink} U. and {Larson} H.~P. (1975) \emph{{Temperature Dependence of the
  Water-Ice Spectrum between 1 and 4 Microns: Application to Europa, Ganymede
  and Saturn's Rings}}, \emph{\icarus}, \emph{24}, 411--420.

\bibitem[{\emph{{Fulle} et~al.}(2019)\emph{{Fulle}, {Blum}, {Green},
  {Gundlach}, {Herique}, {Moreno}, {Mottola}, {Rotundi}, and
  {Snodgrass}}}]{Fulle19}
{Fulle} M., {Blum} J., {Green} S.~F., {Gundlach} B., {Herique} A., {Moreno} F.,
  {Mottola} S., {Rotundi} A., and {Snodgrass} C. (2019) \emph{{The
  refractory-to-ice mass ratio in comets}}, \emph{\mnras}, \emph{482},
  3326--3340.

\bibitem[{\emph{{Ghormley}}(1968)}]{Ghormley1968}
{Ghormley} J.~A. (1968) \emph{{Enthalpy Changes and Heat-Capacity Changes in
  the Transformations from High-Surface-Area Amorphous Ice to Stable Hexagonal
  Ice}}, \emph{\jcp}, \emph{48}, 503--508.

\bibitem[{\emph{{Giauque} and {Stout}}(1936)}]{Giauque1936}
{Giauque} W.~F. and {Stout} J.~W. (1936) \emph{{The entropy of water and third
  law of thermodynamics. The heat capacity of ice from 15 to 273K}},
  \emph{Journal of the American Chemical Society}, \emph{58}, 1144--1144.

\bibitem[{\emph{{Golabek} and {Jutzi}}(2021)}]{Golabek21}
{Golabek} G.~J. and {Jutzi} M. (2021) \emph{{Modification of icy planetesimals
  by early thermal evolution and collisions: Constraints for formation time and
  initial size of comets and small KBOs}}, \emph{\icarus}, \emph{363}, 114437.

\bibitem[{\emph{{Gonz{\'a}lez} et~al.}(2008)\emph{{Gonz{\'a}lez},
  {Guti{\'e}rrez}, {Lara}, and {Rodrigo}}}]{Gonzalez2008}
{Gonz{\'a}lez} M., {Guti{\'e}rrez} P.~J., {Lara} L.~M., and {Rodrigo} R. (2008)
  \emph{{Evolution of the crystallization front in cometary models. Effect of
  the net energy released during crystallization}}, \emph{\aap}, \emph{486},
  331--340.

\bibitem[{\emph{{Gourgeot} et~al.}(2016)\emph{{Gourgeot}, {Carry}, {Dumas},
  {Vachier}, {Merlin}, {Lacerda}, {Barucci}, and {Berthier}}}]{Gourgeot16}
{Gourgeot} F., {Carry} B., {Dumas} C., {Vachier} F., {Merlin} F., {Lacerda} P.,
  {Barucci} M.~A., and {Berthier} J. (2016) \emph{{Near-infrared spatially
  resolved spectroscopy of (136108) Haumea's multiple system}}, \emph{\aap},
  \emph{593}, A19.

\bibitem[{\emph{{Grundy} et~al.}(1999)\emph{{Grundy}, {Buie}, {Stansberry},
  {Spencer}, and {Schmitt}}}]{Grundy99}
{Grundy} W.~M., {Buie} M.~W., {Stansberry} J.~A., {Spencer} J.~R., and
  {Schmitt} B. (1999) \emph{{Near-Infrared Spectra of Icy Outer Solar System
  Surfaces: Remote Determination of H $_{2}$O Ice Temperatures}},
  \emph{\icarus}, \emph{142}, 536--549.

\bibitem[{\emph{{Grundy} and {Schmitt}}(1998)}]{Grundy98}
{Grundy} W.~M. and {Schmitt} B. (1998) \emph{{The temperature-dependent
  near-infrared absorption spectrum of hexagonal H$_{2}$O ice}}, \emph{\jgr},
  \emph{103}, 25809--25822.

\bibitem[{\emph{{Grundy} et~al.}(2006)\emph{{Grundy}, {Young}, {Spencer},
  {Johnson}, {Young}, and {Buie}}}]{Grundy06}
{Grundy} W.~M., {Young} L.~A., {Spencer} J.~R., {Johnson} R.~E., {Young} E.~F.,
  and {Buie} M.~W. (2006) \emph{{Distributions of H $_{2}$O and CO $_{2}$ ices
  on Ariel, Umbriel, Titania, and Oberon from IRTF/SpeX observations}},
  \emph{\icarus}, \emph{184}, 543--555.

\bibitem[{\emph{{Gudipati} et~al.}(2015{\natexlab{a}})\emph{{Gudipati}, {Abou
  Mrad}, {Blum}, {Charnley}, {Chiavassa}, {Cordiner}, {Mousis}, {Danger},
  {Duvernay}, {Gundlach}, {Hartogh}, {Marboeuf}, {Simonia}, {Simonia},
  {Theul{\'e}}, and {Yang}}}]{Gudipati15}
{Gudipati} M.~S., {Abou Mrad} N., {Blum} J., {Charnley} S.~B., {Chiavassa} T.,
  {Cordiner} M.~A., {Mousis} O., {Danger} G., {Duvernay} F., {Gundlach} B.,
  {Hartogh} P., {Marboeuf} U., {Simonia} I., {Simonia} T., {Theul{\'e}} P., and
  {Yang} R. (2015{\natexlab{a}}) \emph{{Laboratory Studies Towards
  Understanding Comets}}, \emph{\ssr}, \emph{197}, 101--150.

\bibitem[{\emph{{Gudipati} et~al.}(2015{\natexlab{b}})\emph{{Gudipati}, {Abou
  Mrad}, {Blum}, {Charnley}, {Chiavassa}, {Cordiner}, {Mousis}, {Danger},
  {Duvernay}, {Gundlach}, {Hartogh}, {Marboeuf}, {Simonia}, {Simonia},
  {Theul{\'e}}, and {Yang}}}]{Gudipati2015}
{Gudipati} M.~S., {Abou Mrad} N., {Blum} J., {Charnley} S.~B., {Chiavassa} T.,
  {Cordiner} M.~A., {Mousis} O., {Danger} G., {Duvernay} F., {Gundlach} B.,
  {Hartogh} P., {Marboeuf} U., {Simonia} I., {Simonia} T., {Theul{\'e}} P., and
  {Yang} R. (2015{\natexlab{b}}) \emph{{Laboratory Studies Towards
  Understanding Comets}}, \emph{\ssr}, \emph{197}, 101--150.

\bibitem[{\emph{{Guilbert} et~al.}(2009)\emph{{Guilbert}, {Alvarez-Candal},
  {Merlin}, {Barucci}, {Dumas}, {de Bergh}, and {Delsanti}}}]{Guilbert09}
{Guilbert} A., {Alvarez-Candal} A., {Merlin} F., {Barucci} M.~A., {Dumas} C.,
  {de Bergh} C., and {Delsanti} A. (2009) \emph{{ESO-Large Program on TNOs:
  Near-infrared spectroscopy with SINFONI}}, \emph{\icarus}, \emph{201},
  272--283.

\bibitem[{\emph{{Guilbert-Lepoutre}}(2012)}]{Guilbert12}
{Guilbert-Lepoutre} A. (2012) \emph{{Survival of Amorphous Water Ice on
  Centaurs}}, \emph{\aj}, \emph{144}, 97.

\bibitem[{\emph{{Guilbert-Lepoutre} and {Jewitt}}(2011)}]{Guilbert2011}
{Guilbert-Lepoutre} A. and {Jewitt} D. (2011) \emph{{Thermal Shadows and
  Compositional Structure in Comet Nuclei}}, \emph{\apj}, \emph{743}, 31.

\bibitem[{\emph{{Guilbert-Lepoutre} et~al.}(2016)\emph{{Guilbert-Lepoutre},
  {Rosenberg}, {Prialnik}, and {Besse}}}]{Guilbert2016}
{Guilbert-Lepoutre} A., {Rosenberg} E.~D., {Prialnik} D., and {Besse} S. (2016)
  \emph{{Modelling the evolution of a comet subsurface: implications for
  67P/Churyumov-Gerasimenko}}, \emph{\mnras}, \emph{462}, S146--S155.

\bibitem[{\emph{{Gunnarsson} et~al.}(2008)\emph{{Gunnarsson},
  {Bockel{\'e}e-Morvan}, {Biver}, {Crovisier}, and {Rickman}}}]{Gunnarsson08}
{Gunnarsson} M., {Bockel{\'e}e-Morvan} D., {Biver} N., {Crovisier} J., and
  {Rickman} H. (2008) \emph{{Mapping the carbon monoxide coma of comet
  29P/Schwassmann-Wachmann 1}}, \emph{\aap}, \emph{484}, 537--546.

\bibitem[{\emph{{Halukeerthi} et~al.}(2020)\emph{{Halukeerthi}, {Shephard},
  {Talewar}, {Evans}, {Rosu-Finsen}, and {Salzmann}}}]{Hakukeerthi20}
{Halukeerthi} S.~O., {Shephard} J.~J., {Talewar} S.~K., {Evans} J. S.~O.,
  {Rosu-Finsen} A., and {Salzmann} C.~G. (2020) \emph{{Amorphous Mixtures of
  Ice and C60 Fullerene}}, \emph{Journal of Physical Chemistry A}, \emph{124},
  5015--5022.

\bibitem[{\emph{{Hansen} and {McCord}}(2004)}]{Hansen04}
{Hansen} G.~B. and {McCord} T.~B. (2004) \emph{{Amorphous and crystalline ice
  on the Galilean satellites: A balance between thermal and radiolytic
  processes}}, \emph{Journal of Geophysical Research (Planets)}, \emph{109},
  E01012.

\bibitem[{\emph{{Haruyama} et~al.}(1993)\emph{{Haruyama}, {Yamamoto},
  {Mizutani}, and {Greenberg}}}]{Haruyama1993}
{Haruyama} J., {Yamamoto} T., {Mizutani} H., and {Greenberg} J.~M. (1993)
  \emph{{Thermal history of comets during residence in the Oort cloud: Effect
  of radiogenic heating in combination with the very low thermal conductivity
  of amorphous ice}}, \emph{\jgr}, \emph{98}, 15079--15090.

\bibitem[{\emph{{He} et~al.}(2019)\emph{{He}, {Clements}, {Emtiaz}, {Toriello},
  {Garrod}, and {Vidali}}}]{He2019}
{He} J., {Clements} A.~R., {Emtiaz} S., {Toriello} F., {Garrod} R.~T., and
  {Vidali} G. (2019) \emph{{The Effective Surface Area of Amorphous Solid Water
  Measured by the Infrared Absorption of Carbon Monoxide}}, \emph{\apj},
  \emph{878}, 94.

\bibitem[{\emph{{Herman} and {Weissman}}(1987)}]{Herman1987}
{Herman} G. and {Weissman} P.~R. (1987) \emph{{Numerical simulation of cometary
  nuclei III. Internal temperatures of cometary nuclei}}, \emph{\icarus},
  \emph{69}, 314--328.

\bibitem[{\emph{{Hillman} and {Prialnik}}(2012)}]{Hillman2012}
{Hillman} Y. and {Prialnik} D. (2012) \emph{{A quasi 3-D model of an outburst
  pattern that explains the behavior of Comet 17P/Holmes}}, \emph{\icarus},
  \emph{221}, 147--159.

\bibitem[{\emph{{Hsieh} et~al.}(2010)\emph{{Hsieh}, {Fitzsimmons}, {Joshi},
  {Christian}, and {Pollacco}}}]{Hsieh10}
{Hsieh} H.~H., {Fitzsimmons} A., {Joshi} Y., {Christian} D., and {Pollacco}
  D.~L. (2010) \emph{{SuperWASP observations of the 2007 outburst of Comet
  17P/Holmes}}, \emph{\mnras}, \emph{407}, 1784--1800.

\bibitem[{\emph{{Ishiguro} et~al.}(2014)\emph{{Ishiguro}, {Jewitt}, {Hanayama},
  {Usui}, {Sekiguchi}, {Yanagisawa}, {Kuroda}, {Yoshida}, {Ohta}, {Kawai},
  {Miyaji}, {Fukushima}, and {Watanabe}}}]{Ishiguro14}
{Ishiguro} M., {Jewitt} D., {Hanayama} H., {Usui} F., {Sekiguchi} T.,
  {Yanagisawa} K., {Kuroda} D., {Yoshida} M., {Ohta} K., {Kawai} N., {Miyaji}
  T., {Fukushima} H., and {Watanabe} J.-i. (2014) \emph{{Outbursting Comet
  P/2010 V1 (Ikeya-Murakami): A Miniature Comet Holmes}}, \emph{\apj},
  \emph{787}, 55.

\bibitem[{\emph{{Jenniskens} and {Blake}}(1994)}]{Jenniskens94}
{Jenniskens} P. and {Blake} D.~F. (1994) \emph{{Structural Transitions in
  Amorphous Water Ice and Astrophysical Implications}}, \emph{Science},
  \emph{265}, 753--756.

\bibitem[{\emph{{Jewitt}}(1990)}]{Jewitt90}
{Jewitt} D. (1990) \emph{{The Persistent Coma of Comet P/Schwassmann-Wachmann
  1}}, \emph{\apj}, \emph{351}, 277.

\bibitem[{\emph{{Jewitt}}(2015)}]{Jewitt15}
{Jewitt} D. (2015) \emph{{Color Systematics of Comets and Related Bodies}},
  \emph{\aj}, \emph{150}, 201.

\bibitem[{\emph{{Jewitt} et~al.}(2017)\emph{{Jewitt}, {Hui}, {Mutchler},
  {Weaver}, {Li}, and {Agarwal}}}]{Jewitt17}
{Jewitt} D., {Hui} M.-T., {Mutchler} M., {Weaver} H., {Li} J., and {Agarwal} J.
  (2017) \emph{{A Comet Active Beyond the Crystallization Zone}}, \emph{\apjl},
  \emph{847}, L19.

\bibitem[{\emph{{Jewitt} and {Kim}}(2020)}]{Jewitt20}
{Jewitt} D. and {Kim} Y. (2020) \emph{{Outbursting Quasi-Hilda Asteroid P/2010
  H2 (Vales)}}, \emph{Plan. Sci. Journal}, \emph{1}, 77.

\bibitem[{\emph{{Jewitt} et~al.}(2016)\emph{{Jewitt}, {Mutchler}, {Weaver},
  {Hui}, {Agarwal}, {Ishiguro}, {Kleyna}, {Li}, {Meech}, {Micheli},
  {Wainscoat}, and {Weryk}}}]{Jewitt16}
{Jewitt} D., {Mutchler} M., {Weaver} H., {Hui} M.-T., {Agarwal} J., {Ishiguro}
  M., {Kleyna} J., {Li} J., {Meech} K., {Micheli} M., {Wainscoat} R., and
  {Weryk} R. (2016) \emph{{Fragmentation Kinematics in Comet
  332P/Ikeya-Murakami}}, \emph{\apjl}, \emph{829}, L8.

\bibitem[{\emph{{Jewitt} et~al.}(1996)\emph{{Jewitt}, {Senay}, and
  {Matthews}}}]{Jewitt1996}
{Jewitt} D., {Senay} M., and {Matthews} H. (1996) \emph{{Observations of Carbon
  Monoxide in Comet Hale-Bopp}}, \emph{Science}, \emph{271}, 1110--1113.

\bibitem[{\emph{{Jewitt} and {Luu}}(2004)}]{Jewitt04}
{Jewitt} D.~C. and {Luu} J. (2004) \emph{{Crystalline water ice on the Kuiper
  belt object (50000) Quaoar}}, \emph{\nat}, \emph{432}, 731--733.

\bibitem[{\emph{{Jorda} et~al.}(2016)\emph{{Jorda}, {Gaskell}, {Capanna},
  {Hviid}, {Lamy}, {{\v{D}}urech}, {Faury}, {Groussin}, {Guti{\'e}rrez},
  {Jackman}, {Keihm}, {Keller}, {Knollenberg}, {K{\"u}hrt}, {Marchi},
  {Mottola}, {Palmer}, {Schloerb}, {Sierks}, {Vincent}, {A'Hearn}, {Barbieri},
  {Rodrigo}, {Koschny}, {Rickman}, {Barucci}, {Bertaux}, {Bertini},
  {Cremonese}, {Da Deppo}, {Davidsson}, {Debei}, {De Cecco}, {Fornasier},
  {Fulle}, {G{\"u}ttler}, {Ip}, {Kramm}, {K{\"u}ppers}, {Lara}, {Lazzarin},
  {Lopez Moreno}, {Marzari}, {Naletto}, {Oklay}, {Thomas}, {Tubiana}, and
  {Wenzel}}}]{Jorda2016}
{Jorda} L., {Gaskell} R., {Capanna} C., {Hviid} S., {Lamy} P., {{\v{D}}urech}
  J., {Faury} G., {Groussin} O., {Guti{\'e}rrez} P., {Jackman} C., {Keihm}
  S.~J., {Keller} H.~U., {Knollenberg} J., {K{\"u}hrt} E., {Marchi} S.,
  {Mottola} S., {Palmer} E., {Schloerb} F.~P., {Sierks} H., {Vincent} J.~B.,
  {A'Hearn} M.~F., {Barbieri} C., {Rodrigo} R., {Koschny} D., {Rickman} H.,
  {Barucci} M.~A., {Bertaux} J.~L., {Bertini} I., {Cremonese} G., {Da Deppo}
  V., {Davidsson} B., {Debei} S., {De Cecco} M., {Fornasier} S., {Fulle} M.,
  {G{\"u}ttler} C., {Ip} W.~H., {Kramm} J.~R., {K{\"u}ppers} M., {Lara} L.~M.,
  {Lazzarin} M., {Lopez Moreno} J.~J., {Marzari} F., {Naletto} G., {Oklay} N.,
  {Thomas} N., {Tubiana} C., and {Wenzel} K.~P. (2016) \emph{{The global shape,
  density and rotation of Comet 67P/Churyumov-Gerasimenko from preperihelion
  Rosetta/OSIRIS observations}}, \emph{\icarus}, \emph{277}, 257--278.

\bibitem[{\emph{{Kawakita} et~al.}(2006)\emph{{Kawakita}, {Ootsubo}, {Furusho},
  and {Watanabe}}}]{Kawakita06}
{Kawakita} H., {Ootsubo} T., {Furusho} R., and {Watanabe} J.-I. (2006)
  \emph{{Crystallinity and temperature of icy grains in comet C/2002 T7
  (LINEAR)}}, \emph{Advances in Space Research}, \emph{38}, 1968--1971.

\bibitem[{\emph{{Kawakita} et~al.}(2004)\emph{{Kawakita}, {Watanabe},
  {Ootsubo}, {Nakamura}, {Fuse}, {Takato}, {Sasaki}, and
  {Sasaki}}}]{Kawakita04}
{Kawakita} H., {Watanabe} J.-i., {Ootsubo} T., {Nakamura} R., {Fuse} T.,
  {Takato} N., {Sasaki} S., and {Sasaki} T. (2004) \emph{{Evidence of Icy
  Grains in Comet C/2002 T7 (LINEAR) at 3.52 AU}}, \emph{\apjl}, \emph{601},
  L191--L194.

\bibitem[{\emph{{Klinger}}(1980)}]{Klinger1980}
{Klinger} J. (1980) \emph{{Influence of a Phase Transition of Ice on the Heat
  and Mass Balance of Comets}}, \emph{Science}, \emph{209}, 271--272.

\bibitem[{\emph{{Klinger}}(1981)}]{Klinger1981}
{Klinger} J. (1981) \emph{{Some consequences of a phase transition of water ice
  on the heat balance of comet nuclei}}, \emph{\icarus}, \emph{47}, 320--324.

\bibitem[{\emph{{Kossacki} and {Szutowicz}}(2010)}]{Kossacki2010}
{Kossacki} K.~J. and {Szutowicz} S. (2010) \emph{{Crystallization of ice in
  Comet 17P/Holmes: Probably not responsible for the explosive 2007
  megaburst}}, \emph{\icarus}, \emph{207}, 320--340.

\bibitem[{\emph{{Kouchi} et~al.}(1992)\emph{{Kouchi}, {Greenberg}, {Yamamoto},
  {Mukai}, and {Xing}}}]{Kouchi1992}
{Kouchi} A., {Greenberg} J.~M., {Yamamoto} T., {Mukai} T., and {Xing} Z.~F.
  (1992) in \emph{Asteroids, Comets, Meteors 1991} (A.~W. {Harris} and
  E.~{Bowell}, eds.), p. 325.

\bibitem[{\emph{{Kouchi} et~al.}(2016)\emph{{Kouchi}, {Hama}, {Kimura},
  {Hidaka}, {Escribano}, and {Watanabe}}}]{Kouchi2016}
{Kouchi} A., {Hama} T., {Kimura} Y., {Hidaka} H., {Escribano} R., and
  {Watanabe} N. (2016) \emph{{Matrix sublimation method for the formation of
  high-density amorphous ice}}, \emph{Chemical Physics Letters}, \emph{658},
  287--292.

\bibitem[{\emph{{Kouchi} and {Sirono}}(2001)}]{Kouchi2001}
{Kouchi} A. and {Sirono} S.-i. (2001) \emph{{Crystallization heat of impure
  amorphous H$_{2}$O ice}}, \emph{\grl}, \emph{28}, 827--830.

\bibitem[{\emph{{Li} et~al.}(2011)\emph{{Li}, {Jewitt}, {Clover}, and
  {Jackson}}}]{Li11}
{Li} J., {Jewitt} D., {Clover} J.~M., and {Jackson} B.~V. (2011)
  \emph{{Outburst of Comet 17P/Holmes Observed with the Solar Mass Ejection
  Imager}}, \emph{\apj}, \emph{728}, 31.

\bibitem[{\emph{{Li} et~al.}(2020)\emph{{Li}, {Jewitt}, {Mutchler}, {Agarwal},
  and {Weaver}}}]{Li20}
{Li} J., {Jewitt} D., {Mutchler} M., {Agarwal} J., and {Weaver} H. (2020)
  \emph{{Hubble Space Telescope Search for Activity in High-perihelion
  Objects}}, \emph{\aj}, \emph{159}, 209.

\bibitem[{\emph{{Ligier} et~al.}(2016)\emph{{Ligier}, {Poulet}, {Carter},
  {Brunetto}, and {Gourgeot}}}]{ligier16}
{Ligier} N., {Poulet} F., {Carter} J., {Brunetto} R., and {Gourgeot} F. (2016)
  \emph{{VLT/SINFONI Observations of Europa: New Insights into the Surface
  Composition}}, \emph{\aj}, \emph{151}, 163.

\bibitem[{\emph{{Lignell} and {Gudipati}}(2015)}]{Lignell2015}
{Lignell} A. and {Gudipati} M.~S. (2015) \emph{{Mixing of the Immiscible:
  Hydrocarbons in Water-Ice near the Ice Crystallization Temperature}},
  \emph{Journal of Physical Chemistry A}, \emph{119}, 2607--2613.

\bibitem[{\emph{{Luspay-Kuti} et~al.}(2016)\emph{{Luspay-Kuti}, {Mousis},
  {H{\"a}ssig}, {Fuselier}, {Lunine}, {Marty}, {Mandt}, {Wurz}, and
  {Rubin}}}]{Luspay16}
{Luspay-Kuti} A., {Mousis} O., {H{\"a}ssig} M., {Fuselier} S.~A., {Lunine}
  J.~I., {Marty} B., {Mandt} K.~E., {Wurz} P., and {Rubin} M. (2016) \emph{{The
  presence of clathrates in comet 67P/Churyumov-Gerasimenko}}, \emph{Science
  Advances}, \emph{2}, 1501781.

\bibitem[{\emph{{Manca} and {Allouche}}(2001)}]{Manca2001}
{Manca} C. and {Allouche} A. (2001) \emph{{Quantum study of the adsorption of
  small molecules on ice: The infrared frequency of the surface hydroxyl group
  and the vibrational stark effect}}, \emph{\jcp}, \emph{114}, 4226--4234.

\bibitem[{\emph{{Marboeuf} et~al.}(2012)\emph{{Marboeuf}, {Schmitt}, {Petit},
  {Mousis}, and {Fray}}}]{Marboeuf2012}
{Marboeuf} U., {Schmitt} B., {Petit} J.~M., {Mousis} O., and {Fray} N. (2012)
  \emph{{A cometary nucleus model taking into account all phase changes of
  water ice: amorphous, crystalline, and clathrate}}, \emph{\aap}, \emph{542},
  A82.

\bibitem[{\emph{{Martin} et~al.}(2002)\emph{{Martin}, {Manca}, and
  {Roubin}}}]{Martin2002}
{Martin} C., {Manca} C., and {Roubin} P. (2002) \emph{{Adsorption of small
  molecules on amorphous ice: volumetric and FT-IR isotherm co-measurements.
  Part II. The case of CO}}, \emph{Surface Science}, \emph{502-503}, 280--284.

\bibitem[{\emph{{Mastrapa} and {Brown}}(2006)}]{Mastrapa06}
{Mastrapa} R. M.~E. and {Brown} R.~H. (2006) \emph{{Ion irradiation of
  crystalline H $_{2}$O-ice: Effect on the 1.65-{\ensuremath{\mu}}m band}},
  \emph{\icarus}, \emph{183}, 207--214.

\bibitem[{\emph{{Mastrapa} et~al.}(2013)\emph{{Mastrapa}, {Grundy}, and
  {Gudipati}}}]{Mastrapa13}
{Mastrapa} R. M.~E., {Grundy} W.~M., and {Gudipati} M.~S. (2013) in
  \emph{Astrophysics and Space Science Library} (M.~S. {Gudipati} and
  J.~{Castillo-Rogez}, eds.), vol. 356 of \emph{Astrophysics and Space Science
  Library}, p. 371.

\bibitem[{\emph{{Mayer} and {Pletzer}}(1986)}]{Mayer1986}
{Mayer} E. and {Pletzer} R. (1986) \emph{{Astrophysical implications of
  amorphous ice-a microporous solid}}, \emph{\nat}, \emph{319}, 298--301.

\bibitem[{\emph{{Mazzotta Epifani} et~al.}(2009)\emph{{Mazzotta Epifani},
  {Palumbo}, and {Colangeli}}}]{Mazzota2009}
{Mazzotta Epifani} E., {Palumbo} P., and {Colangeli} L. (2009) \emph{{A survey
  on the distant activity of short period comets}}, \emph{\aap}, \emph{508},
  1031--1044.

\bibitem[{\emph{{Meech} et~al.}(2017)\emph{{Meech}, {Kleyna}, {Hainaut},
  {Micheli}, {Bauer}, {Denneau}, {Keane}, {Stephens}, {Jedicke}, {Wainscoat},
  {Weryk}, {Flewelling}, {Schunov{\'a}-Lilly}, {Magnier}, and
  {Chambers}}}]{Meech2017}
{Meech} K.~J., {Kleyna} J.~T., {Hainaut} O., {Micheli} M., {Bauer} J.,
  {Denneau} L., {Keane} J.~V., {Stephens} H., {Jedicke} R., {Wainscoat} R.,
  {Weryk} R., {Flewelling} H., {Schunov{\'a}-Lilly} E., {Magnier} E., and
  {Chambers} K.~C. (2017) \emph{{CO-driven Activity in Comet C/2017 K2
  (PANSTARRS)}}, \emph{\apjl}, \emph{849}, L8.

\bibitem[{\emph{{Meech} et~al.}(2009)\emph{{Meech}, {Pittichov{\'a}},
  {Bar-Nun}, {Notesco}, {Laufer}, {Hainaut}, {Lowry}, {Yeomans}, and
  {Pitts}}}]{Meech2009}
{Meech} K.~J., {Pittichov{\'a}} J., {Bar-Nun} A., {Notesco} G., {Laufer} D.,
  {Hainaut} O.~R., {Lowry} S.~C., {Yeomans} D.~K., and {Pitts} M. (2009)
  \emph{{Activity of comets at large heliocentric distances pre-perihelion}},
  \emph{\icarus}, \emph{201}, 719--739.

\bibitem[{\emph{{Merk} and {Prialnik}}(2003)}]{Merk2003}
{Merk} R. and {Prialnik} D. (2003) \emph{{Early Thermal and Structural
  Evolution of Small Bodies in the Trans-Neptunian Zone}}, \emph{Earth Moon and
  Planets}, \emph{92}, 359--374.

\bibitem[{\emph{{Merk} and {Prialnik}}(2006)}]{Merk2006}
{Merk} R. and {Prialnik} D. (2006) \emph{{Combined modeling of thermal
  evolution and accretion of trans-neptunian objects{\textemdash}Occurrence of
  high temperatures and liquid water}}, \emph{\icarus}, \emph{183}, 283--295.

\bibitem[{\emph{{Merlin} et~al.}(2007)\emph{{Merlin}, {Guilbert}, {Dumas},
  {Barucci}, {de Bergh}, and {Vernazza}}}]{Merlin07}
{Merlin} F., {Guilbert} A., {Dumas} C., {Barucci} M.~A., {de Bergh} C., and
  {Vernazza} P. (2007) \emph{{Properties of the icy surface of the TNO 136108
  (2003 EL\{61\})}}, \emph{\aap}, \emph{466}, 1185--1188.

\bibitem[{\emph{{Mousis} et~al.}(2015)\emph{{Mousis}, {Guilbert-Lepoutre},
  {Brugger}, {Jorda}, {Kargel}, {Bouquet}, {Auger}, {Lamy}, {Vernazza},
  {Thomas}, and {Sierks}}}]{Mousis2015}
{Mousis} O., {Guilbert-Lepoutre} A., {Brugger} B., {Jorda} L., {Kargel} J.~S.,
  {Bouquet} A., {Auger} A.~T., {Lamy} P., {Vernazza} P., {Thomas} N., and
  {Sierks} H. (2015) \emph{{Pits Formation from Volatile Outgassing on
  67P/Churyumov-Gerasimenko}}, \emph{\apjl}, \emph{814}, L5.

\bibitem[{\emph{Nelson et~al.}(2016)\emph{Nelson, Dee, Gudipati, HorÃ¡nyi,
  James, Kempf, Munsat, Sternovsky, and Ulibarri}}]{Nelson16}
Nelson A.~O., Dee R., Gudipati M.~S., HorÃ¡nyi M., James D., Kempf S., Munsat
  T., Sternovsky Z., and Ulibarri Z. (2016) \emph{New experimental capability
  to investigate the hypervelocity micrometeoroid bombardment of cryogenic
  surfaces}, \emph{Review of Scientific Instruments}, \emph{87}, 024502.

\bibitem[{\emph{{Notesco} et~al.}(2003)\emph{{Notesco}, {Bar-Nun}, and
  {Owen}}}]{Notesco03}
{Notesco} G., {Bar-Nun} A., and {Owen} T. (2003) \emph{{Gas trapping in water
  ice at very low deposition rates and implications for comets}},
  \emph{\icarus}, \emph{162}, 183--189.

\bibitem[{\emph{{{\"O}berg} et~al.}(2011)\emph{{{\"O}berg}, {Boogert},
  {Pontoppidan}, {van den Broek}, {van Dishoeck}, {Bottinelli}, {Blake}, and
  {Evans}}}]{Oberg2011}
{{\"O}berg} K.~I., {Boogert} A.~C.~A., {Pontoppidan} K.~M., {van den Broek} S.,
  {van Dishoeck} E.~F., {Bottinelli} S., {Blake} G.~A., and {Evans} I., Neal~J.
  (2011) \emph{{The Spitzer Ice Legacy: Ice Evolution from Cores to
  Protostars}}, \emph{\apj}, \emph{740}, 109.

\bibitem[{\emph{{Patashnick}}(1974)}]{Patashnick1974}
{Patashnick} H. (1974) \emph{{Energy source for comet outbursts}}, \emph{\nat},
  \emph{250}, 313--314.

\bibitem[{\emph{{Porter} et~al.}(2010)\emph{{Porter}, {Desch}, and
  {Cook}}}]{Porter10}
{Porter} S.~B., {Desch} S.~J., and {Cook} J.~C. (2010) \emph{{Micrometeorite
  impact annealing of ice in the outer Solar System}}, \emph{\icarus},
  \emph{208}, 492--498.

\bibitem[{\emph{{Prialnik}}(1993)}]{Prialnik1993}
{Prialnik} D. (1993) \emph{{A Two-Zone Steady State Crystallization Model for
  Comets}}, \emph{\apjl}, \emph{418}, L49.

\bibitem[{\emph{{Prialnik}}(1997)}]{Prialnik1997}
{Prialnik} D. (1997) \emph{{A Model for the Distant Activity of Comet
  Hale-Bopp}}, \emph{\apjl}, \emph{478}, L107--L110.

\bibitem[{\emph{{Prialnik} and {Bar-Nun}}(1987)}]{Prialnik1987}
{Prialnik} D. and {Bar-Nun} A. (1987) \emph{{On the Evolution and Activity of
  Cometary Nuclei}}, \emph{\apj}, \emph{313}, 893.

\bibitem[{\emph{{Prialnik} and {Bar-Nun}}(1990)}]{Prialnik1990}
{Prialnik} D. and {Bar-Nun} A. (1990) \emph{{Gas Release in Comet Nuclei}},
  \emph{\apj}, \emph{363}, 274.

\bibitem[{\emph{{Prialnik} and {Bar-Nun}}(1992)}]{Prialnik1992}
{Prialnik} D. and {Bar-Nun} A. (1992) \emph{{Crystallization of amorphous ice
  as the cause of Comet P/Halley's outburst at 14 AU}}, \emph{\aap},
  \emph{258}, L9--L12.

\bibitem[{\emph{{Prialnik} et~al.}(1995)\emph{{Prialnik}, {Brosch}, and
  {Ianovici}}}]{Prialnik95}
{Prialnik} D., {Brosch} N., and {Ianovici} D. (1995) \emph{{Modelling the
  activity of 2060 Chiron}}, \emph{\mnras}, \emph{276}, 1148--1154.

\bibitem[{\emph{{Prialnik} et~al.}(1993)\emph{{Prialnik}, {Egozi}, {Ban-Nun},
  {Podolak}, and {Greenzweig}}}]{Prialnik1993a}
{Prialnik} D., {Egozi} U., {Ban-Nun} A., {Podolak} M., and {Greenzweig} Y.
  (1993) \emph{{On Pore Size and Fracture in Gas-Laden Comet Nuclei}},
  \emph{\icarus}, \emph{106}, 499--507.

\bibitem[{\emph{{Prialnik} and {Podolak}}(1995)}]{Prialnik1995}
{Prialnik} D. and {Podolak} M. (1995) \emph{{Radioactive heating of porous
  comet nuclei.}}, \emph{\icarus}, \emph{117}, 420--430.

\bibitem[{\emph{{Prialnik} et~al.}(2008)\emph{{Prialnik}, {Sarid}, {Rosenberg},
  and {Merk}}}]{Prialnik2008}
{Prialnik} D., {Sarid} G., {Rosenberg} E.~D., and {Merk} R. (2008)
  \emph{{Thermal and Chemical Evolution of Comet Nuclei and Kuiper Belt
  Objects}}, \emph{\ssr}, \emph{138}, 147--164.

\bibitem[{\emph{{Prialnik} and {Sierks}}(2017)}]{Prialnik2017}
{Prialnik} D. and {Sierks} H. (2017) \emph{{A mechanism for comet surface
  collapse as observed by Rosetta on 67P/Churyumov-Gerasimenko}},
  \emph{\mnras}, \emph{469}, S217--S221.

\bibitem[{\emph{{Reach} et~al.}(2000)\emph{{Reach}, {Sykes}, {Lien}, and
  {Davies}}}]{Reach00}
{Reach} W.~T., {Sykes} M.~V., {Lien} D., and {Davies} J.~K. (2000) \emph{{The
  Formation of Encke Meteoroids and Dust Trail}}, \emph{\icarus}, \emph{148},
  80--94.

\bibitem[{\emph{{Reach} et~al.}(2010)\emph{{Reach}, {Vaubaillon}, {Lisse},
  {Holloway}, and {Rho}}}]{Reach10}
{Reach} W.~T., {Vaubaillon} J., {Lisse} C.~M., {Holloway} M., and {Rho} J.
  (2010) \emph{{Explosion of Comet 17P/Holmes as revealed by the Spitzer Space
  Telescope}}, \emph{\icarus}, \emph{208}, 276--292.

\bibitem[{\emph{{Rosenberg} and {Prialnik}}(2007)}]{Rosenberg2007}
{Rosenberg} E.~D. and {Prialnik} D. (2007) \emph{{A fully 3-dimensional thermal
  model of a comet nucleus}}, \emph{\na}, \emph{12}, 523--532.

\bibitem[{\emph{{Samarasinha}}(2001)}]{Samarasinha01}
{Samarasinha} N.~H. (2001) \emph{{NOTE: A Model for the Breakup of Comet LINEAR
  (C/1999 S4)}}, \emph{\icarus}, \emph{154}, 540--544.

\bibitem[{\emph{{Sandford} and {Allamandola}}(1988)}]{Sanford1988}
{Sandford} S.~A. and {Allamandola} L.~J. (1988) \emph{{The condensation and
  vaporization behavior of H $_{2}$O: CO ices and implications for interstellar
  grains and cometary activity}}, \emph{\icarus}, \emph{76}, 201--224.

\bibitem[{\emph{{Sandford} and {Allamandola}}(1990)}]{Sandford1990}
{Sandford} S.~A. and {Allamandola} L.~J. (1990) \emph{{The Physical and
  Infrared Spectral Properties of CO 2 in Astrophysical Ice Analogs}},
  \emph{\apj}, \emph{355}, 357.

\bibitem[{\emph{{Schambeau} et~al.}(2021)\emph{{Schambeau}, {Fern{\'a}ndez},
  {Samarasinha}, {Womack}, {Bockel{\'e}e-Morvan}, {Lisse}, and
  {Woodney}}}]{Schambeau21}
{Schambeau} C.~A., {Fern{\'a}ndez} Y.~R., {Samarasinha} N.~H., {Womack} M.,
  {Bockel{\'e}e-Morvan} D., {Lisse} C.~M., and {Woodney} L.~M. (2021)
  \emph{{Characterization of Thermal-infrared Dust Emission and Refinements to
  the Nucleus Properties of Centaur 29P/Schwassmann-Wachmann 1}}, \emph{\pasj},
  \emph{2}, 126.

\bibitem[{\emph{{Schmitt} et~al.}(1989)\emph{{Schmitt}, {Espinasse}, {Grim},
  {Greenberg}, and {Klinger}}}]{Schmitt1989b}
{Schmitt} B., {Espinasse} S., {Grim} R.~J.~A., {Greenberg} J.~M., and {Klinger}
  J. (1989) in \emph{Physics and Mechanics of Cometary Materials} (J.~J. {Hunt}
  and T.~D. {Guyenne}, eds.), vol. 302 of \emph{ESA Special Publication}, pp.
  65--69.

\bibitem[{\emph{{Schmitt} et~al.}(1991)\emph{{Schmitt}, {Espinasse}, and
  {Klinger}}}]{Schmitt1991}
{Schmitt} B., {Espinasse} S., and {Klinger} J. (1991) \emph{{A possible
  mechanism for outbursts of comet P/Halley at large heliocentric distances}},
  \emph{Meteoritics}, \emph{26}, 392.

\bibitem[{\emph{{Schmitt} and {Klinger}}(1987)}]{Schmitt87}
{Schmitt} B. and {Klinger} J. (1987) in \emph{Diversity and Similarity of
  Comets} (E.~J. {Rolfe}, B.~{Battrick}, M.~{Ackerman}, M.~{Scherer}, and
  R.~{Reinhard}, eds.), vol. 278 of \emph{ESA Special Publication}, pp.
  613--619.

\bibitem[{\emph{{Senay} and {Jewitt}}(1994)}]{Senay94}
{Senay} M.~C. and {Jewitt} D. (1994) \emph{{Coma formation driven by carbon
  monoxide release from comet Schwassmann-Wachmann 1}}, \emph{\nat},
  \emph{371}, 229--231.

\bibitem[{\emph{{Smoluchowski}}(1981)}]{Smoluchowski1981}
{Smoluchowski} R. (1981) \emph{{Amorphous ice and the behavior of cometary
  nuclei}}, \emph{\apjl}, \emph{244}, L31--L34.

\bibitem[{\emph{{Stevenson} et~al.}(2010)\emph{{Stevenson}, {Kleyna}, and
  {Jewitt}}}]{Stevenson10}
{Stevenson} R., {Kleyna} J., and {Jewitt} D. (2010) \emph{{Transient Fragments
  in Outbursting Comet 17P/Holmes}}, \emph{\aj}, \emph{139}, 2230--2240.

\bibitem[{\emph{{Tancredi} et~al.}(1994)\emph{{Tancredi}, {Rickman}, and
  {Greenberg}}}]{Tancredi1994}
{Tancredi} G., {Rickman} H., and {Greenberg} J.~M. (1994)
  \emph{{Thermochemistry of cometary nuclei I. The Jupiter family case.}},
  \emph{\aap}, \emph{286}, 659--682.

\bibitem[{\emph{{Terai} et~al.}(2016)\emph{{Terai}, {Itoh}, {Oasa}, {Furusho},
  and {Watanabe}}}]{Terai16}
{Terai} T., {Itoh} Y., {Oasa} Y., {Furusho} R., and {Watanabe} J. (2016)
  \emph{{Photometric Measurements of H$_{2}$O Ice Crystallinity on
  Trans-Neptunian Objects}}, \emph{\apj}, \emph{827}, 65.

\bibitem[{\emph{{Trigo-Rodr{\'\i}guez}
  et~al.}(2008)\emph{{Trigo-Rodr{\'\i}guez}, {Garc{\'\i}a-Melendo},
  {Davidsson}, {S{\'a}nchez}, {Rodr{\'\i}guez}, {Lacruz}, {de Los Reyes}, and
  {Pastor}}}]{Trigo08}
{Trigo-Rodr{\'\i}guez} J.~M., {Garc{\'\i}a-Melendo} E., {Davidsson} B.~J.~R.,
  {S{\'a}nchez} A., {Rodr{\'\i}guez} D., {Lacruz} J., {de Los Reyes} J.~A., and
  {Pastor} S. (2008) \emph{{Outburst activity in comets. I. Continuous
  monitoring of comet 29P/Schwassmann-Wachmann 1}}, \emph{\aap}, \emph{485},
  599--606.

\bibitem[{\emph{{Trujillo} et~al.}(2007)\emph{{Trujillo}, {Brown}, {Barkume},
  {Schaller}, and {Rabinowitz}}}]{Trujillo07}
{Trujillo} C.~A., {Brown} M.~E., {Barkume} K.~M., {Schaller} E.~L., and
  {Rabinowitz} D.~L. (2007) \emph{{The Surface of 2003 EL$_{61}$ in the
  Near-Infrared}}, \emph{\apj}, \emph{655}, 1172--1178.

\bibitem[{\emph{{van Dishoeck} et~al.}(2013)\emph{{van Dishoeck}, {Herbst}, and
  {Neufeld}}}]{VanDishoeck2013}
{van Dishoeck} E.~F., {Herbst} E., and {Neufeld} D.~A. (2013)
  \emph{{Interstellar Water Chemistry: From Laboratory to Observations}},
  \emph{Chemical Reviews}, \emph{113}, 9043--9085.

\bibitem[{\emph{{Vernazza} et~al.}(2021)\emph{{Vernazza}, {Beck}, and
  {Ruesch}}}]{Vernazza21}
{Vernazza} P., {Beck} P., and {Ruesch} O. (2021) \emph{{Sample return of
  primitive matter from the outer Solar System}}, \emph{Experimental
  Astronomy}.

\bibitem[{\emph{{Wallis}}(1980)}]{Wallis1980}
{Wallis} M.~K. (1980) \emph{{Radiogenic melting of primordial comet
  interiors}}, \emph{\nat}, \emph{284}, 431--433.

\bibitem[{\emph{{Weissman}}(1991)}]{Weissman1991}
{Weissman} P. (1991) \emph{{Why did Halley hiccup?}}, \emph{\nat}, \emph{353},
  793--794.

\bibitem[{\emph{{West} et~al.}(1991)\emph{{West}, {Hainaut}, and
  {Smette}}}]{West1991}
{West} R.~M., {Hainaut} O., and {Smette} A. (1991) \emph{{Post-perihelion
  observations of P/Halley. III - an outburst at R = 14.3 AU}}, \emph{\aap},
  \emph{246}, L77--L80.

\bibitem[{\emph{{Westphal} et~al.}(2020)\emph{{Westphal}, {Nittler}, {Stroud},
  {Zolensky}, {Chabot}, {Dello Russo}, {Elsila}, {Sandford}, {Glavin}, {Evans},
  {Nuth}, {Sunshine}, {Vervack}, and {Weaver}}}]{Westphal20}
{Westphal} A.~J., {Nittler} L.~R., {Stroud} R., {Zolensky} M.~E., {Chabot}
  N.~L., {Dello Russo} N., {Elsila} J.~E., {Sandford} S.~A., {Glavin} D.~P.,
  {Evans} M.~E., {Nuth} J.~A., {Sunshine} J., {Vervack} J., Ronald~J., and
  {Weaver} H.~A. (2020) \emph{{Cryogenic Cometary Sample Return}}, \emph{arXiv
  e-prints}, arXiv:2009.00101.

\bibitem[{\emph{{Whipple} and {Huebner}}(1976)}]{Whipple1976}
{Whipple} F.~L. and {Huebner} W.~F. (1976) \emph{{Physical processes in
  comets.}}, \emph{\araa}, \emph{14}, 143--172.

\bibitem[{\emph{{Wierzchos} and {Womack}}(2020)}]{Wierzchos20}
{Wierzchos} K. and {Womack} M. (2020) \emph{{CO Gas and Dust Outbursts from
  Centaur 29P/Schwassmann-Wachmann}}, \emph{\aj}, \emph{159}, 136.

\bibitem[{\emph{{Wong} et~al.}(2019)\emph{{Wong}, {Brown}, {Blacksberg},
  {Ehlmann}, and {Mahjoub}}}]{Wong19}
{Wong} I., {Brown} M.~E., {Blacksberg} J., {Ehlmann} B.~L., and {Mahjoub} A.
  (2019) \emph{{Hubble Ultraviolet Spectroscopy of Jupiter Trojans}},
  \emph{\aj}, \emph{157}, 161.

\bibitem[{\emph{{Yabushita}}(1993)}]{Yabushita1993}
{Yabushita} S. (1993) \emph{{Thermal Evolution of Cometary Nuclei by
  Radioactive Heating and Possible Formation of Organic Chemicals}},
  \emph{\mnras}, \emph{260}, 819--825.

\bibitem[{\emph{{Yang} and {Sarid}}(2010)}]{Yang10}
{Yang} B. and {Sarid} G. (2010) in \emph{AAS/Division for Planetary Sciences
  Meeting Abstracts \#42}, vol.~42 of \emph{AAS/Division for Planetary Sciences
  Meeting Abstracts}, p. 5.09.

\bibitem[{\emph{{Yokochi} et~al.}(2012)\emph{{Yokochi}, {Marboeuf}, {Quirico},
  and {Schmitt}}}]{Yokochi12}
{Yokochi} R., {Marboeuf} U., {Quirico} E., and {Schmitt} B. (2012)
  \emph{{Pressure dependent trace gas trapping in amorphous water ice at 77 K:
  Implications for determining conditions of comet formation}}, \emph{\icarus},
  \emph{218}, 760--770.

\bibitem[{\emph{{Zheng} et~al.}(2009)\emph{{Zheng}, {Jewitt}, and
  {Kaiser}}}]{Zheng09}
{Zheng} W., {Jewitt} D., and {Kaiser} R.~I. (2009) \emph{{On the State of Water
  Ice on Saturn's Moon Titan and Implications to Icy Bodies in the Outer Solar
  System}}, \emph{Journal of Physical Chemistry A}, \emph{113}, 11174--11181.

\bibitem[{\emph{{Zubko} et~al.}(2017)\emph{{Zubko}, {Videen}, {Shkuratov}, and
  {Hines}}}]{Zubko17}
{Zubko} E., {Videen} G., {Shkuratov} Y., and {Hines} D.~C. (2017) \emph{{On the
  reflectance of dust in comets}}, \emph{\jqsrt}, \emph{202}, 104--113.

\end{thebibliography}

\end{document}